\documentclass{article}

\usepackage{PRIMEarxiv}
\usepackage[numbers]{natbib}
\usepackage[utf8]{inputenc} % allow utf-8 input
\usepackage[T1]{fontenc}    % use 8-bit T1 fonts
\usepackage{hyperref}       % hyperlinks
\usepackage{url}            % simple URL typesetting
\usepackage{booktabs}       % professional-quality tables
\usepackage{amsfonts}       % blackboard math symbols
\usepackage{nicefrac}       % compact symbols for 1/2, etc.
\usepackage{microtype}      % microtypography
\usepackage{lipsum}
\usepackage{fancyhdr}       % header
\usepackage[pdftex]{graphicx}       % graphics
\graphicspath{{media/}}     % organize your images and other figures under media/ folder
\usepackage{physics}
\usepackage{xcolor}
\usepackage{algorithm}
\usepackage{algpseudocode}
\usepackage{tikz}
\usepackage{multirow}
\usepackage{colortbl}
\usepackage{subcaption}
\usepackage[font=footnotesize,labelfont=bf]{caption}
\usepackage{kotex}
\usepackage{float}
\usepackage{caption}
\usepackage{nicematrix}
\usepackage{tabularx}
\usepackage[title]{appendix}
\usepackage{makecell} 
\usepackage{bm}
\usepackage{tablefootnote}
\usepackage{threeparttable}
\usepackage{lineno}
\title{Data-Driven Physics-Informed Neural Networks\\: A Digital Twin Perspective}
\author{
  Sunwoong Yang\\
  Cho Chun Shik Graduate School of Mobility \\
  Korea Advanced Institute of Science and Technology \\
  Daejeon 34051, Republic of Korea \\
  \texttt{sunwoongy@kaist.ac.kr} \\
  \And
  Hojin Kim\\
  Department of Aerospace Engineering\\
  Seoul National University\\
  Seoul 08826, Republic of Korea \\
  \texttt{rhrhak96@snu.ac.kr}\\
  \And
  Yoonpyo Hong\thanks{Current position: Helicopter Department,
Institute of Aerodynamics and Flow Technology, German Aerospace Center (DLR, Deutsches Zentrum für Luft- und Raumfahrt)}\\
  Institute of Advanced Machines and Design \\
  Seoul National University \\
  Seoul 08826, Republic of Korea \\
  \texttt{hyp1227@snu.ac.kr} \\
  \And
  Kwanjung Yee\\
  Department of Aerospace Engineering \\
  Seoul National University \\
  Seoul 08826, Republic of Korea \\
  \texttt{kjyee@snu.ac.kr} \\
  \And
  Romit Maulik\\
  College of Information Sciences and Technology \\
  The Pennsylvania State University \\
  Pennsylvania 16802, USA \\
  \texttt{rmaulik@psu.edu} \\
  \And
  Namwoo Kang†\\
  Cho Chun Shik Graduate School of Mobility \\
  Korea Advanced Institute of Science and Technology \\
  Daejeon 34051, Republic of Korea \\
  \texttt{nwkang@kaist.ac.kr} \\
  †Corresponding author\\
}

\begin{document}
% \begin{linenumbers}
\maketitle
\setcounter{footnote}{0}
\begin{abstract}
This study explores the potential of physics-informed neural networks (PINNs) for the realization of digital twins (DT) from various perspectives. First, various adaptive sampling approaches for collocation points are investigated to verify their effectiveness in the mesh-free framework of PINNs, which allows automated construction of virtual representation without manual mesh generation. Then, the overall performance of the data-driven PINNs (DD-PINNs) framework is examined, which can utilize the acquired datasets in DT scenarios. Its scalability to more general physics is validated within parametric Navier-Stokes equations, where PINNs do not need to be retrained as the Reynolds number varies. In addition, since datasets can be often collected from different fidelity/sparsity in practice, multi-fidelity DD-PINNs are also proposed and evaluated. They show remarkable prediction performance even in the extrapolation tasks, with $42\sim62\%$ improvement over the single-fidelity approach. Finally, the uncertainty quantification performance of multi-fidelity DD-PINNs is investigated by the ensemble method to verify their potential in DT, where an accurate measure of predictive uncertainty is critical. The DD-PINN frameworks explored in this study are found to be more suitable for DT scenarios than traditional PINNs from the above perspectives, bringing engineers one step closer to seamless DT realization.
\end{abstract}

% keywords can be removed

\clearpage
\section{Introduction}

\begin{figure*}[htb!]
    \centering

        \includegraphics[trim=140 70 140 70,clip, width=.6\textwidth]{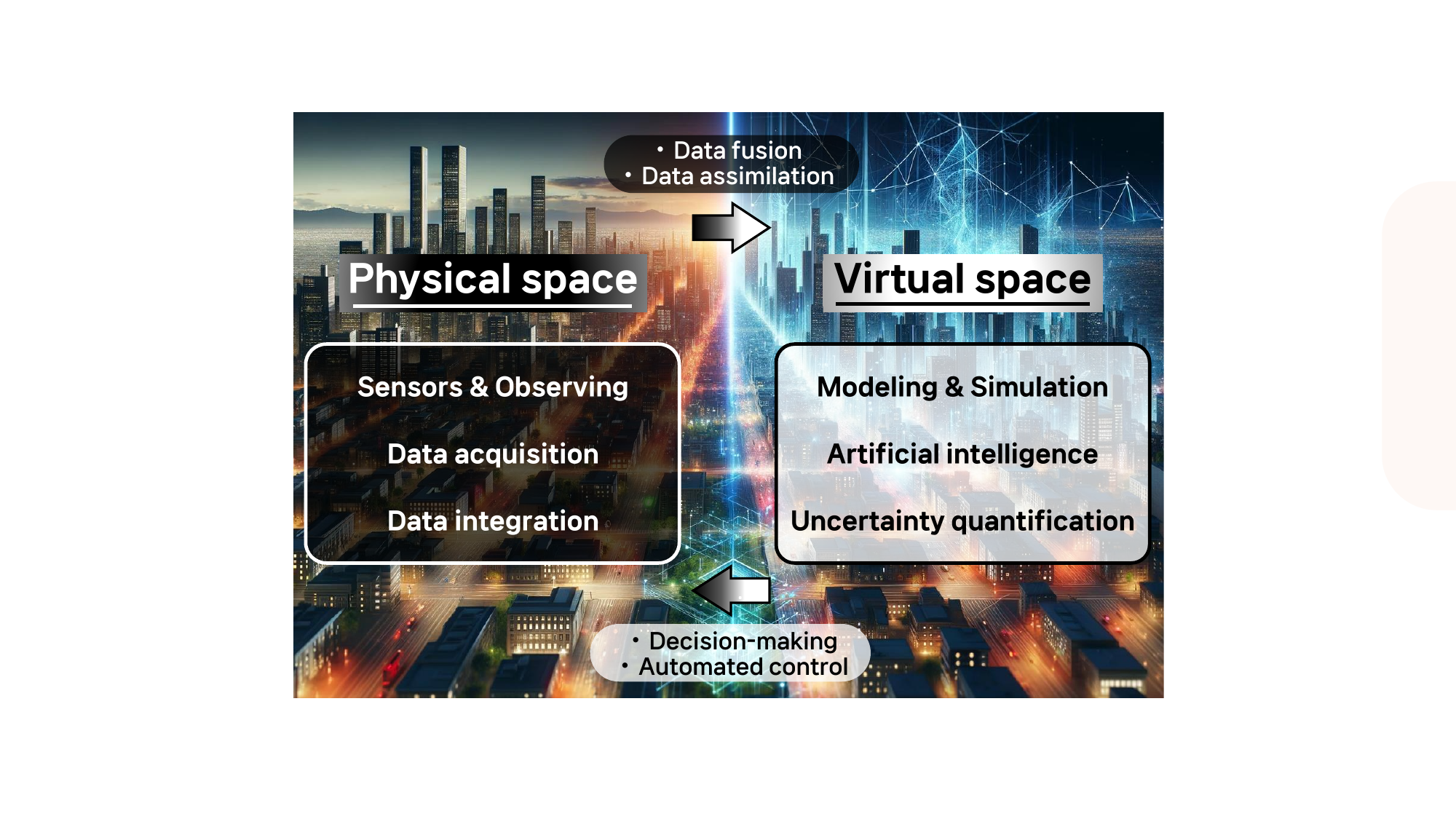}
        
    \caption{Ecosystem of the digital twins. This image is generated largely based on the report by National Academies \cite{national2023foundational}, while some parts have been modified. The background image is generated with the assistance of DALL$\cdot$E 2, an AI model developed by OpenAI.}
    % ·
    \label{fig:DT}
\end{figure*}

We are entering an era of digital twins (DT), where computational models are integrated with a physical space, creating a system that is dynamically updated through bidirectional interaction between the physical and virtual spaces (Fig. \ref{fig:DT}) \cite{kapteyn2020toward, rasheed2020digital, vinuesa2023transformative, attaran2023digital, tao2019digitaltwin, national2023foundational}. Due to its strong potential in decision-making during the design and development process, DT has been adopted by major companies such as IBM, GE, and Siemens \cite{li2022digitaltwin}. Especially, its ability to provide real-time guidance plays a crucial role in integrating the physical and digital worlds seamlessly \cite{niederer2021scaling, san2021digital, goh2021regulating,zhang2022digital}. In this context, physics-informed neural networks (PINNs), novel paradigms for solving partial differential equations (PDEs) leveraging deep learning framework \cite{raissi2019physics}, have been attracting engineers interested in alternatives to resource-intensive numerical simulations. While conventional surrogate models such as Gaussian process regression specialize in scalar output predictions, PINNs can predict high-dimensional flow fields efficiently, improving scalability and efficiency in complex simulations \cite{yang2022design, yang2023towards}. Given that PINNs can replace the demanding physical space of DT with real-time virtual representation processes, their potential seems boundless.

PINNs aim to provide approximate solutions to PDEs at each spatial/temporal collocation point utilizing the architecture of neural networks (NNs). They can be realized by updating model parameters of NNs in both supervised/unsupervised manner: the initial or boundary conditions are leveraged as labels of each point and PDEs residuals are utilized as unsupervised loss functions. These simple but intuitive approaches have advantages over traditional numerical simulations, especially in the field of computational fluid dynamics (CFD) where the following long-standing challenges exist. First, PINNs are theoretically mesh-free techniques. This is because the PDEs are solved directly at each collocation point and there is no need to exploit the connectivity between their points. Their property eliminates the need for expensive mesh generation, which has been a tedious and cumbersome procedure for fluid engineers, and therefore allows PINNs to construct virtual space of the DT without human intervention due to manual mesh generation \cite{wang2021deep, aygun2023physics}. In addition, once a PINN is trained, the prediction of the flow field based on CFD simulations can be replaced by the real-time inference using the trained model \cite{markidis2021old}. Moreover, considering that \citet{sun2020surrogate} successfully validated PINNs can also solve parametric Naver-Stokes (NS) equations, their deployment can be extended to DT scenarios where the modeling capacity for the general physical system is required. Lastly, the use of PINNs can take advantage of recent developments in GPUs, and it is becoming much easier to implement their algorithms and variants thanks to numerous open-source deep learning libraries such as \textsc{DeepXDE} and \textsc{Modulus} \cite{markidis2021old, karniadakis2021physics, lu2021deepxde, nvidiaModulusNeural}.

However, conventional PINNs without labeled data (data-free PINNs) still have limitations to be applied as predictive models for the realization of DT, especially in the field of fluid dynamics. Although there are numerous previous studies using PINNs in flow problems, most of them are based on the low Reynolds numbers ($Re$), which limits further investigation of their potential in real-world problems at higher $Re$. For example, in lid-driven cavity flow, \citet{amalinadhi2022physics, jagtap2020conservative, li2022dynamic, bai2020applying, wang2021understanding} showed that PINNs can work successfully in the forward problem at $Re=100$. Also, \citet{wong2022learning} and \citet{chiu2022can} tested PINNs at $Re=400$ in forward problems. However, \citet{krishnapriyan2021characterizing} found that PINNs work for simple cases in 1D convection and reaction-diffusion problems, but fail when convection or diffusion coefficients become slightly higher, raising alarms about naive cut-and-paste use of PINNs. In a similar context, \citet{chuang2023predictive} discovered that while the data-free approach could accurately capture the flow fields around the cylinder at $Re=40$, it failed at a bit higher $Re$ (200) where vortex shedding exists: the similar failure even at $Re=100$ was also confirmed by \citet{rohrhofer2022role}. From the failures of PINNs that occurred slightly beyond the boundary of simple pedagogical flow problems, additional techniques to mitigate their failures seem necessary.

In this respect, there has been a recent movement towards the use of labeled data for the training of PINNs. \citet{rohrhofer2022role} proposed data-driven PINNs (DD-PINNs) using labeled dataset as a training guide and successfully captured the cylinder vortex shedding at $Re=100$. However, they simply used a high-fidelity CFD dataset, which conflicts with the purpose of training PINNs to replace CFD. Also, they claimed that consideration of this guide dataset leads to tricky optimization problems in terms of the loss landscape. Conversely, \citet{gopakumar2023loss} argued that DD-PINNs make landscapes have more pronounced valleys, which helps the optimizer converge to an optimal point more easily. In addition, they suggest a more pragmatic scenario for using DD-PINNs, where sparse guide data is generated from the low-fidelity simulation. Nevertheless, they still limited their scope to using only sparse guide data (from 1\% to 10\% grid points of the entire flow fields obtained by CFD), which can be considered as an inefficient approach as it does not fully exploit the obtained CFD results. Furthermore, they mainly focused on the change in the loss landscape and concluded that the incorporation of guide data helps the optimizer to find an optimal point more easily. However, in their case study of predicting the flow field around a rectangular block, DD-PINN only detected the presence of objects but failed to predict the detailed flow structure. Finally, \citet{chuang2023predictive} observed that the DD-PINNs were able to predict the vortex shedding at $Re=200$, where data-free PINNs failed. In their study, nevertheless, the DD-PINNs exhibited poor predictive performance in extrapolation when applied to non-parametric Navier-Stokes (NS) equations where the Reynolds number is fixed. This observation has necessitated a more in-depth exploration of the extrapolation capability of DD-PINNs, particularly in the context of solving parametric NS equations with different Reynolds numbers.

The objective of this research is to explore the potential of DD-PINNs from the perspective of DT. First, by thoroughly investigating the effectiveness of different adaptive sampling techniques on data-free PINNs, we aim to further consolidate their mesh-free advantage, which promotes automatic interaction between the physical and virtual spaces of the DT without human intervention. Second, given that the DT environment is continuously updated with collected data \cite{national2023foundational}, this study scrutinizes a DD-PINN framework that can fully exploit the dataset assumed to be obtained from the numerical simulation. We then apply it to the higher $Re$ flow problems where the data-free PINNs fail, aiming to bridge the existing gap between low Reynolds toy problems for academic validation and the higher Reynolds problems often encountered in the real-world. However, in practice, there exists a challenge in incorporating obtained data into the predictive models: datasets can be obtained from different sources \cite{tao2018digital, wang2023digital}. Thus, we also extend the DD-PINN framework to the multi-fidelity approach, which can efficiently leverage datasets from different sources ---even if some of the datasets are assumed to be sparsely collected from physical space--- and thus significantly improve its scalability and flexibility. To validate multi-fidelity DD-PINNs in DT scenarios, the forward problem with parametric NS equations is adopted to verify whether they can be used for real-time prediction of general physics under different Reynolds numbers. Furthermore, we attempt to investigate their applicability to the uncertainty quantification (UQ) procedure, which is also a crucial component for the realization of the DT \cite{national2023foundational}. \clearpage The main contributions of this paper, paired with its main research questions, are summarized as follows:
\begin{enumerate}
    
    \item 
    \textbf{\textit{Is the construction of virtual space automated effectively?}} [Sections \ref{sec:VA} and \ref{sec:100para}]
    \begin{itemize}
        \item Given that the meshless nature of PINNs is a key ingredient for their automated construction, we aim to investigate the practical effectiveness of their mesh-free property. We propose an adaptive sampling method for collocation points that is specifically tailored to fluid dynamics and comprehensively compare it with existing sampling approaches.
    \end{itemize}

    \item 
    \textbf{\textit{Is the virtual space ready for data-driven model updating?}} [Section \ref{sec:remedy}]
    \begin{itemize}
        \item We extensively investigate the performance of DD-PINNs that can fully exploit the guide data set repeatedly acquired in the DT scenario. Then, moving away from the toy problem at low Reynolds numbers, we verify the superiority of the DD-PINNs in higher Reynolds conditions where conventional data-free PINNs fail. Furthermore, by visualizing the loss landscape, the reason for the success of DD-PINN over a data-free approach is demonstrated. See Section \ref{sec:remedy} for the details.
    \end{itemize}
       
    \item 
    \textbf{\textit{Is there any guideline for DD-PINN in DT scenarios?}} [Section \ref{sec:samplingEffect}]
    \begin{itemize}
        \item We discover that random sampling outperforms other adaptive sampling techniques in DD-PINNs. We also conclude that this is because they require uniformly distributed collocation points to compensate for the local regularization effects of the data-driven loss term.
    \end{itemize}
    
    \item 
    \textbf{\textit{Is it scalable to general physics with different PDE parameters?}} [Section \ref{sec:Param}]
    \begin{itemize}
        \item To validate the scalability of DD-PINNs, we apply them to parametric NS equations and show their advantage over data-free PINNs and data-driven NNs: it highlights the versatility of the proposed DD-PINN framework in DT, where real-time prediction under different PDE parameters is required.
    \end{itemize}
    
    \item 
    \textbf{\textit{Is it applicable to datasets of different fidelity/sparsity?}} [Section \ref{sec:Param_MF}]
    \begin{itemize}
        \item We further extend DD-PINNs to leverage heterogeneous data, taking into account the fact that in physical space the dataset is often collected from different sources. Then, their superiority over data-free PINNs, data-driven NNs, and single-fidelity DD-PINNs is effectively verified. Corresponding multi-fidelity DD-PINNs also show remarkable extrapolation performance even beyond the Reynolds numbers used for the train. It highlights the flexibility of the proposed DD-PINNs when different types of data are given, such as some sparse but high-fidelity data from physical space and others from virtual space.
    \end{itemize}
    
    \item 
    \textbf{\textit{Is it able to quantify reasonable predictive uncertainty?}} [Section \ref{sec:Param_UQ}]
    \begin{itemize}
        \item The UQ performance of multi-fidelity DD-PINNs is also evaluated qualitatively by an ensemble approach. Since they are judged to provide reasonable uncertainty according to different Reynolds numbers, their potential in DT scenarios, where an accurate measure of predictive uncertainty is crucial, seems promising.
    \end{itemize}
  % Unlike previous research that has simply shown that DD-PINNs with artificially selected sparse data can outperform data-free PINNs in forward problems \cite{gopakumar2023loss}, we reformulate a DD-PINN framework with a more practical perspective. 
    
    % Their practical utility is then successfully verified by identifying the conditions under which their accuracy improves through a variety of parametric studies. Our work effectively addresses the previous limitations of data-driven PINNs, transforming them from a theoretical concept to a practically viable solution.
\end{enumerate}

The rest of this paper is organized as follows. In Section \ref{sec:method}, the background on data-free and data-driven PINNs is described along with the adaptive sampling techniques for their efficient training. In Section \ref{sec:Re100}, the effects of adaptive sampling techniques on data-free PINNs in predicting lid-driven flow at $Re=100$ are investigated. In Section \ref{sec:fail}, the failure of the data-free PINNs at higher Reynolds numbers is observed, and in Section \ref{sec:remedy}, DD-PINNs are proposed as a remedy and the reason for their successful training is discussed. In Section \ref{sec:Param}, DD-PINNs are extended to parametric NS equations and further improved with multi-fidelity approaches. Section \ref{sec:conclu} presents the conclusions and future work of this study.

% \newpage
\section{Methodologies}\label{sec:method}
\subsection{Physics-informed neural networks (PINNs)}

PINN, whose algorithm is to constrain NNs by incorporating PDEs as loss functions, is one of the most fascinating concepts for engineers of all disciplines. Its goal is to solve PDEs using NNs, and therefore the loss function of NNs includes PDE losses, which are computed as zero when the PDEs are exactly satisfied. In most cases, PDEs are usually accompanied by boundary conditions (BC) or initial conditions (IC) that make their problem closed and have a unique solution. The corresponding PINN framework is called data-free PINN in that it does not require any labeled data such as velocity fields at specific spatial/temporal locations. There is another PINN framework called data-driven PINN (DD-PINN) \cite{chuang2023predictive}, which has an additional data-driven loss term. This term takes into account the deviation from the labeled data and the predicted values of PINNs, driving them to predict the specific values (output labels) at specific locations (input labels). Data-free and DD-PINNs are explained in more detail in Section \ref{sec:DFPINN} and \ref{sec:DDPINN}, respectively.

\begin{figure*}[htb!]
    \centering

        \includegraphics[width=.9\textwidth]{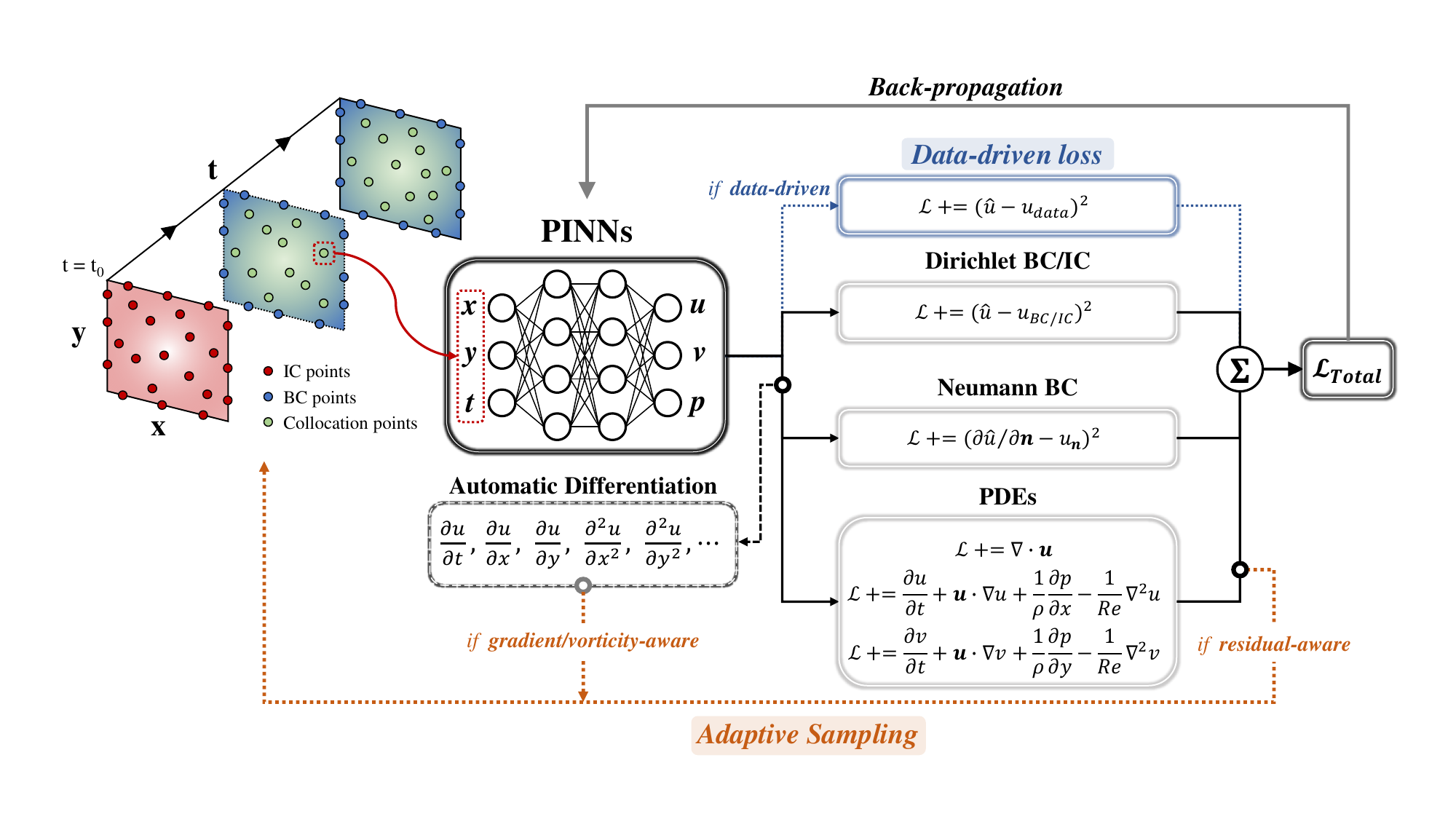}
        
    \caption{Overall architectures of PINN. The data-free PINN does not include data-driven loss term whereas data-driven PINN does.}
    \label{fig:PINNarchit}
\end{figure*} 

\subsubsection{Data-free PINNs}\label{sec:DFPINN}
As in Fig. \ref{fig:PINNarchit}, the input to the PINN consists of spatial ($x, y$ for 2D) and temporal ($t$) coordinates: it has different input and output structures than conventional convolutional neural network-based frameworks, which construct entire flow fields from a single input set at a time \cite{hasegawa2020machine, maulik2021reduced, kang2022pof}. PINN outputs the quantities of interest at the corresponding coordinates, and in this study, x-velocity ($u$), y-velocity ($v$), and pressure ($p$) are adopted as output. Based on them, PDE and BC/IC losses can be defined. Since the calculation of these loss terms does not require the labeled dataset, it is referred to as data-free PINN. For incompressible NS equations, PDE losses for PINN are defined as follows:

\begin{equation}
\label{eq:PDEloss}
\begin{gathered}
\mathcal{L}_{mass} = ||\nabla \cdot \mathbf{u}|| \\
\mathcal{L}_{x-momentum} = ||\pdv{u}{t} + \mathbf{u} \cdot \nabla u + \cfrac{1}{\rho} \pdv{p}{x} - \cfrac{1}{Re} \nabla^{2} u|| \\
\mathcal{L}_{y-momentum} = ||\pdv{v}{t} + \mathbf{u} \cdot \nabla v + \cfrac{1}{\rho} \pdv{p}{y} - \cfrac{1}{Re} \nabla^{2} v|| \\
x, y \in \Omega, t \in [0,T]
\end{gathered}
\end{equation}

where $\mathbf{u}$ is velocity vector consists of $u, v$; and $\rho, Re$ each denotes density and Reynolds number. However, since PDEs alone cannot represent the unique solution of the system, additional information about domain boundaries is needed \cite{sun2020surrogate}. BCs and ICs are considered to this end, and assume they are given as follows:

\begin{equation}
\label{eq:BCIC}
% \begin{gathered}
\begin{cases}
\mathcal{NN}(x,y,t) = \mathcal{F}_{Dirichlet BC}, & [x, y] \in \Gamma_{Dirichlet BC}, \,\,\,t \in [0,T]\\[1ex]
\pdv{\mathcal{NN}(x,y,t)}{n} = \mathcal{F}_{Neumann BC}, & [x, y] \in \Gamma_{Neumann BC}, \,\,\,t \in [0,T]\\[1ex]
\mathcal{NN}(x,y,0) = \mathcal{G}_{IC}, & [x, y] \in \Gamma_{IC}\\
\end{cases}
% \end{gathered}
\end{equation}

Therefore, the loss functions related to them can be written as:

% and when Dirichlet BC of $\mathcal{NN}(x,y,t) = \mathcal{F}_{Dirichlet}$, Neumann BC of $\pdv{\mathcal{NN}(x,y,t)}{n}=\mathcal{F}_{Neumann}$, and IC of $\mathcal{NN}(x,y,0)=\mathcal{G}_{IC}$ are given, the loss functions with respect to them can be written as:

\begin{equation}
\label{eq:BCICloss}
\begin{gathered}
\mathcal{L}_{BC} = ||\mathcal{NN}(x,y,t)-\mathcal{F}_{Dirichlet}||_2 + ||\pdv{\mathcal{NN}(x,y,t)}{n}-\mathcal{F}_{Neumann}||_2 \\
\mathcal{L}_{IC} = ||\mathcal{NN}(x,y,0)-\mathcal{G}_{IC}||_2
\end{gathered}
\end{equation}

where $||\cdot||_2$ denotes $L_2$-norm. In Eq. \ref{eq:PDEloss} and \ref{eq:BCICloss}, partial derivative terms such as $\pdv{u}{t}, \pdv{p}{x}, \pdv{\mathcal{NN}(x,y,t)}{n}$ can be calculated by using automatic differentiation (AD) within NNs \cite{raissi2019physics, baydin2018automatic}. AD enables one to compute the partial derivative of the output with respect to the input of the NNs leveraging back-propagation operation. Finally, PINN loss for NS equations is calculated as below:

\begin{equation}
\label{eq:PINNloss}
\mathcal{L}_{PINN} = \mathcal{L}_{mass} + \mathcal{L}_{x-momentum} + \mathcal{L}_{y-momentum} + \mathcal{L}_{BC} + \mathcal{L}_{IC}
\end{equation}

Note that each loss term in Eq. \ref{eq:PINNloss} is evaluated using collocation points. Specifically, $\mathcal{L}_{mass}$, $\mathcal{L}_{x-momentum}$, $\mathcal{L}_{y-momentum}$ are evaluated within the collocation points sampled in $\Omega$, and $\mathcal{L}_{BC}$ within points in $\Gamma_{Dirichlet BC}$ or $\Gamma_{Neumann BC}$, and $\mathcal{L}_{IC}$ within $\Gamma_{IC}$. In this context, data-free PINN means that the training of the PINN does not require the labeled dataset with input ($x, y, t$) and output ($u, v, p$), while it still requires the locations of the collocation points ($x, y, t$) to be explored. Based on this fact, one can easily infer that the sampling strategy of these collocation points would greatly affect the performance of PINN, just as CFD results are highly dependent on the mesh quality.

\subsubsection{Data-driven PINNs (DD-PINNs)}\label{sec:DDPINN}
Though data-free PINN has the decisive advantage that it does not require any labeled dataset for its training, as will be figured out later in this study, the data-free approach can cause training failures in that PDEs themselves are solved directly without any data for guidance. With this regard, another approach to leverage the labeled dataset is also being adopted \cite{chuang2023predictive, gopakumar2023loss}, and it is referred to as DD-PINNs in this study. It has been frequently utilized in inverse problems, which aim to predict unknown PDE coefficients or BC/IC from sparse measurements of the PDE solutions \cite{lu2021physics}. Its approach does not modify the architecture of the data-free PINN; it only adds one more loss term as below:

\begin{equation}
\label{eq:DDPINNloss}
\mathcal{L}_{DD-PINN} = \mathcal{L}_{mass} + \mathcal{L}_{x-momentum} + \mathcal{L}_{y-momentum} + \mathcal{L}_{BC} + \mathcal{L}_{IC} + w_{data}*\mathcal{L}_{data}
\end{equation}

where $\mathcal{L}_{data}$ denotes the loss term of the labeled guide data (as in Fig. \ref{fig:PINNarchit}, $\mathcal{L}_{data}$ is considered if PINN adopts data-driven concept). For example, when the velocity fields of some locations are given, $\mathcal{L}_{data} = ||\hat{\mathbf{u}} - \mathbf{u}_{data}||_2$, where $\hat{\mathbf{u}}$ and $\mathbf{u}_{data}$ respectively indicate the predicted velocity by PINNs and the velocity of the labeled data. By including only this $\mathcal{L}_{data}$ additional term in the loss function, \citet{gopakumar2023loss} argued that DD-PINNs can be biased toward the solution in forward problems, reducing both the approximation error and optimization error. They added that the former error can be reduced due to the regularization effects of the leveraged data points, and the latter due to the morphing effect on the loss landscape towards a simpler optimization task. In this study, to control the effectiveness of this $\mathcal{L}_{data}$ term, the hyperparameter $w_{data}$ is additionally considered: the larger its value, the greater the impact of the data-driven loss term.

The guide data used in the data-driven approach for PINNs can be regarded as the measurements from the sparse sensor data in the experimental domain \cite{go2023physics, eivazi2022physics, wang2022dense, hasanuzzaman2023enhancement}. However, for the simulation domain, using data from numerical simulation in forward problems can be paradoxical \cite{gopakumar2023loss}; the purpose of PINNs in forward problems is to replace numerical simulation, but for DD-PINNs, we need to perform numerical simulation to obtain data that will be used to train PINNs. In fact, \citet{gopakumar2023loss} have already raised this point and suggested using part ($1\%$, $5\%$, and $10\%$) of the data obtained from the low-fidelity simulation, which was called \textit{coarse regulated PINN} in their work. Their main focus was on the loss landscapes of PINNs: landscapes of the DD-PINNs regulated with sparse simulation data have more pronounced valleys, which helps the optimizer to converge to an optimal point more easily (thus reducing the optimization error). However, they failed to achieve reasonable accuracy when solving 2D NS equations for flow around a rectangular block: DD-PINNs were only able to detect the presence of the block, not the detailed structure of the flow field. In this context, this study aims to verify their practical utility by comprehensively identifying the conditions under which their accuracy improves through a variety of parametric studies, which is one of our main contributions.

% In this context, this study aims to verify their practical utility by comprehensively identifying the conditions under which their accuracy improves through a variety of parametric studies, which is one of our main contributions. The parametric studies of DD-PINN using direct numerical simulation (DNS) results from various grids are performed. Then, its data-driven concept is further fused with the multi-fidelity approach: different grid resolutions for CFD are used to solve parametric NS equations, which is the first attempt to the best of the authors' knowledge.

\subsection{Adaptive sampling techniques}\label{sec:VA}

\subsubsection{Existing adaptive sampling techniques}\label{sec:VA_exist}

As mentioned in Section \ref{sec:DFPINN}, the results of PINN strongly depend on the sampling strategy of the collocation points, since the collocation distribution acts like the grid in the CFD simulation. However, the conventional PINN approach often uses the uniform sampling strategies \cite{sun2020surrogate, chuang2023predictive, ang2022physics}. Such strategies are easy to implement, but have a critical aspect in that they cannot consider the viscous effects of the flow field in an efficient way as CFD, where the mesh is generated to be dense near the inner objects or wall \cite{salim2009wall}. In addition, they sample the collocation points only once so that their distribution does not change during the entire training procedure, which leads to the inefficient training of PINN. To overcome this inefficiency, \citet{lu2021deepxde} proposed the adaptive sampling strategy where the collocation points are re-sampled at the region of high PDE residual. This approach is intuitive in that since PINN aims to reduce the residuals of the PDEs, additional sample points are added where the residuals are high. Specifically, it first randomly samples $M$ evaluation points to estimate the residuals in the entire region, and then sorts the $N$ largest residual points among them, which will be the newly added collocation points. This deterministic nature makes the points to be added highly dependent on the size of $M$. More specifically, if $M$ is too large, the adaptive points will be concentrated in a small region, while a small $M$ would lead to even sampling, which is not consistent with its residual-aware concept. Therefore, a stochastic approach using a probability density function (PDF) has been proposed \cite{nabian2021efficient}. It samples the adaptive points based on a PDF proportional to the residuals, thus removing the deterministic nature of the previous approach. There has been another attempt to add new samples where the gradient of the outputs (e.g. $u$, $v$, $p$ for flow field prediction) is large \cite{mao2020physics}. This gradient-aware approach is well suited to the field of fluid dynamics, where the steep gradient of the flow quantities often plays a critical role in the flow phenomena. However, its algorithm was implemented manually with prior knowledge of the solution in the work by \citet{mao2020physics}, intentionally adding points around the discontinuous region.

\subsubsection{Vorticity-aware adaptive sampling}\label{sec:VA_VA}
Existing adaptive sampling approaches in the previous section can be used universally across all engineering disciplines because they have focused on the general properties of the PINN architecture, namely residuals of PDEs or gradients of outputs. However, when it comes to the specific domain, there may be an alternative way to realize adaptive sampling that works more efficiently than the approaches designed for general domains. To explore this potential possibility, we attempt to develop an adaptive sampling technique specialized for fluid problems. In fact, numerous conventional CFD studies have already attempted to accurately capture the vortices by adding meshes in regions of high vorticity (see Appendix \ref{sec:vorticity} for more details). To extend these approaches in the CFD community to the PINN architecture filling their academic gap, this study proposes a novel algorithm that adaptively adds new collocation points where high vorticity exists. This method that focuses on the region of high vorticity magnitude is referred to as the ``vorticity-aware'' approach throughout this manuscript, and its algorithm can be summarized as Algorithm \ref{alg:VA}. First, sample the locations where the vorticity magnitude is to be evaluated, $X_{eval}$ (line $5$). Note that the calculation of the vorticity ($\omega$) requires a partial derivative of velocities with respect to the spatial coordinates ($\pdv{u}{y}$, $\pdv{v}{x}$), which can be easily performed owing to the AD property of the NNs. Then we define the PDF, $p_{V}(\mathbf{x})$, which is proportional to the vorticity magnitudes computed at $X_{eval}$ as follows (line $6$) \cite{wu2023comprehensive}:

\begin{equation}
\label{eq:PDF}
p_{V}(\mathbf{x}) = \frac{|\omega|^k(\mathbf{x})}{\mathbb{E}[|\omega|^k(\mathbf{x})]}
\end{equation}

where $\mathbf{x}=(x,y)\in\Omega$ for the 2D spatial domain, and $k$ is a hyperparameter for adjusting the intensity of the correlation between new sample points and $p_{V}(\mathbf{x})$. More specifically, a large value of $k$ indicates that points will be sampled more intensely according to the $p_{V}(\mathbf{x})$ distribution (i.e., more on the high vorticity region), while a small value leads to uniform sampling. Therefore, it is referred to as ``stochastic intensity'' throughout this study. Back to Algorithm \ref{alg:VA}, as in line $7$, $X_{adapt}$ of size $N$ are newly sampled according to $p_{V}(\mathbf{x})$. Since the original size of the whole collocation points is $M$, the new points $X_{random}$ of size $(M-N)$ are additionally randomly sampled to keep the original dataset size (line $8$). Finally, the next collocation dataset is defined as $X_{adapt} \cup X_{random}$, and retraining of the PINN continues (line $9, 10$) --- this retraining can be considered as transfer learning, which makes the repetitive training of NNs efficient by using the pre-trained model parameters \cite{yang2023inverse}, and it can also be considered as a way to quickly incorporate the data obtained from the physical space into the virtual space model reducing the time delay in DT environment. This process is repeated until the predefined maximum iteration is reached (line $4, 11$). Note that the size ratio of the newly sampled points to the total collocation points, $\frac{N}{M}$, is the hyperparameter for the adaptive sampling approaches. The effects of the mentioned hyperparameters, $k$ and $\frac{N}{M}$, will be discussed later in Section \ref{sec:100para}.

\begin{algorithm}[htb!]
\caption{Vorticity-aware adaptive sampling}\label{alg:VA}
    \begin{algorithmic}[1]
        \State Prepare the initial collocation points $X_{train}$, size of $M$
        \State Set $N$, which is the size of the points to be updated
        \State Train PINN with $k_{init}$ epochs
        % \While{$i < i_{max}$} %\Comment{Loop for NN (parallelizable)}
        \Repeat
            \State Randomly sample evaluation points size of $M$ (=$X_{eval}$)
            \State Calculate PDF $p_{V}(\mathbf{x})$ that is proportional to the vorticity magnitudes calculated on $X_{eval}$
            \State Sample $N$ points from $X_{eval}$ according to $p_{V}(\mathbf{x})$ distribution (=$X_{adapt}$)
            \State Randomly sample points size of $M-N$ (=$X_{random}$)
            \State Update collocation points \Comment{$X_{train} = X_{adapt} \cup X_{random}$}
            \State Train PINN with $k_{adapt}$ epochs
        % \EndWhile
        \Until{\textit{Max epoch is reached;}}
    \end{algorithmic}
\end{algorithm}

\subsubsection{Fusion of all sampling techniques}\label{sec:VA_fusion}
Though this study proposes the new technique named vortcity-aware adaptive sampling, existing adaptive sampling approaches based on PDE residuals and output gradients cannot simply be considered inefficient for flow problems. Rather, we fuse the proposed vorticity-aware concept (Section \ref{sec:VA_VA}) with existing approaches (Section \ref{sec:VA_exist}), adopting all their ideas. In particular, the residual-aware concept based on mass conservation PDE, the gradient-aware concept in terms of freestream direction ($\pdv{u}{x}$), and the vorticity-aware concept are all fused: the corresponding algorithm is given in Algorithm \ref{alg:RGV}. Note that the only difference with the vorticity-aware algorithm in Algorithm \ref{alg:VA} is that it adds $N/3$ points each for all three criteria, whereas vorticity-aware adds the whole $N$ points according to the vorticity magnitude only. Also, $P_R(x)$ and $P_G(x)$, the PDFs used for the residual and gradient criteria in lines $6, 7$, can be computed by replacing the vorticity magnitude $|\omega|$ in Eq. \ref{eq:PDF} by the PDE residual and output gradient, respectively. From now on, ``RGV-xyz'' refers to the adaptive sampling technique where $N$ new points to be added consist of residual, gradient, and vorticity-aware points with the ratio x:y:z. For example, vorticity-aware approach in Algorithm \ref{alg:VA} can be referred to as ``RGV-001'' (vorticity-aware only), while Algorithm \ref{alg:RGV} as ``RGV-111'' (all three criteria are considered equally). Similarly, ``RGV-110'' denotes a sampling that adopts residual and gradient-aware sampling without consideration of vorticity, by adding $N/2$ points for each of these two criteria.
 
\begin{algorithm}[htb!]
\caption{Residual-Gradient-Vorticity-aware adaptive sampling}\label{alg:RGV}
    \begin{algorithmic}[1]
        \State Prepare the initial collocation points $X_{train}$, size of $M$
        \State Set $N$, which is the size of the points to be updated
        \State Train PINN with $k_{init}$ epochs
        \Repeat %{$i < i_{max}$} 
            \State Randomly sample evaluation points size of $M$ (=$X_{eval}$)
            \State Calculate PDF $p_{R}(\mathbf{x})$ based on the residual of PDEs on $X_{eval}$
            \State Calculate PDF $p_{G}(\mathbf{x})$ based on the gradient of outputs on $X_{eval}$
            \State Calculate PDF $p_{V}(\mathbf{x})$ based on the vorticity magnitude on $X_{eval}$
            \State Sample $N/3$ points from $X_{eval}$ according to $p_{R}(\mathbf{x})$ distribution (=$X_{adapt\_R}$)
            \State Sample $N/3$ points from $X_{eval}$ according to $p_{G}(\mathbf{x})$ distribution (=$X_{adapt\_G}$)
            \State Sample $N/3$ points from $X_{eval}$ according to $p_{V}(\mathbf{x})$ distribution (=$X_{adapt\_V}$)
            \State Randomly sample points size of $M-N$ (=$X_{random}$)
            \State Update collocation points \Comment{$X_{train} = X_{adapt\_R} \cup X_{adapt\_G} \cup X_{adapt\_V} \cup X_{random}$}
            \State Train PINN with $k_{adapt}$ epochs
        \Until{\textit{Max epoch is reached;}}
        % \EndWhile
    \end{algorithmic}
\end{algorithm}

\section{Preliminary study: lid-driven cavity flow at \textit{Re=100}} \label{sec:Re100}

In this section, the lid-driven cavity flow problem at Reynolds number 100 is investigated using data-free PINNs. DNS is also performed using OpenFOAM, the open-source CFD framework with the finite volume method , to compare the results of PINNs with conventional CFD approaches. IcoFoam solver in OpenFOAM is utilized, which is a transient solver for incompressible laminar Newtonian fluid without a turbulence model \cite{liu2016instability}. Uniform grids with different resolution levels (20$\times$20, 40$\times$40, 80$\times$80, and 160$\times$160) are adopted and their details can be found in Fig. \ref{fig:grid_visualization}, Appendix \ref{sec:openfoam}. The The simulations are run on a single CPU core of the Intel i7-1165G7 processor until the residual dropped below $10^{-6}$, and simulation wall times for each grid are 2, 8, 80, and 875 seconds respectively. The validation of the DNS results is also performed as shown in Fig. \ref{fig:DNSvali}, Appendix \ref{sec:openfoam}. To estimate the error of the PINNs compared to DNS, the root mean squared error (RMSE) between PINN prediction ($\hat{u}$) and DNS results from the $160\times160$ grid ($u_{data}$) is leveraged as below:

\begin{equation}
\label{eq:RMSE}
RMSE = \sqrt{\frac{\sum\limits_{i=1}^N(\hat{u}-u_{data})^2}{N}}
\end{equation}

where $N$ is the test dataset size.

\subsection{Hyperparameter tuning}\label{sec:100hyper}

Before investigating the overall performance of data-free PINNs at $Re=100$, hyperparameter tuning is performed to adopt the best model architecture. For this purpose, the number of hidden layers ($N_{layer} \in \{4, 6, 8\}$), the number of nodes at each hidden layer ($N_{node} \in \{32, 64, 128\}$), and the learning rate ($lr \in \{1e-3, 1e-4, 1e-5\}$) are set as hyperparameters. Note that for this tuning, the fixed learning rate is used without the decaying strategy (the decaying technique will be introduced in the higher Reynolds cases, Section \ref{sec:Param}). For the activation function, a hyperbolic tangent is applied as recommended by \citet{wang2023expert}: while there are adaptive activation functions proposed by \citet{jagtap2020locally, jagtap2020adaptive} to improve the training speed, we decided to choose the hyperbolic tangent since the acceleration of PINNs is beyond the scope of this study. Each PINN model is trained for 20,000 iterations with an Adam optimizer, and 2,000 collocation points and 500 boundary points are selected. Only vanilla PINNs without special adaptive sampling techniques (hereafter called ``Van'') are trained to briefly explore the hyperparameter space. Considering the stochastic nature of the training, each PINN is trained at least four times, and their averages are presented throughout all experiments in this study. Finally, errors with respect to the velocity components ($u$, $v$) between PINN and DNS are calculated as Table \ref{tab:100hyp_results} in Appendix \ref{sec:app_hyp}, and they indicate that $N_{layer}=6, N_{node}=32, lr=1e-3$ and $N_{layer}=8, N_{node}=32, lr=1e-3$ cases have the best results. The former one has been selected for further investigation due to its simpler architecture: note that the corresponding model requires 810 seconds for the training with NVIDIA 3080 GPU, and the training time for each hyperparameter tuning case averages 900 seconds (therefore about 24,300 seconds for all cases).

\subsection{Parametric studies for adaptive sampling techniques}\label{sec:100para}
Based on the PINN architecture selected in Section \ref{sec:100hyper}, this section aims to scrutinize the practical implications of the various adaptive samplings that promote automatic interaction between the physical and virtual spaces of the DT without human intervention. To this end, effects of the stochastic intensity ($k$) and the ratio of the newly updated collocation points ($\frac{N}{M}$) are investigated in Section \ref{sec:100k} and \ref{sec:100ratio}, respectively, effects of the number of collocation points can be found in Section \ref{sec:num_collo}, and the final remarks on their results are given in Section \ref{sec:remark}. %\textcolor{red}{Please note that though the initial sampling techniques of the collocation points can have an significant impact on the results of PINNs \cite{wu2023comprehensive}, this study focuses on the adaptive sampling of the collocation: we mainly aim to investigate the real-time update of the PINNs in virtual space using the new dataset obtained in the physical counterpart.} 

\subsubsection{Effects of stochastic intensity ($k$)}\label{sec:100k}

As mentioned in Section \ref{sec:VA_VA}, the hyperparameter $k$ determines the intensity of the correlation between the collocation points to be newly added and residual/gradient/vorticity PDF. When $k \rightarrow \infty$, adaptive sampling no longer adopts stochastic but deterministic selection, which means that the top $N$ points with respect to residual/gradient/vorticity values will be selected \cite{wu2023comprehensive}. On the other hand, when $k = 0$, the PDF of the residual/gradient/vorticity ($p_{R}(\mathbf{x})$ / $p_{G}(\mathbf{x})$ / $p_{V}(\mathbf{x})$) becomes constant over $\mathbf{x}$, indicating that the uniform sampling will be performed. 

This section aims to investigate the effects of the above stochastic intensity ($k$) for various adaptive sampling techniques, and therefore a total of 8 techniques are adopted: Van (random sampling), RGV-100, RGV-010, RGV-001, RGV-110, RGV-101, RGV-011, RGV-111 (see Section \ref{sec:VA_fusion} for abbreviations). To focus on analyzing the effects of $k$, another hyperparameter $\frac{N}{M}$ is set to 0.7, where $M$ is 2,000. All PINNs with different adaptive sampling are trained for 20,000 epochs, and the adaptive sampling procedure is repeated every 4,000 epochs (i.e., after the initial sampling, the adaptive sampling procedure is performed four times per model). Here, note again that this repetition of training with updated collocation points can be considered as transfer learning approach mentioned in Section \ref{sec:VA_VA}.

\begin{figure*}[htb!]
    \begin{minipage}[t]{\textwidth}
        \centering
        \begin{subfigure}[t]{0.35\textwidth}
            \centering
            \includegraphics[width=\linewidth]{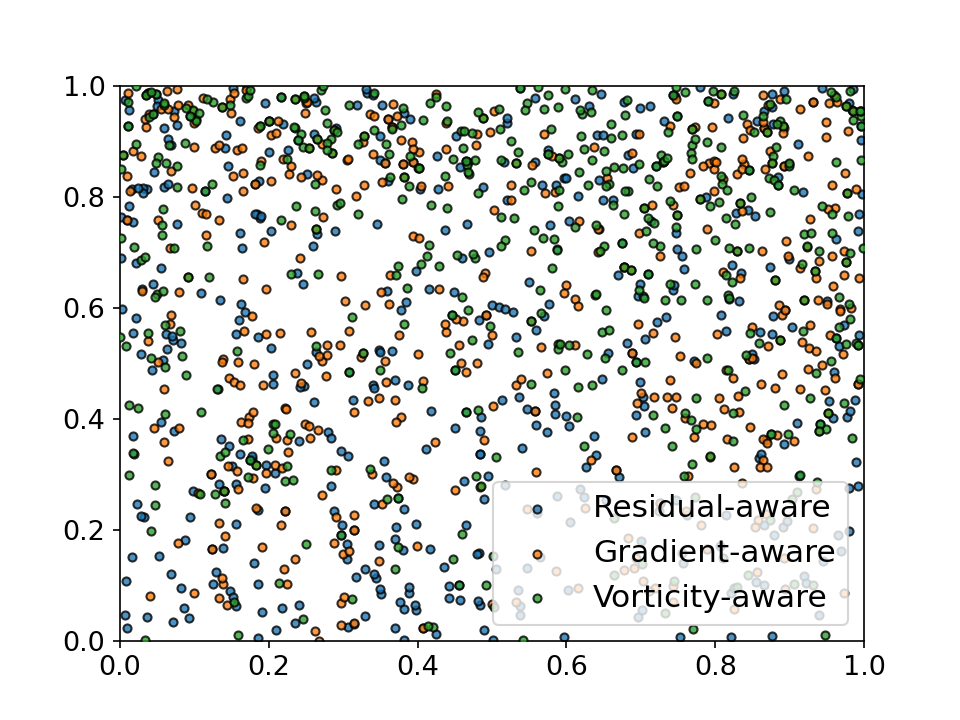}
            \caption{$k=0.5$}
            \label{fig:k_a}
        \end{subfigure}
        \hspace{0.02\textwidth}
        \begin{subfigure}[t]{0.35\textwidth}
            \centering
            \includegraphics[width=\linewidth]{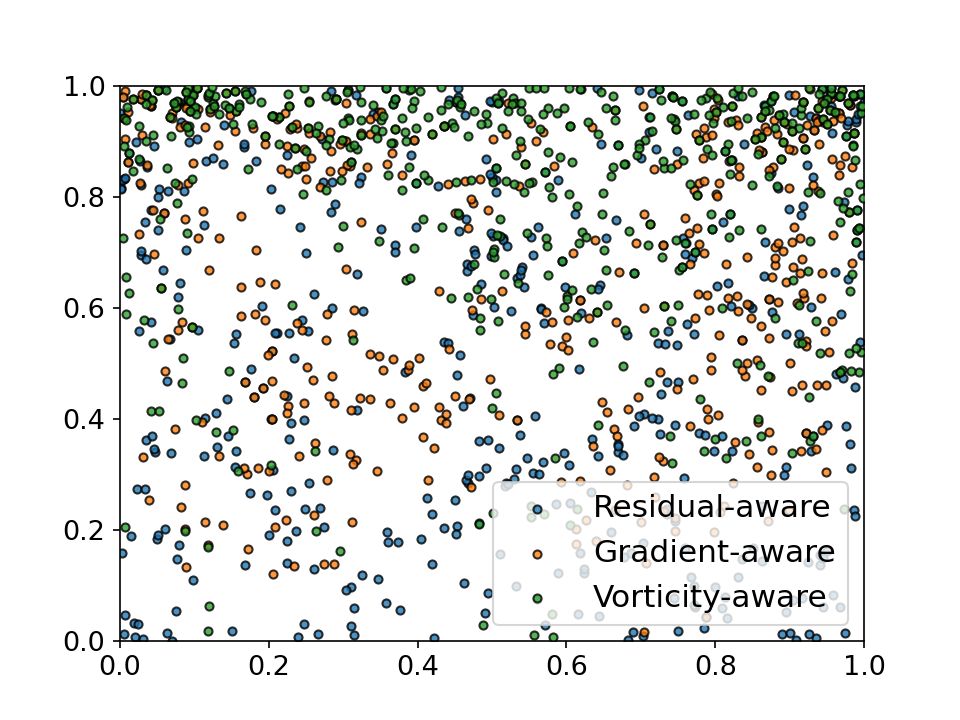}
            \caption{$k=1$}
            \label{fig:k_b}
        \end{subfigure}
        
        % \vspace{\baselineskip} % Adjust the vertical space between top and bottom
        \vfill
        
        \begin{subfigure}[t]{0.35\textwidth}
            \centering
            \includegraphics[width=\linewidth]{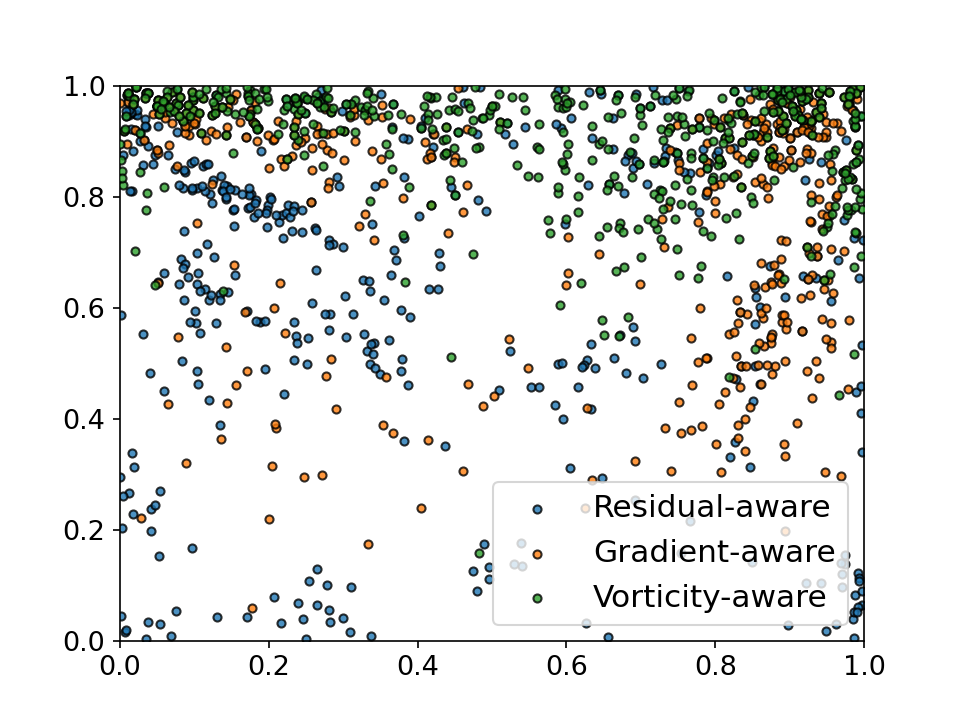}
            \caption{$k=2$}
            \label{fig:k_c}
        \end{subfigure}
        \hspace{0.02\textwidth}
        \begin{subfigure}[t]{0.35\textwidth}
            \centering
            \includegraphics[width=\linewidth]{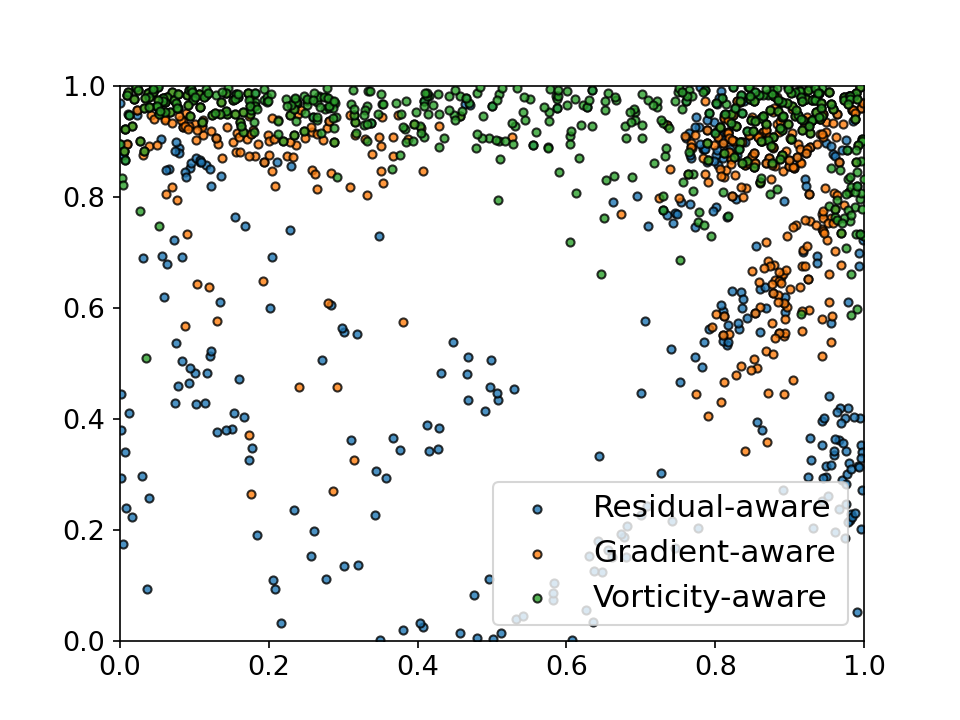}
            \caption{$k=3$}
            \label{fig:k_d}
        \end{subfigure}
    \end{minipage}
    
    \caption{Newly added adaptive points according to each criterion for different $k$ values. Note that all figures are the results of the last iteration in the RGV-111 model.}
    \label{fig:pts}
\end{figure*}

A parametric study is conducted for $k\in[0.5, 1, 2, 3]$. Fig. \ref{fig:pts} shows the $N$ updated collocation points at the last adaptive sampling iteration in RGV-111. There are three labels for the points, and each group shows the newly added points according to the high residual/gradient/vorticity value. As expected, the lower the $k$, the updated points are evenly distributed as in the random sampling technique, while the higher the $k$, the more collocation points are added to the specific regions. The quantitative RMSE between DNS from $160\times160$ grid and PINNs are summarized in Table \ref{tab:k_effect}. The average and minimum RMSE values among all trained PINNs within the same $k$ value are also presented: again, since the four models are trained for each case, the minimum RMSE value is derived through all trained models. The interesting point is that although vorticity-aware adaptive sampling is the most indirect algorithm for the training of PINN compared to residual/gradient-aware sampling, it manages to achieve moderate or even better performance. For example, sampling with only vorticity-aware technique (RGV-001) is the best in $k=0.5$ cases, and the residual-gradient-vorticity-aware technique (RGV-111) is the best in $k=1$ cases. To further investigate if $k$ affects the accuracy of PINNs, one-way analysis of variance (ANOVA) is performed \cite{heiman2001understanding}: within the same sampling approach, 100 y-velocity RMSE values at different locations are extracted for every $k$ value. The results are shown in the right part of Table \ref{tab:k_effect}, and it can be confirmed that only the RGV-001 approach has a statistically significant difference (p-value < 0.05) due to varying $k$ --- which indicates that the appropriate selection of $k$ value is critical in vorticity-aware sampling. Since the $k=3$ case has the minimum RMSE and to provide at least some contrast between the different sampling algorithms, it is adopted throughout this paper.

\renewcommand{\arraystretch}{1.2}
\begin{table}[htb!]
\centering
\caption{RMSE of y-velocity for different $k$ values. Four models are trained for each case, therefore mean and minimum values of all trained models are shown in the last two rows. Results of one-way ANOVA are also presented to examine the effect of $k$ in each sampling method.}\label{tab:k_effect}
    \begin{NiceTabular*}{0.65\columnwidth}{@{\extracolsep{\fill}}c|cccc|cc}
    % \extracolsep{\fill}
    \cline{1-7}
    % \Xhline{pt}
    \Block[c]{2-1}{} & \Block[c]{1-4}{$k$} &&&& \Block[c]{1-2}{ANOVA}\\ %\cline{5}
    & 0.5 & 1 & 2 & 3 & F statistic & p-value \\ \cline{1-7}
    % Van & 0.0360 	&	0.0360 	&	0.0360 	&	0.0360 \\
    Van & 0.0360 	&	- 	&	- 	&	- & - & - \\
    RGV-100 & 0.0440 	&	0.0532 	&	0.0646 	&	0.0810 & 2.3676 & 0.0704\\
    RGV-010 &	0.0349 	&	0.0393 	&	0.0308 	&	\textbf{0.0273} & 1.9975 & 0.1138 \\
    RGV-001 & \textbf{0.0298} 	&	0.0490 	&	0.0644 	&	0.0388 & 4.4502 & \textbf{0.0043}\\
    RGV-110 &	0.0531 	&	0.0501 	& \textbf{0.0301} 	&	0.0336 & 2.2290 & 0.0843\\
    RGV-101 &	0.0550 	&	0.0424 	& 0.0378 	&	0.0422 & 1.2376 & 0.2957\\
    RGV-011 &	0.0398 	&	0.0441 	& 0.0332 	&	0.0490 & 1.3829 & 0.2475\\
    RGV-111 & 0.0319	&	\textbf{0.0332} 	&	0.0361 	&	0.0422 & 0.8059 & 0.4911\\ \cline{1-7}
    mean &	\textbf{0.0406}	&	0.0434 	&	0.0416 	&	0.0437 & - & - \\
    min &	0.0186 	&	0.0202 	&	0.0230 	&	\textbf{0.0177} & - & - \\
    \cline{1-7}

    \end{NiceTabular*}
\end{table}
% \begin{table}[htb!]
% \centering
% \caption{RMSE of y-velocity for different $k$ values. Four models are trained for each case, therefore mean and minimum values of all trained models are shown in the last two rows.}\label{tab:k_effect}
%     \begin{NiceTabular*}{0.5\columnwidth}{@{\extracolsep{\fill}}c|cccc}
%     % \extracolsep{\fill}
%     \cline{1-5}
%     % \Xhline{pt}
%     \Block[c]{2-1}{} & \Block[c]{1-4}{$k$} \\ %\cline{5}
%     & 0.5 & 1 & 2 & 3 \\ \cline{1-5}
%     % Van & 0.0360 	&	0.0360 	&	0.0360 	&	0.0360 \\
%     Van & 0.0360 	&	- 	&	- 	&	- \\
%     RGV-100 & 0.0440 	&	0.0532 	&	0.0646 	&	0.0810 \\
%     RGV-010 &	0.0349 	&	0.0393 	&	0.0308 	&	\textbf{0.0273} \\
%     RGV-001 & \textbf{0.0298} 	&	0.0490 	&	0.0644 	&	0.0388 \\
%     RGV-110 &	0.0531 	&	0.0501 	& \textbf{0.0301} 	&	0.0336 \\
%     RGV-101 &	0.0550 	&	0.0424 	& 0.0378 	&	0.0422 \\
%     RGV-011 &	0.0398 	&	0.0441 	& 0.0332 	&	0.0490 \\
%     RGV-111 & 0.0319	&	\textbf{0.0332} 	&	0.0361 	&	0.0422 \\ \cline{1-5}
%     mean &	\textbf{0.0406}	&	0.0434 	&	0.0416 	&	0.0437 \\
%     min &	0.0186 	&	0.0202 	&	0.0230 	&	\textbf{0.0177} \\
%     \cline{1-5}

%     \end{NiceTabular*}
% \end{table}

\subsubsection{Effects of the updated collocation sampling ratio ($\frac{N}{M}$)}\label{sec:100ratio}

\begin{figure*}[htb!]
    \centering
    \begin{subfigure}[h]{0.9\textwidth}
        \includegraphics[trim=0 0 0 21,clip, width=\linewidth]{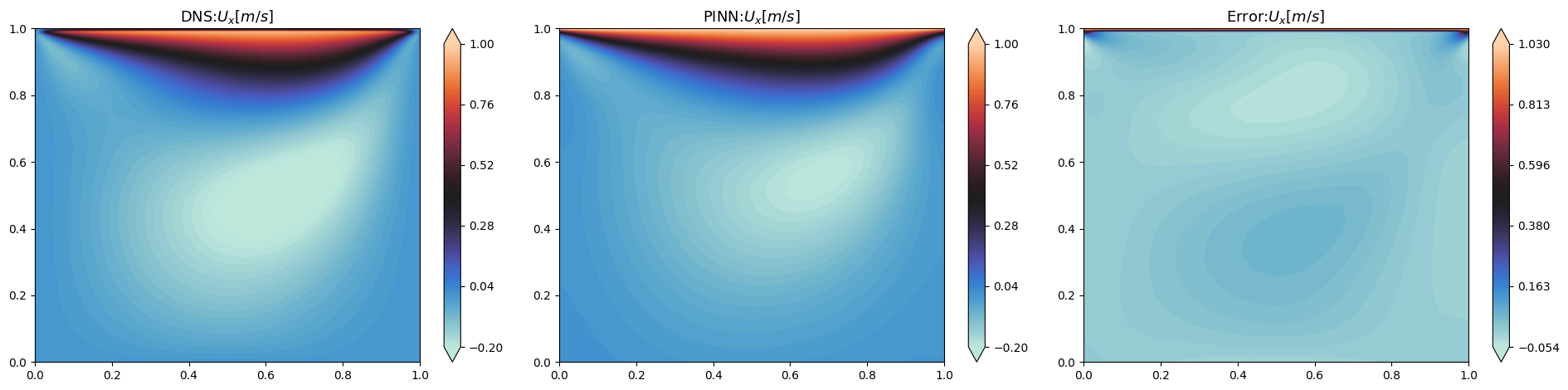}
    \caption{Flow fields of x-velocity: (left) DNS, (middle) PINN, (right) error.}
    \label{fig:re100a}
    \end{subfigure}
    
    \vfill
    \begin{subfigure}[h]{0.9\textwidth}
    \centering
        \includegraphics[trim=0 0 0 21.6,clip, width=\linewidth]{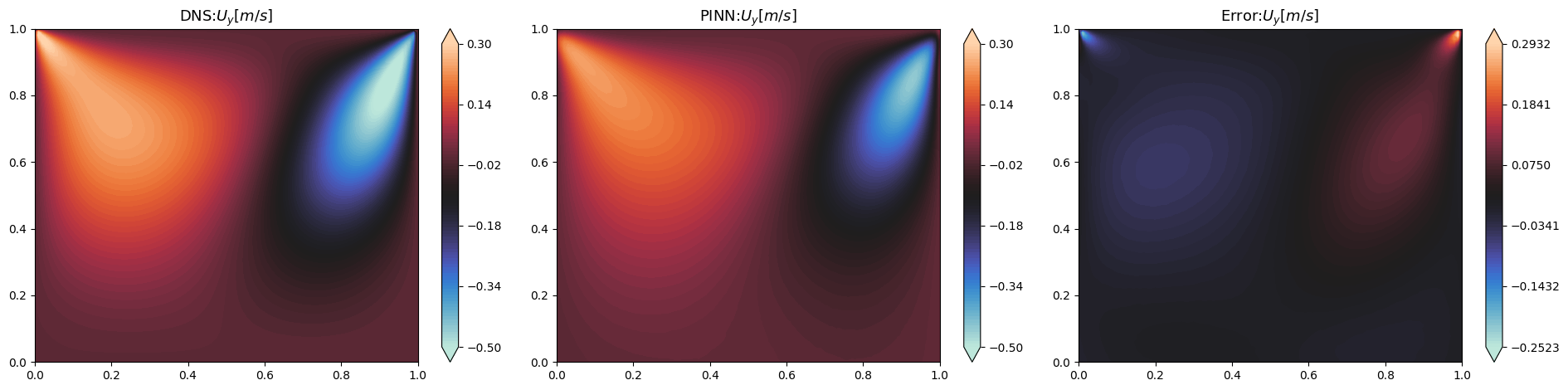}
    \caption{Flow fields of y-velocity: (left) DNS, (middle) PINN, (right) error.}
    \label{fig:re100b}
    \end{subfigure}
    \caption{Comparison of velocity components between DNS with $160\times160$ grid and PINN: (a) x-velocity, (b) y-velocity. The flow fields of PINNs are obtained from the best model among the four trained RGV-111 models.}
    \label{fig:re100}
\end{figure*}

With $k=3$ chosen in Section \ref{sec:100k}, this section focuses on the effects of the updated collocation sampling ratio, $\frac{N}{M}$. Parametric studies are conducted for $\frac{N}{M} \in [0.3, 0.4, 0.5, 0.6, 0.7, 0.8, 0.9]$. Herein, note that out of 2,000 collocation points, $\frac{N}{M}=0.3$ indicates 600 points are updated according to the residual/gradient/vorticity criteria, and the other 1,400 points are updated randomly. The RMSE results of y-velocity are shown in Table \ref{tab:ratio_effect}. Again, the vorticity-aware approaches (RGV-001 and RGV-101) achieve comparable performance in certain cases: $\frac{N}{M}=0.3, 0.6, 0.8$. This highlights the potential of user-defined adaptive sampling when the appropriate criterion is chosen, considering that existing approaches such as residual-aware or gradient-aware are much more intuitive and direct ways than focusing on the high vorticity magnitude regions. Throughout all sampling techniques, the mean values of RMSE are best when $\frac{N}{M}=0.3, 0.6$. A noteworthy point in Tables \ref{tab:k_effect} and \ref{tab:ratio_effect} is that, depending on the sampling technique and the $k$ and $\frac{N}{M}$ values, adaptive sampling techniques can have worse accuracy than the random sampling technique, Van. These results indicate that the use of adaptive sampling does not always improve PINN performance, but that careful selection of the adaptive sampling type and its hyperparameters is crucial. Again, one-way ANOVA is conducted and except for RGV-011, all sampling approaches are affected by the value of $\frac{N}{M}$ in that they have p-values less than 0.05. Finally, together with the results in Section \ref{sec:100k}, hyperparameters of $k=3$ and $\frac{N}{M}=0.6$ are adopted for the rest of this study. Before moving on to the next section for higher Reynolds numbers, the flow fields with respect to both velocity components for the RGV-111 model are visualized in Fig. \ref{fig:re100}: it can be confirmed that the data-free PINNs can predict the overall trends of the flow fields, although some details are captured at low resolutions.

\renewcommand{\arraystretch}{1.2}
\begin{table}[htb!]
\centering
\caption{RMSE of y-velocity for different $\frac{N}{M}$ values. Four models are trained for each case, therefore mean and minimum values of all trained models are shown in the last two rows. Results of one-way ANOVA are also presented to examine the effect of $\frac{N}{M}$ in each sampling method.}\label{tab:ratio_effect}
    \begin{NiceTabular*}{0.85\columnwidth}{@{\extracolsep{\fill}}c|ccccccc|cc}
    % \extracolsep{\fill}
    \cline{1-10}
    \Block{2-1}{} & \Block{1-7}{$N/M$} &&&&&&& \Block{1-2}{ANOVA}\\ %\cline{2-8}
    & 0.3 & 0.4 & 0.5 & 0.6 & 0.7 & 0.8 & 0.9 & F statistic & p-value\\ \cline{1-10}
    % Van &	0.0360 	&	0.0360 	&	0.0360 	&	0.0360 	&	0.0360 	&	0.0360 	&	0.0360\\
    Van &	0.0360 	&	- 	&	- 	&	- 	&	- 	&	- 	&	- & - & - \\
    RGV-100 &	0.0318 	&	0.0563 	&	0.0364 	&	0.0547 	&	0.0810 	&	0.0372 	&	\textbf{0.0328} & 2.2381 & \textbf{0.0380}\\
    RGV-010 &	0.0619 	&	0.0443 	&	\textbf{0.0310} 	&	0.0371 	&	\textbf{0.0273} 	&	0.0335 	&	0.0391 & 3.0807 & \textbf{0.0055}\\
    RGV-001 &	\textbf{0.0287} 	&	0.0354 	&	0.0342 	&	\textbf{0.0309} 	&	0.0388 	&	0.0524 	&	0.0378 & 3.0903 & \textbf{0.0054}\\
    RGV-110 &	0.0353 	&	\textbf{0.0334} 	&	0.0574 	&	0.0329 	&	0.0336 	&	0.0403 	&	0.0596 & 3.0241 & \textbf{0.0063}\\
    RGV-101 &	0.0488 	&	0.0368 	&	0.0390 	&	0.0421 	&	0.0422 	&	\textbf{0.0332} 	&	0.0703 & 2.8077 & \textbf{0.0105}\\
    RGV-011 &	0.0307 	&	0.0489 	&	0.0397 	&	0.0352 	&	0.0490 	&	0.0358 	&	0.0413 & 1.2672 & 0.2702\\
    RGV-111 &	0.0354 	&	0.0650 	&	0.0456 	&	0.0397 	&	0.0422 	&	0.0498 	&	0.0501 & 3.3599 & \textbf{0.0029} \\ \cline{1-10}
    mean &	\textbf{0.0386} 	&	0.0445 	&	0.0399 	&	\textbf{0.0386} 	&	0.0437 	&	0.0398 	&	0.0459 & - & -\\
    min &	0.0208 	&	0.0232 	&	0.0243 	&	0.0226 	&	\textbf{0.0177} 	&	0.0207 	&	0.0227 & - & -\\
    \cline{1-10}

    \end{NiceTabular*}
\end{table}
% \begin{table}[htb!]
% \centering
% \caption{RMSE of y-velocity for different $\frac{N}{M}$ values. Four models are trained for each case, therefore mean and minimum values of all trained models are shown in the last two rows.}\label{tab:ratio_effect}
%     \begin{NiceTabular*}{0.73\columnwidth}{@{\extracolsep{\fill}}c|ccccccc}
%     % \extracolsep{\fill}
%     \cline{1-8}
%     \Block{2-1}{} & \Block{1-7}{$N/M$} \\ %\cline{2-8}
%     & 0.3 & 0.4 & 0.5 & 0.6 & 0.7 & 0.8 & 0.9 \\ \cline{1-8}
%     % Van &	0.0360 	&	0.0360 	&	0.0360 	&	0.0360 	&	0.0360 	&	0.0360 	&	0.0360\\
%     Van &	0.0360 	&	- 	&	- 	&	- 	&	- 	&	- 	&	-\\
%     RGV-100 &	0.0318 	&	0.0563 	&	0.0364 	&	0.0547 	&	0.0810 	&	0.0372 	&	\textbf{0.0328}\\
%     RGV-010 &	0.0619 	&	0.0443 	&	\textbf{0.0310} 	&	0.0371 	&	\textbf{0.0273} 	&	0.0335 	&	0.0391\\
%     RGV-001 &	\textbf{0.0287} 	&	0.0354 	&	0.0342 	&	\textbf{0.0309} 	&	0.0388 	&	0.0524 	&	0.0378\\
%     RGV-110 &	0.0353 	&	\textbf{0.0334} 	&	0.0574 	&	0.0329 	&	0.0336 	&	0.0403 	&	0.0596\\
%     RGV-101 &	0.0488 	&	0.0368 	&	0.0390 	&	0.0421 	&	0.0422 	&	\textbf{0.0332} 	&	0.0703\\
%     RGV-011 &	0.0307 	&	0.0489 	&	0.0397 	&	0.0352 	&	0.0490 	&	0.0358 	&	0.0413\\
%     RGV-111 &	0.0354 	&	0.0650 	&	0.0456 	&	0.0397 	&	0.0422 	&	0.0498 	&	0.0501\\ \cline{1-8}
%     mean &	\textbf{0.0386} 	&	0.0445 	&	0.0399 	&	\textbf{0.0386} 	&	0.0437 	&	0.0398 	&	0.0459\\
%     min &	0.0208 	&	0.0232 	&	0.0243 	&	0.0226 	&	\textbf{0.0177} 	&	0.0207 	&	0.0227\\
%     \cline{1-8}

%     \end{NiceTabular*}
% \end{table}

\subsubsection{Effects of the number of collocation points ($M$)}\label{sec:num_collo}

This section investigates the effects of the number of collocation points ($M$) using the previously selected hyperparameters $k=3$ and $N/M=0.6$. The number of points varies from 500 to 3,000, with an increment of 500. Only the RGV-001 approach is tested, as it was found to be the best approach when $N/M=0.6$ in Table \ref{tab:ratio_effect}. The results are presented in Table \ref{tab:num_collo}. Contrary to expectations, as the number of points increases, there is no clear trend towards the improvement of results. This inconsistency can be attributed to the stochasticity in the initial sampling and adaptive sampling techniques, which can significantly affect the final accuracy of the PINNs. Also, it is noteworthy that, with an adequate combination of the $N/M$ value, a smaller $M$ value can yield superior results compared to larger values. For example, $M=2,000$ performs better than $M=2,500$ and $M=3,000$ when $N/M=0.6$. From another perspective, it can also be inferred that $M=2,500$ and $M=3,000$ can be better than $M=2,000$ if the appropriate $N/M$ values are chosen for their cases.

\begin{table}[htb!]
\centering
\caption{RMSE of y-velocity for different \( M \) values: only RGV-001 is tested.}\label{tab:num_collo}
    \begin{tabular}{c|cccccc}
    \hline
    \( M \) & 500 & 1,000 & 1,500 & 2,000 & 2,500 & 3,000 \\ \hline
    RMSE & 0.0590 & 0.0336 & 0.0487 & \textbf{0.0309} & 0.0430 & 0.0415 \\ \hline
    \end{tabular}
\end{table}

\subsubsection{Remarks on adaptive sampling}\label{sec:remark}

Sections \ref{sec:100k}, \ref{sec:100ratio}, and \ref{sec:num_collo} compare the performances of various sampling techniques in data-free PINNs at low Reynolds number. However, no clear trend is observed throughout parametric studies with respect to the stochastic intensity ($k$), updated sample ratio ($\frac{N}{M}$), and the number of collocation points ($M$). Nevertheless, the more comprehensive investigation of the adaptive sampling approaches is judged to be unnecessary for this study for the following reasons: 1) The first purpose of this section was not to find out which sampling is best, but to show how efficiently PINNs' meshless properties allow them to work automatically with different sampling techniques, 2) the second purpose was to shed a light on the potential of user-defined sampling approach tailored for the specific domain, not to exclusively argue that the proposed vorticity-aware approach is the best.

% \subsubsection{Effects of data-driven approaches}\label{sec:100DD}
% Data-driven approach is not required for Re100 case,,,
% \clearpage
\section{Failures of data-free PINNs at higher Reynolds}\label{sec:fail}

This paper notes the fact that most of the previous studies on lid-driven cavity flow focused on low Reynolds regions. For example, \citet{amalinadhi2022physics, jagtap2020conservative, li2022dynamic, bai2020applying, wang2021understanding} showed that PINNs successfully worked in the forward problem at $Re=100$. Also, \citet{wong2022learning} and \citet{chiu2022can} tested PINNs at $Re=400$ in forward problems. To bridge this gap between toy problems for academic validation and high Reynolds problems encountered in practice, this section investigates the higher Reynolds ($Re=1000, 3200$) using data-free PINN. Again, to compare the results of PINNs with conventional CFD approaches, DNS is performed by OpenFOAM. The settings for DNS including CFD solver, grids, and CPU core are the same as in Section \ref{sec:Re100}: grids of $20\times20$, $40\times40$, $80\times80$, and $160\times160$ are utilized and training time for each grid in $Re=1000$ is 17, 36, 100, and 1,853 seconds, respectively. For $Re=3200$ case, 20, 98, 473, and 4,231 seconds are taken per each grid. Their validation results are shown in Fig. \ref{fig:DNSvali}, Appendix \ref{sec:openfoam}.

\subsection{Experiments at \textit{Re=1000} and \textit{Re=3200}}\label{sec:fail_exp}

\begin{figure*}[htb!]
    \begin{minipage}[t]{\textwidth}
        \centering
        \begin{subfigure}[t]{0.35\textwidth}
            \centering
            \includegraphics[trim=0 0 0 20.9,clip, width=\linewidth]{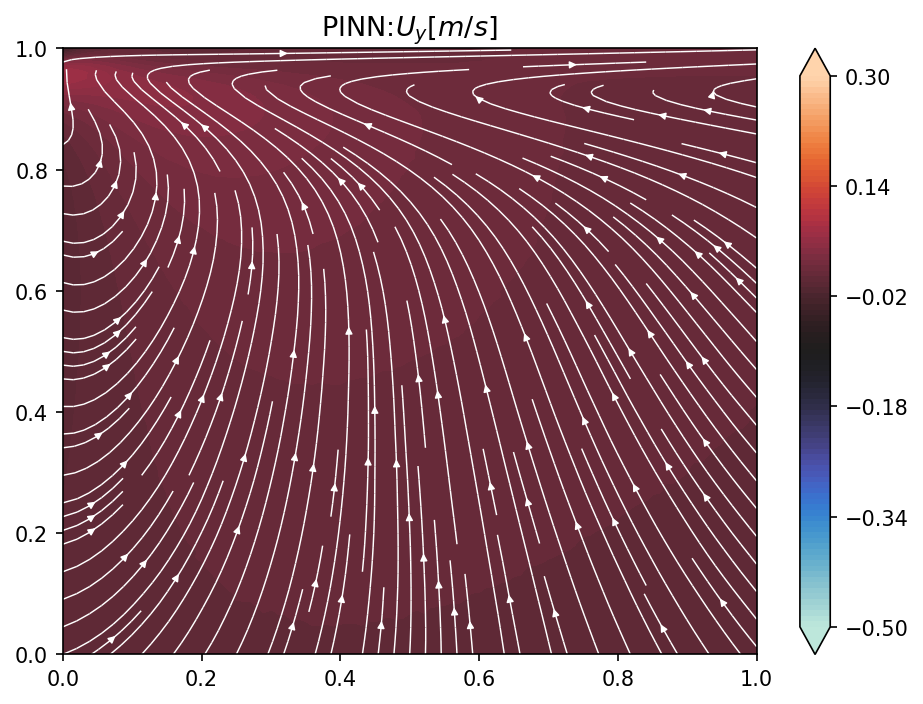}
            \caption{$Re=1000$}
            \label{fig:failed_a}
        \end{subfigure}
        \hspace{0.05\textwidth}
        \begin{subfigure}[t]{0.35\textwidth}
            \centering
            \includegraphics[trim=0 0 0 20.9,clip,width=\linewidth]{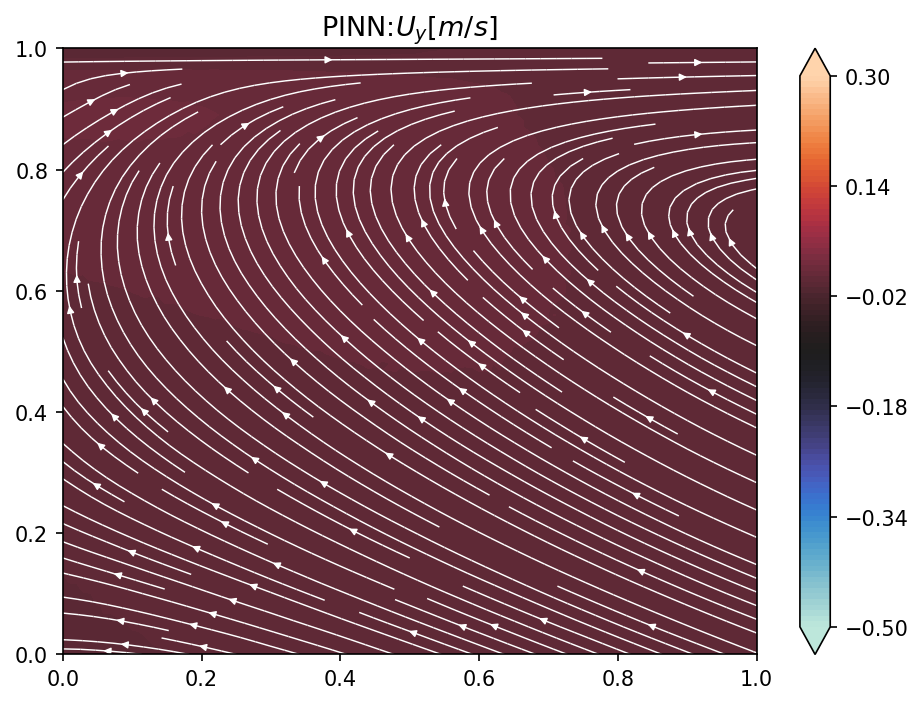}
            \caption{$Re=3200$}
            \label{fig:failed_b}
        \end{subfigure}
    \end{minipage}
    
    \caption{Prediction of y-velocity flow fields using data-free PINNs: (a) \textit{Re=1000}, (b) \textit{Re=3200}. Streamlines are also shown to highlight their failure.}
    \label{fig:failed}
\end{figure*}

Data-free PINNs with $N_{layer}=7, N_{node}=32, k=3, \frac{N}{M}=0.6$ where $M=5,000$ are trained for $Re=1000$ and $Re=3200$ cases. As in Section \ref{sec:100hyper}, only ``Van'' sampling is applied. Finally, both cases require a training wall time of 900 seconds: considering that DNS requires 1,853 and 4,231 seconds for the $160\times160$ grid, the computational efficiency of PINNs can be verified. However, they completely failed to predict y-velocity fields along with unrealistic streamlines as shown in Fig. \ref{fig:failed}. Along with unrealistic streamlines, they completely failed to predict y-velocity fields. More than this case, additional efforts with different hyperparameters ($N_{layer} \in \{5, 7, 9\}$ \& $N_{node} \in \{32, 64, 128\}$) are also tried to find out if the failure is due to insufficient model capacity, but all attempts fail to predict reasonable flow fields.

\subsection{Visualization of loss landscapes}

Since all data-free PINN models failed at higher Reynolds numbers, we then try to investigate the underlying reason by visualizing the loss landscapes of NNs. The loss landscape in deep learning refers to the representation of loss values ($\mathcal{L}$) with respect to the model parameters of NNs ($\theta$). When there are only two model parameters, one can easily visualize the loss landscape of the model with a 3D contour plot: $x$, $y$ axes with two model parameters, and $z$ axis with loss values. However, in practice, there are many more model parameters in NNs, and therefore some technique for the projection of high-dimensional model parameters to 2D space is required. To this end, parameters of the already trained model, $\theta^*$, are projected onto the 2D surface consisting of $\delta$ and $\eta$ axes as below \cite{goodfellow2014qualitatively}:

\begin{equation}
\label{eq:landscape}
f(\alpha, \beta) = \mathcal{L}(\theta^* + \alpha \delta + \beta \eta)
\end{equation}

To utilize Eq. \ref{eq:landscape} to draw high-resolution plots without scale invariance problems, \citet{li2018visualizing} proposed the use of filter-wise normalized directions (refer to Ref. \cite{li2018visualizing} for more details). 

\begin{figure*}[htb!]
    \begin{minipage}[t]{\textwidth}
        \centering
        \begin{subfigure}[t]{0.3\textwidth}
            \centering
            \includegraphics[trim=90 45 60 70,clip, width=\linewidth]{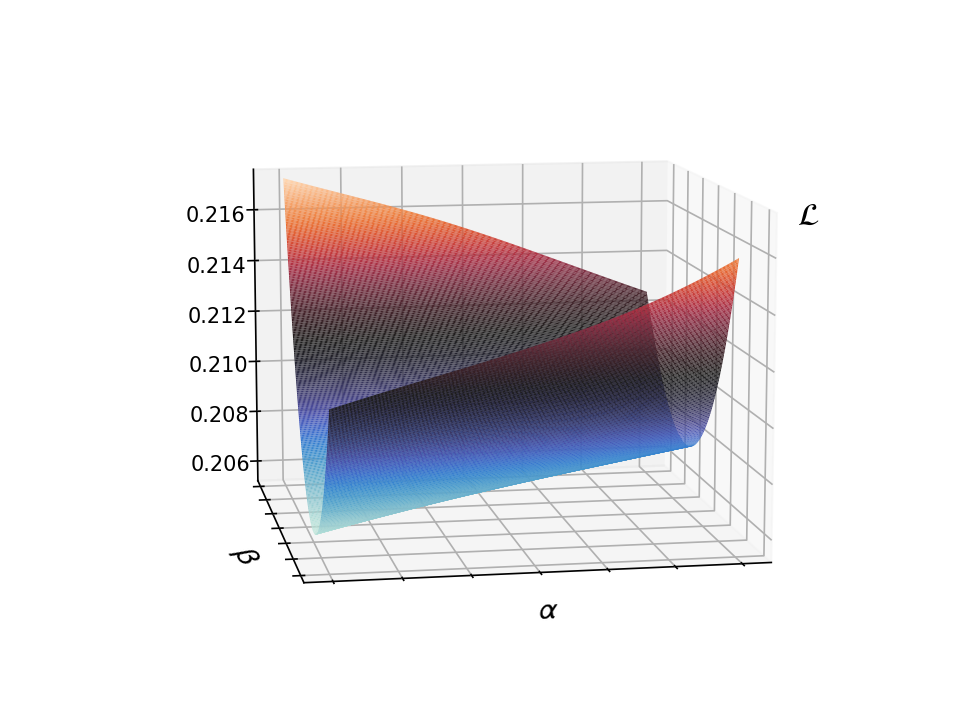}
            % \caption{Data-free (model 1)}
            \label{fig:loss_a}
        \end{subfigure}
        \hfill
        \begin{subfigure}[t]{0.3\textwidth}
            \centering
            \includegraphics[trim=90 45 60 70,clip, width=\linewidth]{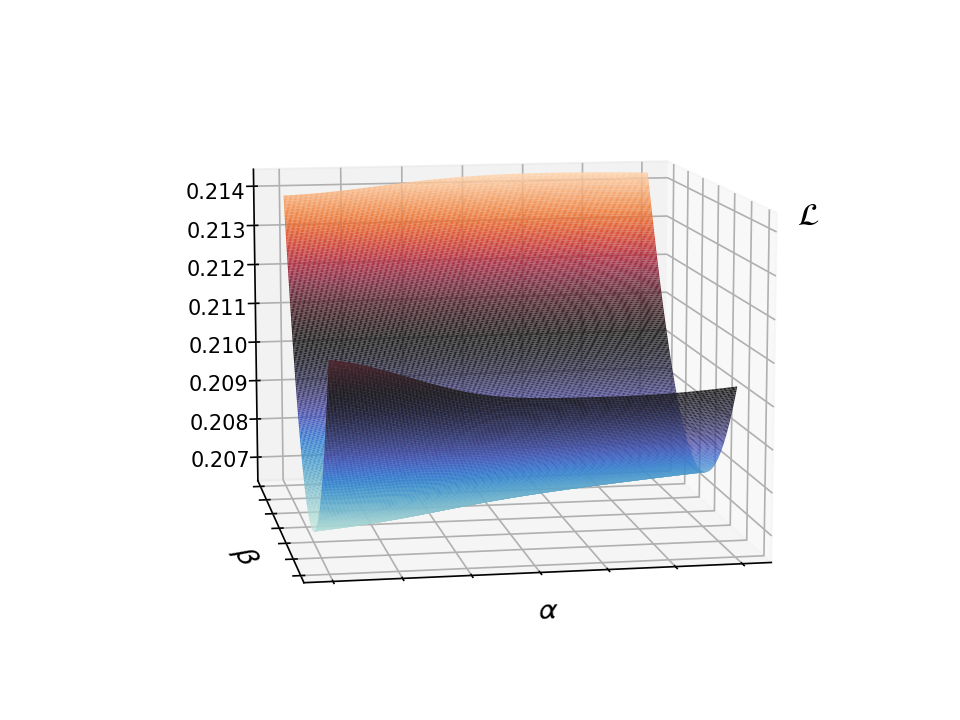}
            \caption{Data-free PINNs}
            \label{fig:loss_b}
        \end{subfigure}
        \hfill
        \begin{subfigure}[t]{0.3\textwidth}
            \centering
            \includegraphics[trim=90 45 60 70,clip, width=\linewidth]{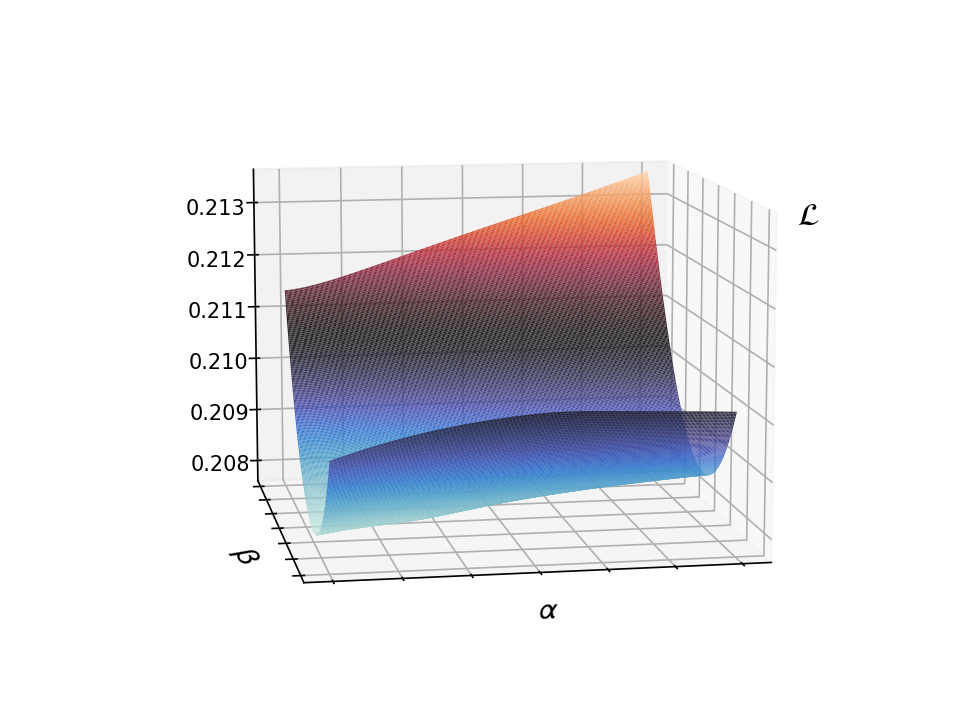}
            % \caption{Data-free (model 3)}
            \label{fig:loss_c}
        \end{subfigure}
        
        \vfill
        
        \centering
        \begin{subfigure}[t]{0.3\textwidth}
            \centering
            \includegraphics[trim=100 45 50 70,clip, width=\linewidth]{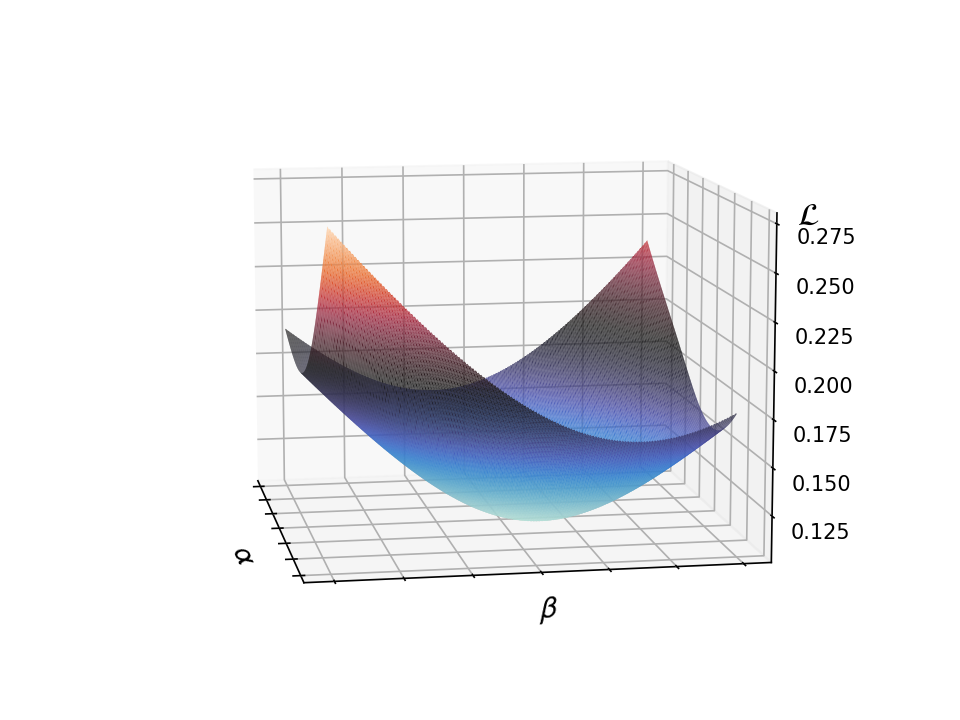}
            % \caption{Data-driven (model 1)}
            \label{fig:loss_d}
        \end{subfigure}
        \hfill
        \begin{subfigure}[t]{0.3\textwidth}
            \centering
            \includegraphics[trim=100 45 50 70,clip, width=\linewidth]{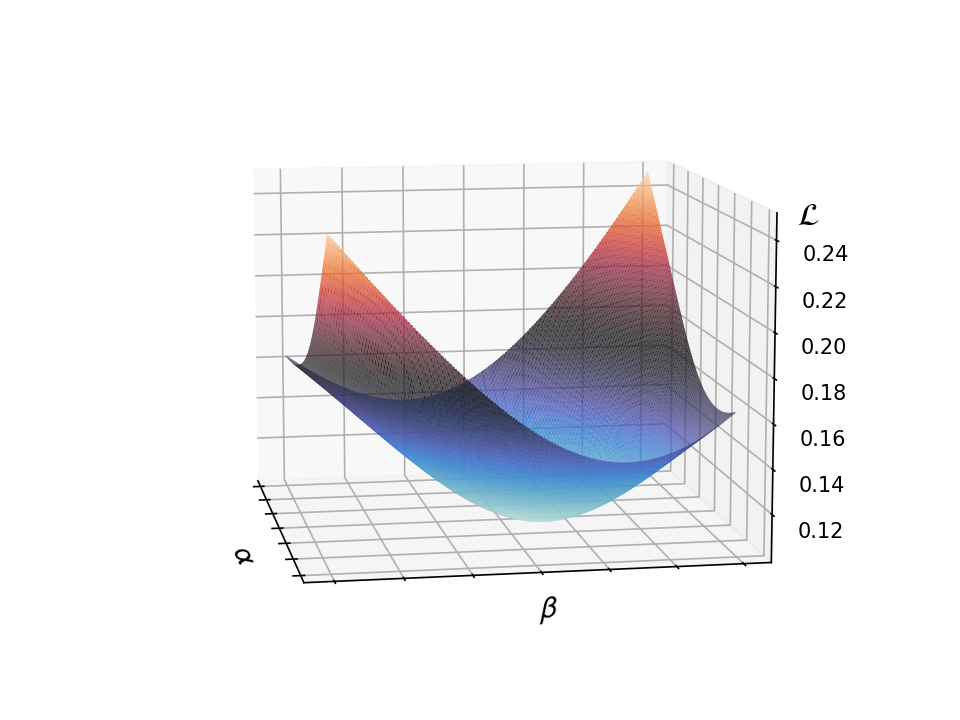}
            \caption{DD-PINNs}
            \label{fig:loss_e}
        \end{subfigure}
        \hfill
        \begin{subfigure}[t]{0.3\textwidth}
            \centering
            \includegraphics[trim=100 45 50 70,clip, width=\linewidth]{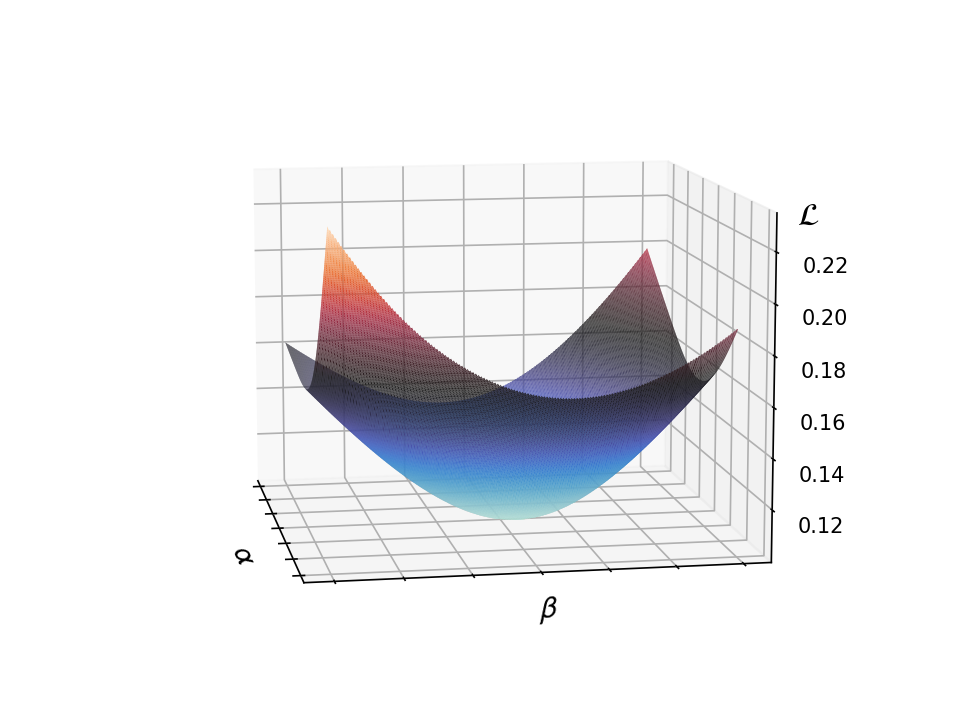}
            % \caption{Data-driven (model 3)}
            \label{fig:loss_f}
        \end{subfigure}
    \end{minipage}
    
    \caption{Loss landscape of PINNs at \textit{Re=3200}: (a) data-free PINNs: from left to right, minimum losses of the trained models are 0.205, 0.207, and 0.208, (b) DD-PINNs: from left to right, minimum losses of the trained models are 0.104, 0.102, and 0.102. Three different models of each type of PINN (with identical settings but trained differently due to the stochastic nature) are visualized.}
    \label{fig:lossland}
\end{figure*}

In fact, \citet{gopakumar2023loss} already deployed the loss landscapes to argue that data-driven PINNs have sharper minima than data-free PINNs. In a similar context, we visualized the loss landscapes to verify the failure of the data-free PINNs at $Re=3200$, as shown in Fig. \ref{fig:loss_b}. As mentioned earlier, all of the experiments in this study train models several times considering the stochastic behavior, and therefore three different data-free PINN models are visualized. Overall, their landscapes seem to have no discernible minima and show flatness (especially along the $\alpha$ axis) compared to the DD-PINNs in \ref{fig:loss_e}, which will be discussed in Section \ref{sec:remedy}. Given that non-convexity in the reduced dimensionality indicates that there must be non-convexity in the full dimension \cite{li2018visualizing}, one can infer that this is the cause of the failure in training data-free PINNs.

% \clearpage
\section{Remedy: data-driven PINNs (DD-PINNs)}\label{sec:remedy}

This section attempts to address the failures of data-free PINNs by leveraging simulation data. To this end, the DD-PINN concept is adopted as a remedy and we focus on $Re = 3200$ case. First, hyperparameter tuning is conducted (Section \ref{sec:3200tuning}). Then, the comprehensive investigations on DD-PINNs in terms of sampling approaches (Section \ref{sec:samplingEffect}), the fidelity of guide data (Section \ref{sec:fidelityeffect}), and number of guide data (Section \ref{sec:numeffect}) are performed.

\subsection{Settings and Hyperparameter tuning for DD-PINNs}\label{sec:3200tuning}

First, the settings required for DD-PINNs should be discussed. Since they leverage the dataset for guidance, we explored the different datasets obtained from DNS. To be more specific, datasets of different grid densities are obtained: grids of $20\times20$ (guide data size of 400), $40\times40$, $80\times80$, and $160\times160$. One can easily expect that as the grid becomes denser, the simulation time for preparing the dataset using CFD would increase while the accuracy of DD-PINN would improve. Note that this study made a simplifying assumption by considering the DNS data as a proxy for the experimental sensor data that would be used as data-driven points in the DD-PINN framework. This assumption allows us to focus on the development and evaluation of the DD-PINN methodology. However, it should be noted that in the realization of DT, a near real-time data acquisition/transmission process should be coupled with DD-PINN, including how the experimental data are acquired in real time and how the acquired data are incorporated into the DD-PINN framework, which is considered outside the scope of our work.

Then, as in the $Re=100$ case study (Section \ref{sec:100hyper}), hyperparameter tuning is performed for the $Re=3200$ case. The number of hidden layers ($N_{layer} \in \{5, 7, 9\}$), the number of nodes at each hidden layer ($N_{node} \in \{32, 64, 128\}$), and the weight of the data-driven loss term in Eq. \ref{eq:DDPINNloss} ($w_{data} \in \{1, 5, 10\}$) are set as the hyperparameters to be explored. PINNs are trained for 20,000 iterations with 5,000 collocation points and 5,000 boundary points. While there are techniques proposed for adaptively adjusting the weights of the loss terms, $w_{data}$ for this study, we decided not to apply them in order to focus on the pure effects of the data-driven loss term without changing its weight during the training procedure \cite{wang2021understanding, mcclenny2020self, wight2020solving}. Among the four grids mentioned above, a $40\times40$ grid is used as a guide dataset for the data-driven approach and the resulting RMSE values are shown in the Table \ref{tab:3200hyp_results} in Appendix \ref{sec:app_hyp}. It indicates that all cases succeed in predicting reasonable flowfields unlike data-free PINNs: among them, $N_{layer}=7, N_{node}=32, w_{data}=10$ with a training time of 960 seconds is one of the best options considering both velocity components and is therefore utilized for further investigation. Note that since both data-free PINN in Section \ref{sec:fail} and DD-PINN here are designed to have the same number of layers and nodes, it can be regarded that the failure of data-free PINN at high Reynolds number is not due to its insufficient model capacity.

To analyze how this DD-PINN model could be well trained compared to data-free PINNs, loss landscapes of DD-PINNs are also shown as Fig. \ref{fig:loss_e}. One can see that the DD-PINNs have much more convex valleys compared to the data-free PINNs in Fig. \ref{fig:loss_b}, which can be considered as the reason for their successful training with about half the loss value of the data-free PINNs.

\subsection{Effects of sampling techniques}\label{sec:samplingEffect}

\begin{figure*}[htb!]
    \begin{minipage}[t]{\textwidth}
        \centering
        \begin{subfigure}[t]{0.33\textwidth}
            \centering
            \includegraphics[width=\linewidth]{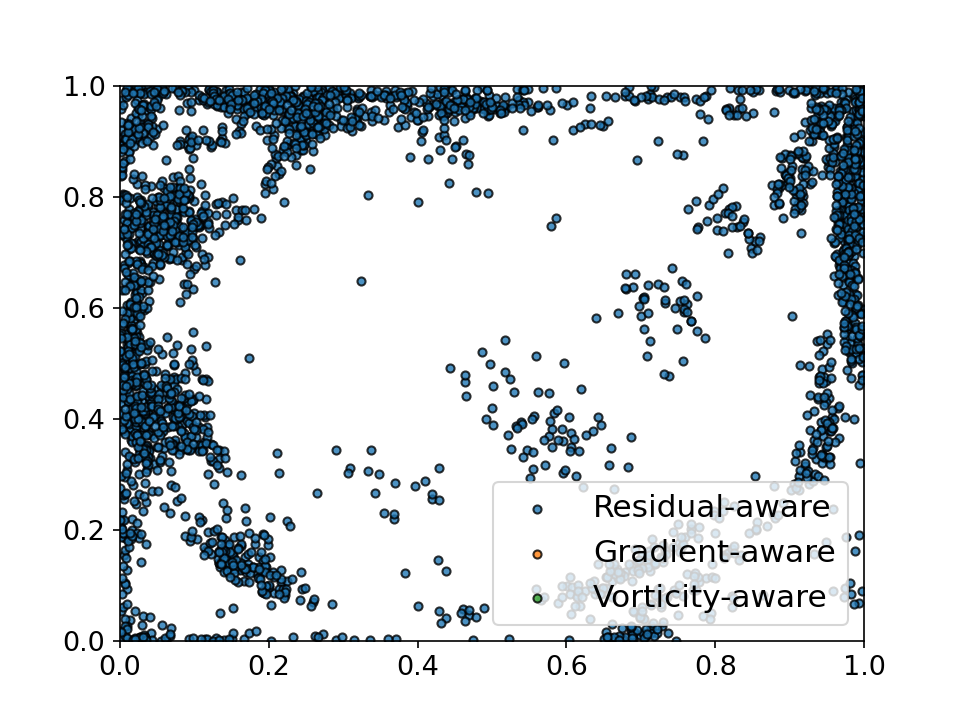}
            \caption{Residual-aware (RGV-100)}
            \label{fig:3200adapt_a}
        \end{subfigure}
        \hfill
        \begin{subfigure}[t]{0.33\textwidth}
            \centering
            \includegraphics[width=\linewidth]{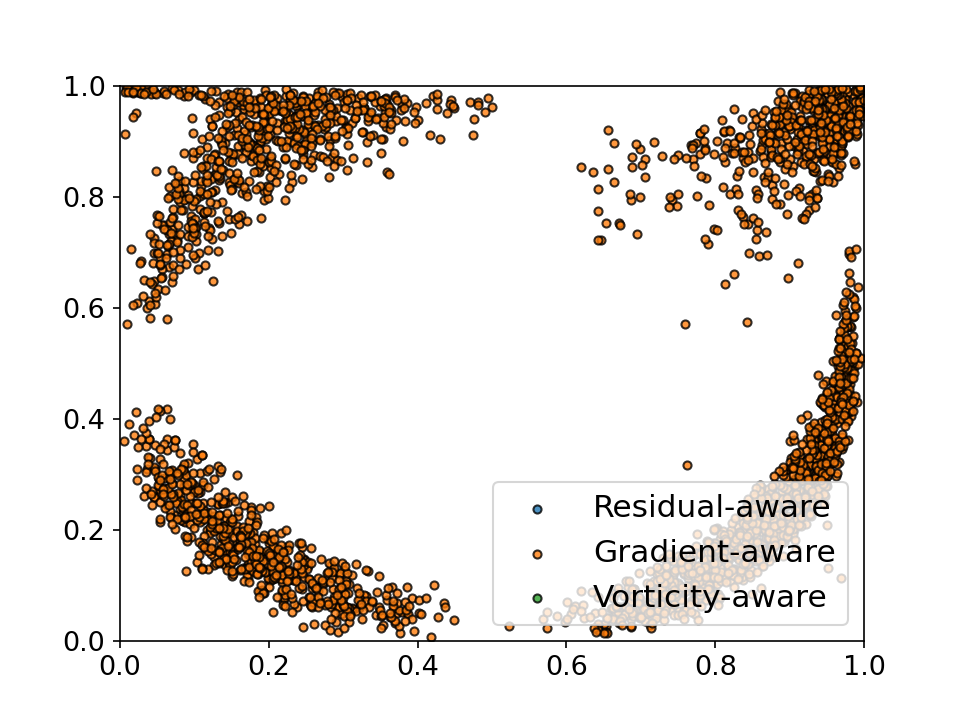}
            \caption{Gradient-aware (RGV-010)}
            \label{fig:3200adapt_b}
        \end{subfigure}
        \hfill
        \begin{subfigure}[t]{0.33\textwidth}
            \centering
            \includegraphics[width=\linewidth]{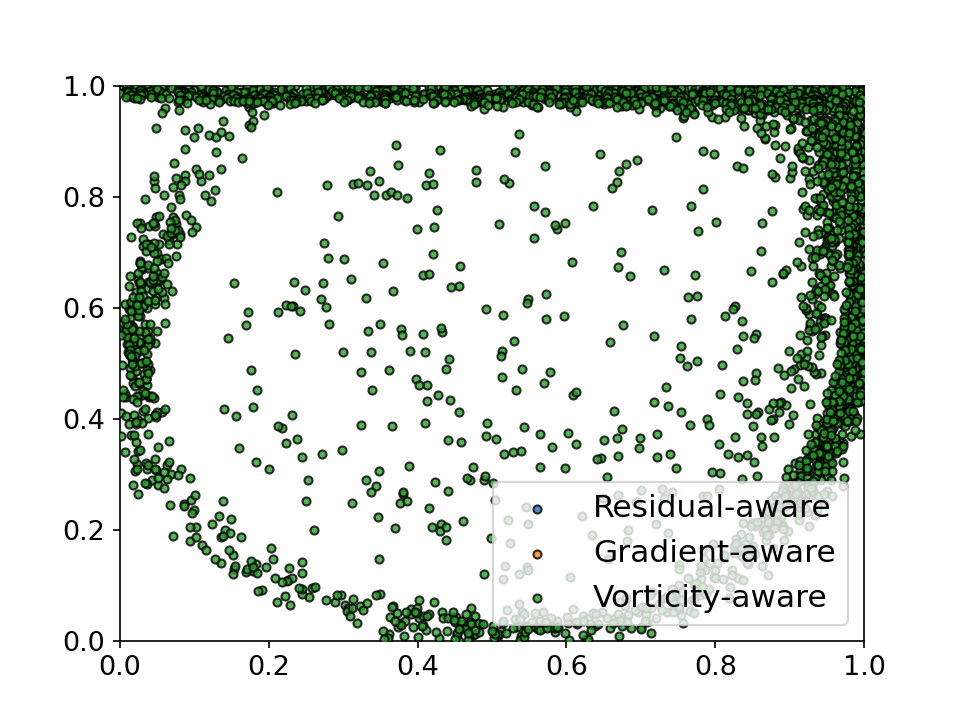}
            \caption{Vorticity-aware (RGV-001)}
            \label{fig:3200adapt_c}
        \end{subfigure}
    \end{minipage}
    
    \caption{Adaptive points added by (a) residual-aware, (b) gradient-aware, and (c) vorticity-aware approaches in DD-PINNs with guide data from $40\times40$ grid.}
    \label{fig:3200adapt}
\end{figure*}

Various adaptive sampling approaches adopted in Section \ref{sec:100para} are also compared for DD-PINNs with different grids. Table \ref{tab:3200RMSE} summarizes their results, and the most interesting point is that for all types of grids, the random sampling approach (Van) shows the best performance, although various adaptive sampling techniques still work properly as shown in Fig. \ref{fig:3200adapt}. Its reason can be inferred that the loss term for the guide data in DD-PINNs ($\mathcal{L}_{data}$) acts as the local regulator, constraining DD-PINNs to predict values similar to the labeled guide data at the specific locations. In this context, PDE loss functions ($\mathcal{L}_{mass} + \mathcal{L}_{x-momentum} + \mathcal{L}_{y-momentum}$) should act as a global calibrator, enabling DD-PINNs to learn the flow fields that globally satisfy the physical conservation laws. However, sampling algorithms working as shown in Fig. \ref{fig:3200adapt} are not consistent with this: residual/gradient/vorticity-aware techniques make DD-PINNs rather focus on the specific regions, which is the opposite of the expected role of PDE losses. In contrast, they could have performed better than random sampling in data-free PINNs, since there was no local regulator, $\mathcal{L}_{data}$. Therefore, the PDE losses in data-free PINNs should act as both local and global calibrators: the former is achieved by residual/gradient/vorticity-aware criteria at $N$ points, while the latter is realized by random sampling at $M-N$ points. This can also be deduced from the fact that the error gap between Van and the other adaptive samplings is most pronounced in the results of the $160\times160$ grid, where the role of the global calibrator is most significant due to the strong local regulator with the largest guide dataset.

In this regard, we additionally analyze the effects of $N/M$ again, to verify the trends in adaptive sampling approaches. $N/M \in [0.2, 0.4, 0.6, 0.8]$ are explored and their results are shown in Table \ref{tab:NM3200} in Appendix \ref{sec:app_adapt}: RMSE results between DD-PINNs and DNS with $160\times160$ grid are shown. Since the ``Van'' method is always the best regardless of the $N/M$, it can be concluded that for DD-PINNs, local refinement during adaptive sampling techniques (except Van) does not increase the accuracy of the PINNs. They rather decrease the accuracy due to their overlapping role of local refinement with a data-driven loss term. Therefore, only random sampling is used for further investigation on DD-PINNs.

\renewcommand{\arraystretch}{1.2}
\begin{table}[htb!]
\centering
\caption{RMSE of y-velocity for different guide datasets used in DD-PINNs.}\label{tab:3200RMSE}
    \begin{NiceTabular*}{0.5\columnwidth}{@{\extracolsep{\fill}}c|cccc}
    % \extracolsep{\fill}
    \cline{1-5}
    % \Xhline{pt}
    \Block[c]{2-1}{} & \Block[c]{1-4}{Grids} \\ %\cline{5}
    & $20\times20$ & $40\times40$ & $80\times80$ & $160\times160$ \\ \cline{1-5}
    Van & \textbf{0.1032} 	&	\textbf{0.0509} 	&	\textbf{0.0223} 	&	\textbf{0.0120} \\
    RGV-100 & 0.1056 	&	0.0534 	&	0.0237 	&	0.0152 \\
    RGV-010 &	0.1062 	&	0.0529 	&	0.0237 	&	0.0156 \\
    RGV-001 & 0.1075 	&	0.0536 	&	0.0272 	&	0.0176  \\
    RGV-110 &	0.1061 	&	0.0518 	&	0.0272 	&	0.0226 \\
    RGV-101 &	0.1067 	&	0.0533 	&	0.0280 	&	0.0208 \\
    RGV-011 &	0.1074 	&	0.0560 	&	0.0282 	&	0.0207 \\
    RGV-111 & 0.1079 	&	0.0554 	&	0.0261 	&	0.0195 \\ \cline{1-5}
    % mean &	0.1063 	&	0.0534 	&	0.0258 	&	0.0180 \\
    % \cline{1-5}

    \end{NiceTabular*}
\end{table}

\subsection{Effects of guide data fidelity}\label{sec:fidelityeffect}

\begin{figure*}[htb!]
    \begin{minipage}[t]{\textwidth}
        \centering
        \begin{subfigure}[t]{0.8\textwidth}
            \centering
            \includegraphics[trim=0 0 0 21.9,clip, width=\linewidth]{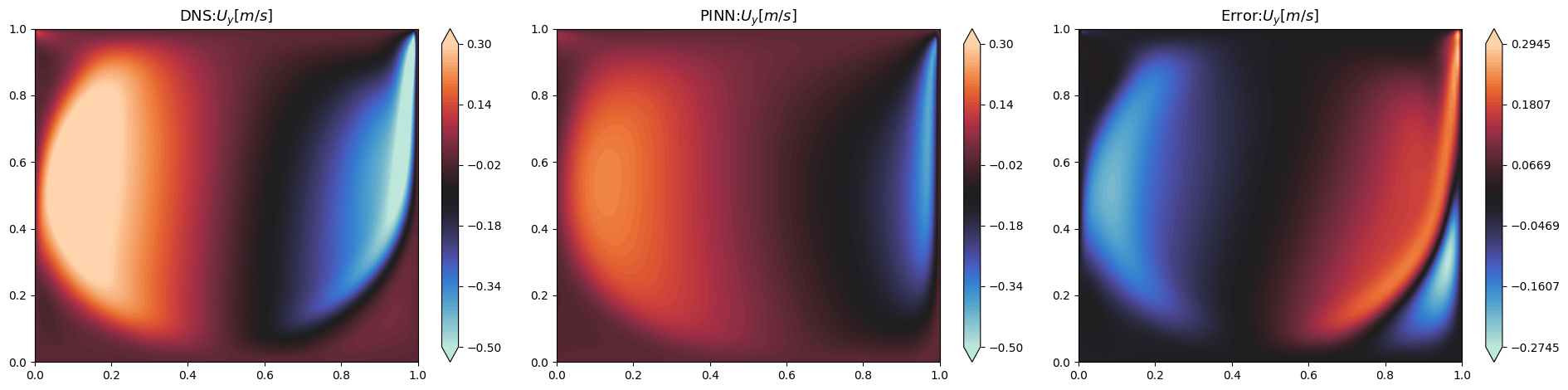}
            \caption{$20\times20$ grid ($RMSE=0.1032$)}
            \label{fig:DGff_a}
        \end{subfigure}
        
        \vfill
        
        \begin{subfigure}[t]{0.8\textwidth}
            \centering
            \includegraphics[trim=0 0 0 22.2,clip, width=\linewidth]{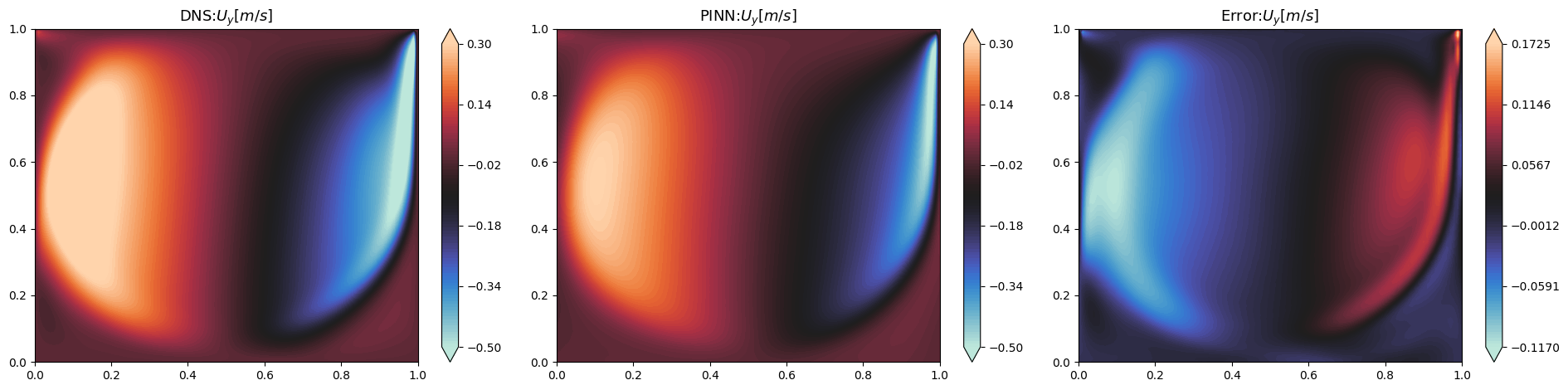}
            \caption{$40\times40$ grid ($RMSE=0.0509$)}
            \label{fig:DGff_b}
        \end{subfigure}
       
        \vfill
        
        \centering
        \begin{subfigure}[t]{0.8\textwidth}
            \centering
            \includegraphics[trim=0 0 0 22.4,clip,  width=\linewidth]{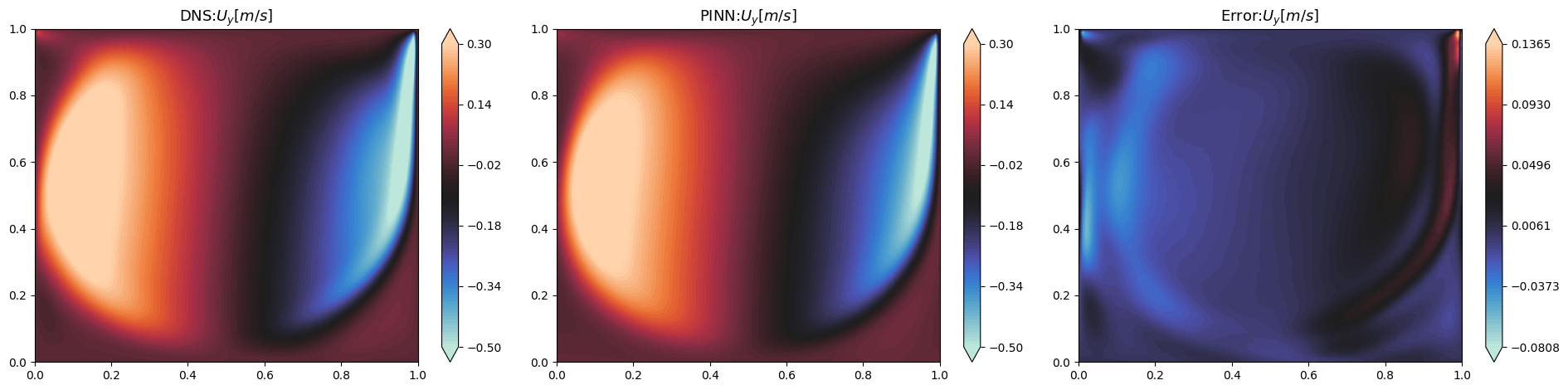}
            \caption{$80\times80$ grid ($RMSE=0.0223$)}
            \label{fig:DGff_c}
        \end{subfigure}
        
        \vfill
        
        \begin{subfigure}[t]{0.8\textwidth}
            \centering
            \includegraphics[trim=0 0 0 21.9,clip, width=\linewidth]{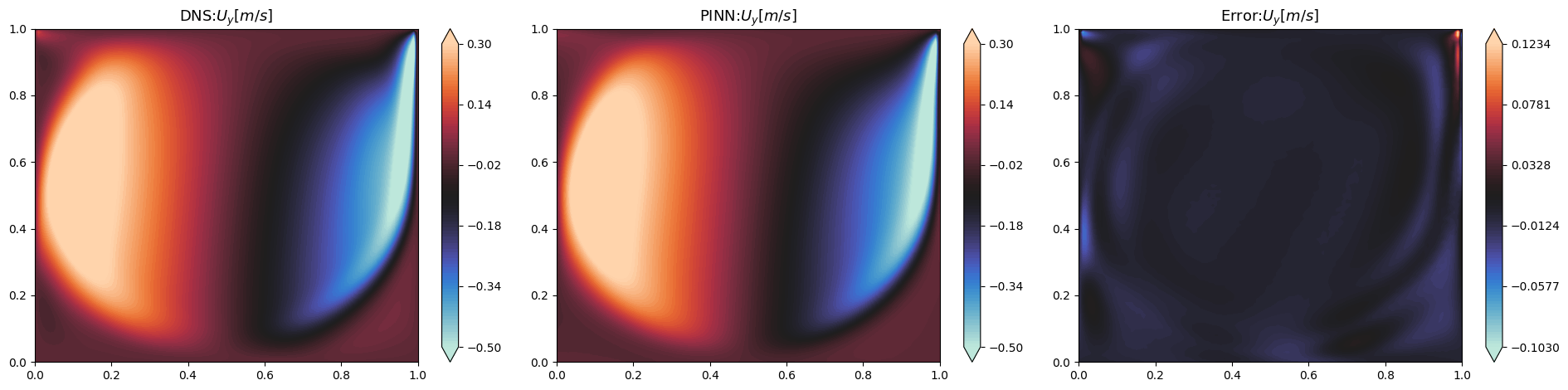}
            \caption{$160\times160$ grid ($RMSE=0.0120$)}
            \label{fig:DGff_d}
        \end{subfigure}
    \end{minipage}
    
    \caption{Flow fields with respect to y-velocity predicted by DD-PINNs with various fidelities of guide data: (left column) DNS with $160\times160$ grid, (middle column) PINN, (right column) error.}
    \label{fig:DGff}
\end{figure*}

\begin{figure*}[htb!]
    \begin{minipage}[t]{\textwidth}
        \centering
        \begin{subfigure}[t]{0.65\textwidth}
            \centering
            \includegraphics[width=\linewidth]{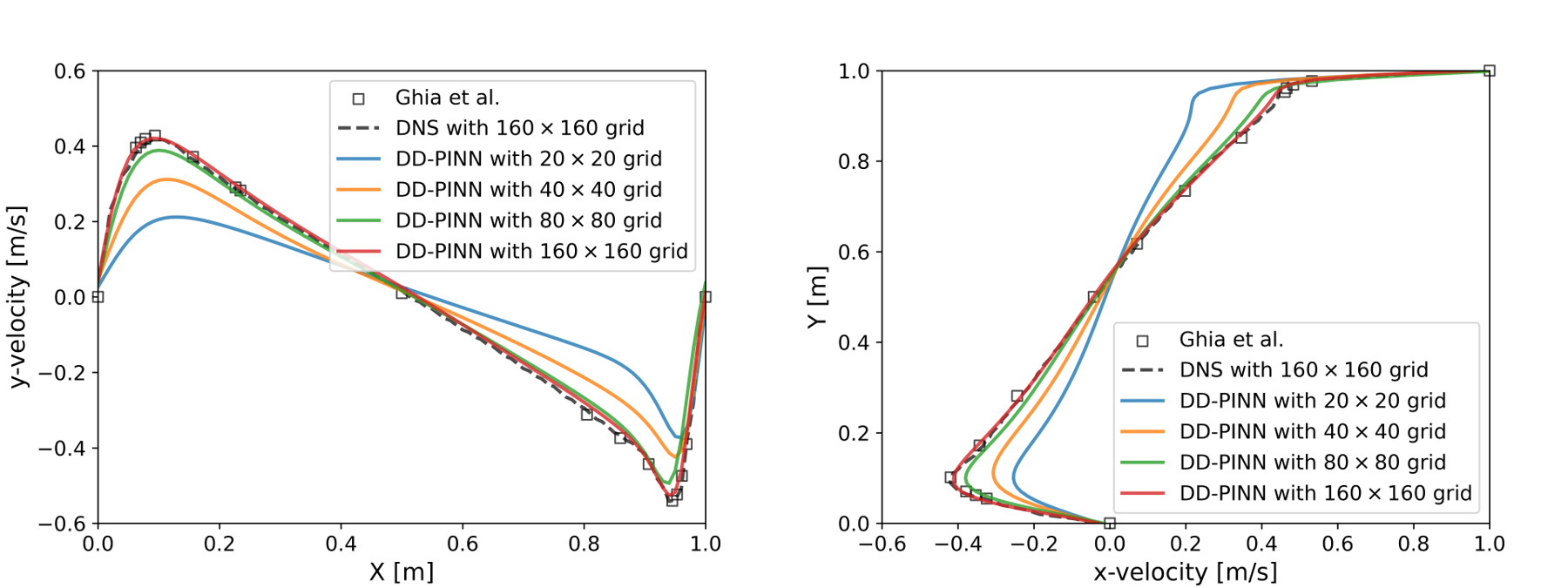}

        \end{subfigure}
        
    \end{minipage}
    
    \caption{Comparison of DD-PINNs with results from OpenFOAM DNS and \citet{ghia1982high}: (left) y-velocity profile along $y=0.5$, (right) x-velocity along $x=0.5$.}
    \label{fig:profile_20to160}
\end{figure*}

In Table \ref{tab:3200RMSE}, as would be expected, there is a clear trend toward improved accuracies as the guide data is obtained from higher fidelity simulations (i.e., as the mesh becomes denser). However, since the computation time of a denser grid is higher than coarser ones, the fidelity of the data used for the guidance should be selected by the user considering their trade-off between time for obtaining the dataset and the accuracy of DD-PINNs. To see how the accuracies differ between DD-PINNs with various grids, Fig. \ref{fig:DGff} visualizes the predicted y-velocity flow fields. As shown in Fig. \ref{fig:DGff_a}, DD-PINN fails to reconstruct a realistic flow field with a $20\times20$ grid, while flow fields with reasonable quality are predicted from a $40\times40$ grid. Although the results from $40\times40$ grid are not as accurate as those from $80\times80$, considering that \citet{ghia1982high} used $129\times129$ grid for the numerical simulation at $Re=3200$ for the same domain, the performance of DD-PINN to capture the overall flow trends using only $40\times40$ grid with the help of PINN framework can be considered as remarkable. Fig. \ref{fig:profile_20to160} shows more quantitative results: DD-PINNs with $80\times80$ and $160\times160$ give almost the same velocity profiles with \citet{ghia1982high} and DNS results along the $x=0.5$ and $y=0.5$ lines, while the $40\times40$ grid-based DD-PINN only follows the global trends of the flow fields with unsatisfactory local fitting.

% \subsection{Effects of applying hard boundary conditions}
% \textcolor{red}{flow fields after hardBC: Fig. \ref{fig:lossland}}

% \begin{figure*}[htb!]
%     \begin{minipage}[t]{\textwidth}
%         \centering
%         \begin{subfigure}[t]{0.8\textwidth}
%             \centering
%             \includegraphics[width=\linewidth]{grid20_fidelity_hard.png}
%             \caption{$20\times20$ grid}
%             \label{fig:DGff_a}
%         \end{subfigure}
        
%         \vfill
        
%         \begin{subfigure}[t]{0.8\textwidth}
%             \centering
%             \includegraphics[width=\linewidth]{grid40_fidelity_hard.png}
%             \caption{$40\times40$ grid}
%             \label{fig:DGff_b}
%         \end{subfigure}
       
%         \vfill
        
%         \centering
%         \begin{subfigure}[t]{0.8\textwidth}
%             \centering
%             \includegraphics[ width=\linewidth]{grid80_fidelity_hard.png}
%             \caption{$80\times80$ grid}
%             \label{fig:DGff_c}
%         \end{subfigure}
        
%         \vfill
        
%         \begin{subfigure}[t]{0.8\textwidth}
%             \centering
%             \includegraphics[width=\linewidth]{grid160_fidelity_hard.png}
%             \caption{$160\times160$ grid}
%             \label{fig:DGff_d}
%         \end{subfigure}
%     \end{minipage}
    
%     \caption{Y-velocity flow fields predicted by various DD-PINNs after applying hard boundary conditions.}
%     \label{fig:lossland}
% \end{figure*}

\subsection{Effects of guide data size}\label{sec:numeffect}

\begin{figure*}[htb!]
    \begin{minipage}[t]{\textwidth}
        \centering
        \begin{subfigure}[t]{0.35\textwidth}
            \centering
            \includegraphics[trim=100 45 50 70,clip, width=\linewidth]{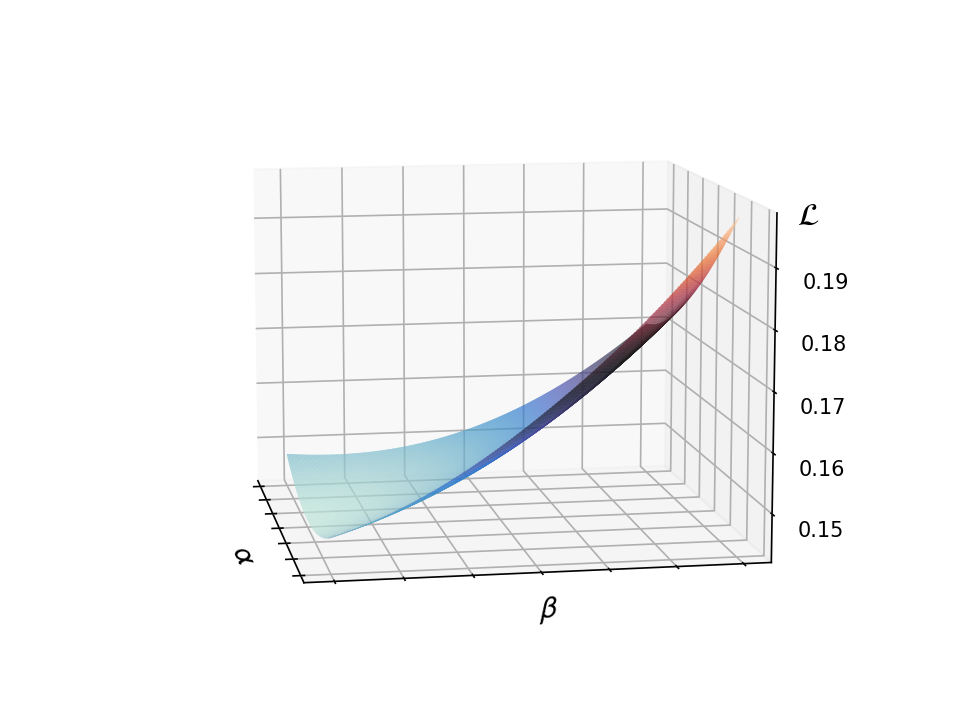}
            \caption{$2*2$ guide data (minimum loss: $0.143$)}
            \label{fig:lossland_a}
        \end{subfigure}
        \hspace{0.05\textwidth}
        \begin{subfigure}[t]{0.35\textwidth}
            \centering
            \includegraphics[trim=100 45 50 70,clip, width=\linewidth]{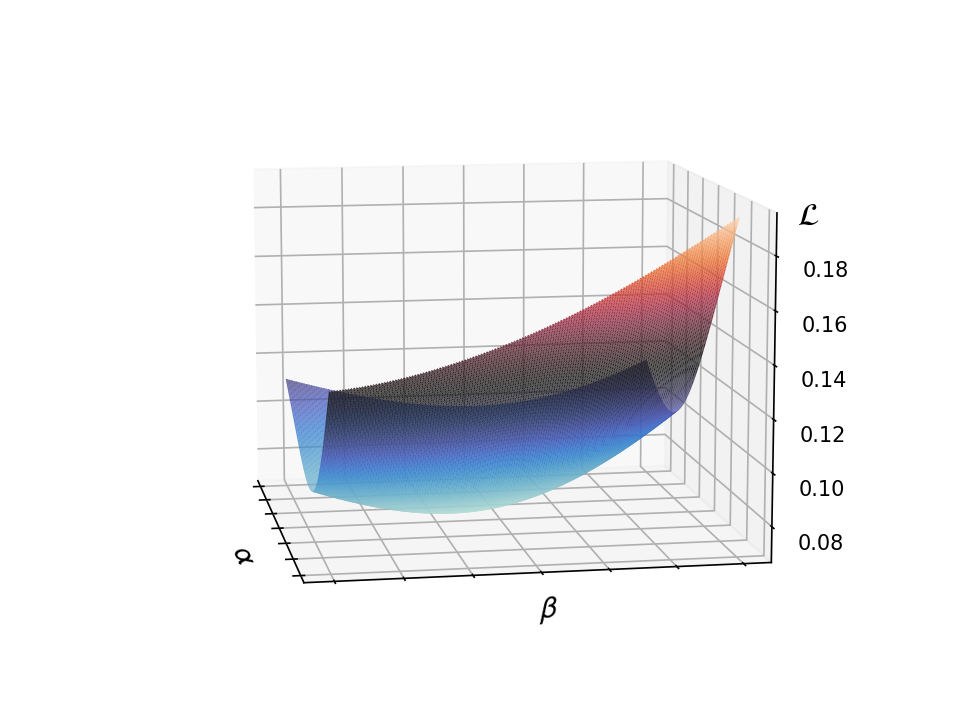}
            \caption{$5*5$ guide data (minimum loss: $0.069$)}
            \label{fig:lossland_b}
        \end{subfigure}
       
        \vfill
        
        \centering
        \begin{subfigure}[t]{0.35\textwidth}
            \centering
            \includegraphics[trim=100 45 50 70,clip, width=\linewidth]{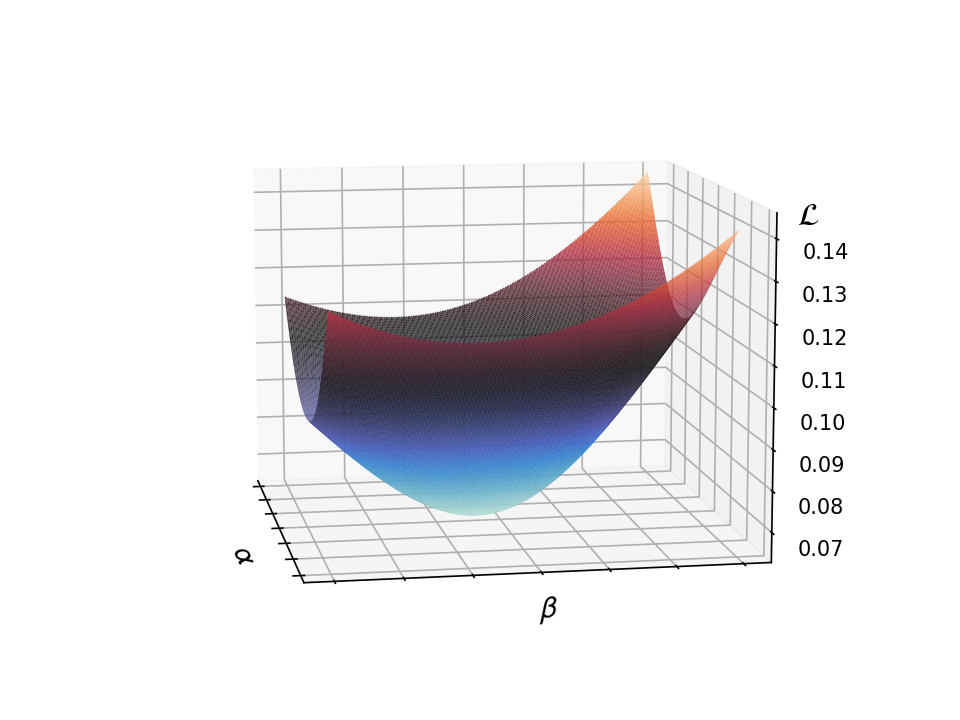}
            \caption{$10*10$ guide data (minimum loss: $0.064$)}
            \label{fig:lossland_c}
        \end{subfigure}
        \hspace{0.05\textwidth}
        \begin{subfigure}[t]{0.35\textwidth}
            \centering
            \includegraphics[trim=100 45 50 70,clip, width=\linewidth]{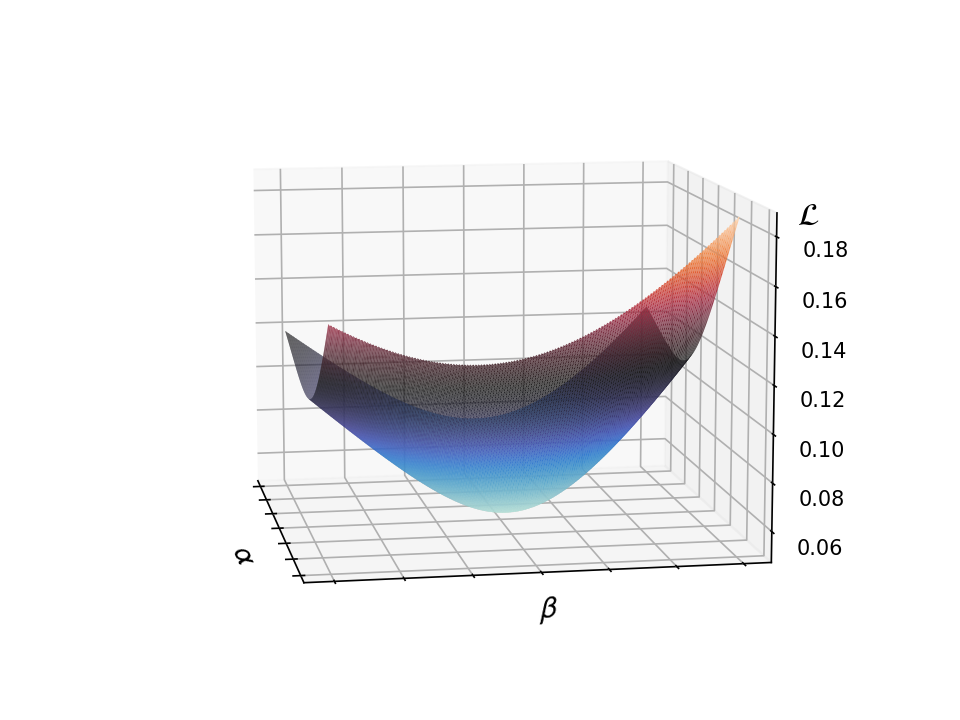}
            \caption{$20*20$ guide data (minimum loss: $0.050$)}
            \label{fig:lossland_d}
        \end{subfigure}
    \end{minipage}
    
    \caption{Loss landscape of $40\times40$ grid-based DD-PINNs with different size of guide data: (a) $2*2$ guide data, (b) $5*5$ guide data, (c) $10*10$ guide data, and (d) $20*20$ guide data. As the number of guide data increases, there is a clear trend toward convexity in loss landscapes.}
    \label{fig:lossland_pts}
\end{figure*}

In previous sections, we select DD-PINN with a $40\times40$ grid as the baseline because it is judged to have the potential to train the global trend of the flow fields with a moderate trade-off between accuracy and time required. Although 1,600 points can be exploited to guide PINNs in the $40\times40$ grid, we further investigate the effect of the number of guide data, considering that the physical space in DT often only allows the acquisition of sparse sensor data. To this end, several candidates of the guide datasets are defined as follows: first, we divide the range of x and y coordinates of the problem domain into $(n_{divide} + 1)$ equal intervals. This division creates a grid of points so that there are $n_{divide} \times n_{divide}$ points in total, evenly distributed within the domain. The reason for using uniform sampling is that we assume there is no prior knowledge of how to effectively locate the sensors (see Appendix \ref{sec:app_prior} when the prior knowledge is given). In this study, $n_{divide}\in[2,5,10,20]$ are explored and their RMSE values for y-velocity are summarized in Table \ref{tab:grid40}. It can be confirmed that in the case of $20*20$, which uses only a quarter of the guide data of the baseline grid $40\times40$, it can perform as well as the baseline $40*40$. Going further, loss landscapes with different $n_{divide}$ values are compared in Fig. \ref{fig:lossland_pts}. The noteworthy point here is that as the number of guide data increases, its regularization effect increases, and thus the convexity of the landscapes increases --- as observed in Ref. \cite{gopakumar2023loss}. This phenomenon supports the points in Section \ref{sec:samplingEffect} that guide data in DD-PINNs act as local regulators, morphing the landscape of the loss function to allow optimizers to traverse it easily \cite{gopakumar2023loss}. It is also important to note that while the convexity of the $10*10$ (Fig. \ref{fig:lossland_c}) and $20*20$ (Fig. \ref{fig:lossland_d}) cases are similar, their minimum losses can be quite different due to the different absolute locations of the valley, despite the similar convexity in neighboring landscapes.

\renewcommand{\arraystretch}{1.2}
\begin{table}[htb!]
\centering
\caption{RMSE of y-velocity predicted by DD-PINNs: different sizes of guide data are extracted from $40\times40$ grid.} \label{tab:grid40}
    \begin{NiceTabular*}{0.6\columnwidth}{@{\extracolsep{\fill}}c|ccccc}
    \cline{1-6}
    \Block[c]{2-1}{} & \Block[c]{1-5}{Guide data for DD-PINN with $40\times40$ grid} \\
    & $2*2$ & $5*5$ & $10*10$ & $20*20$ & $40*40$ \\ \cline{1-6}
    RMSE & 0.1574 & 0.0698 & 0.0658 & \textbf{0.0504} & \textbf{0.0509}\\
    \cline{1-6}

    \end{NiceTabular*}
\end{table}

% \clearpage
\section{Extension of DD-PINNs to parametric NS equations and multi-fidelity approach}\label{sec:Param}

\subsection{DD-PINNs for parametric NS equations}\label{sec:Param_results}

Section \ref{sec:remedy} successfully showed that the proposed DD-PINNs leveraging CFD guide data can remedy the failure in data-free PINNs. However, there is still a serious limitation: it is not scalable to the parametric NS equations as it is. Since conventional PINN models for NS equations are trained for the specific $Re$, flow fields with respect to other Reynolds numbers cannot be predicted; new training is required for new $Re$. To overcome this limitation, \citet{sun2020surrogate} proposed a data-free PINN that has modified inputs consisting not only of spatial and temporal coordinates, but also the $Re$ that appears in the NS equations. They also noted that since the PDE parameter, $Re$ in their study, is trained as an input to the NNs, it can be exploited for uncertainty propagation of $Re$ using Monte Carlo simulation. We extended this parametric setting to our DD-PINNs to verify their scalability and robustness in the context of DT, where general physics should be taken into account.

To comprehensively investigate the performance of the DD-PINNs applied to parametric PDEs compared to existing approaches, we train two additional models. First, conventional data-free PINN is trained. Second, as in the study by \citet{sun2020surrogate}, purely data-driven NNs that utilize only labeled guide data and boundary conditions without PDE information (i.e., without $\mathcal{L}_{mass}$, $\mathcal{L}_{x-momentum}$, $\mathcal{L}_{y-momentum}$ in Eq. \ref{eq:DDPINNloss}) are trained. To train all these models, guide data only at $Re\in[100,1000,3200]$ are used. As in Section \ref{sec:fidelityeffect}, the results of OpenFOAM with a $40\times40$ grid at these $Re$ are leveraged as guide data. Note that the learning rate is adaptively chosen as [1e-3, 5e-4, 1e-4, 5e-5, 1e-5] for every 8,000 epochs (therefore 40,000 epochs in total), with the identical weights on the data-driven loss terms ($w_{data}=10$) for all $Re\in[100, 1000, 3200]$, and hard boundary conditions are applied for $x=0$, $x=1$, $y=0$ walls \cite{lu2021physics} (see Appendix \ref{sec:app_prior}). The model architecture used in Sec. 
\ref{sec:3200tuning} for DD-PINN, with $N_{layer}=7$ and $N_{node}=32$, is also applied here. A total of 15,000 collocation points are utilized in the range of $Re\in[50, 4000]$ with random sampling approach: Table \ref{tab:DGPINN_adapt} in Appendix \ref{sec:app_adapt} shows again that random sampling is still the best among various sampling techniques, especially at high Reynolds number. For training purely data-driven NN, data-free PINN, and DD-PINN models, they require an average of 4,020 seconds, with insignificant differences among them.

\begin{figure*}[htb!]
    \centering
    
        \includegraphics[width=.9\textwidth]{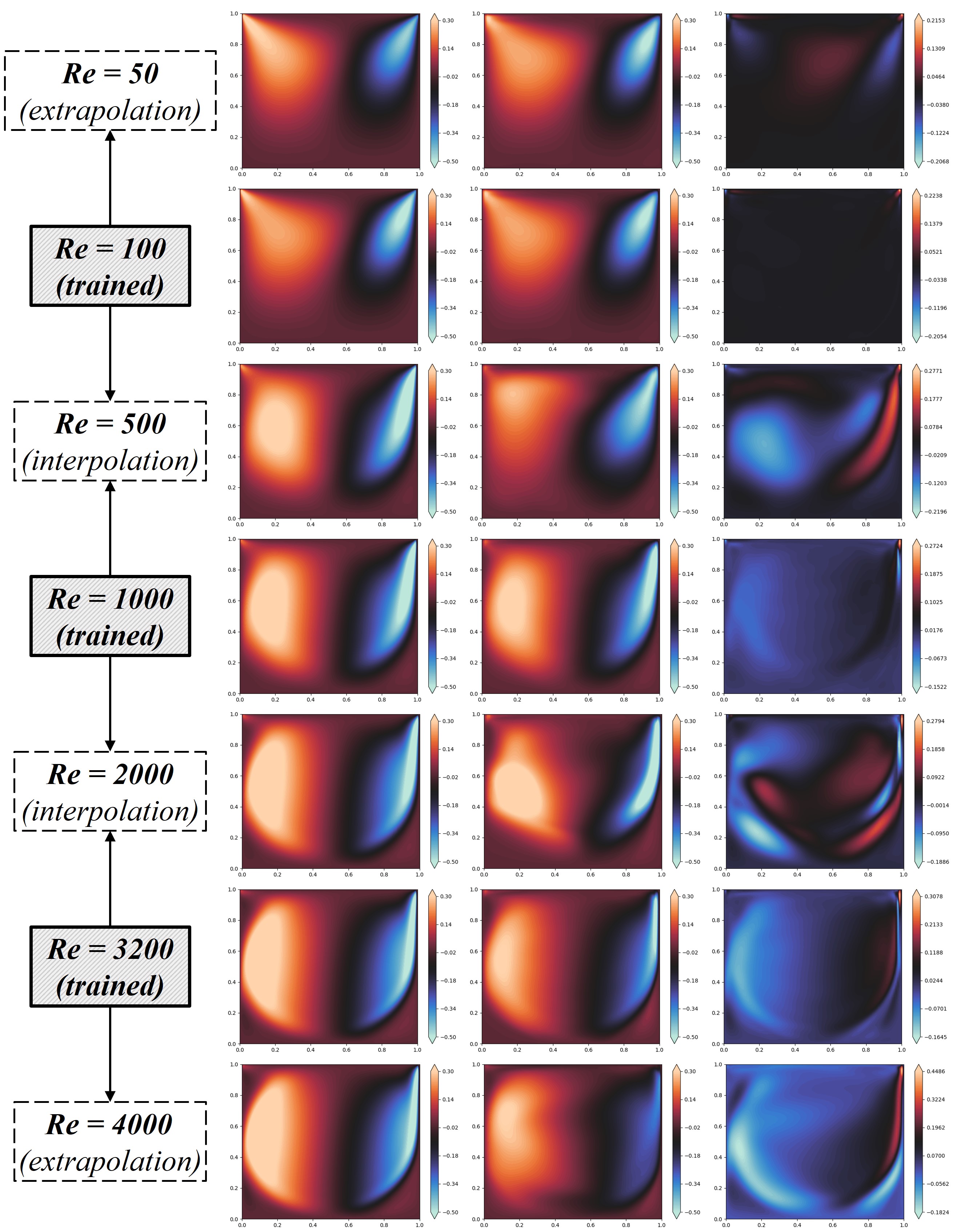}

    \caption{Predicted y-velocity flow fields by purely data-driven NN for parametric NS equations. Note that only $Re=100$, $1000$, $3200$ are trained and other Reynolds numbers are test cases. For three-column figures: (left column) DNS with $160\times160$ grids, (middle column) data-driven NN, (right column) error.}
    \label{fig:ParamResults_nDD}
\end{figure*}

\begin{figure*}[htb!]
    \centering
    
        \includegraphics[width=.9\textwidth]{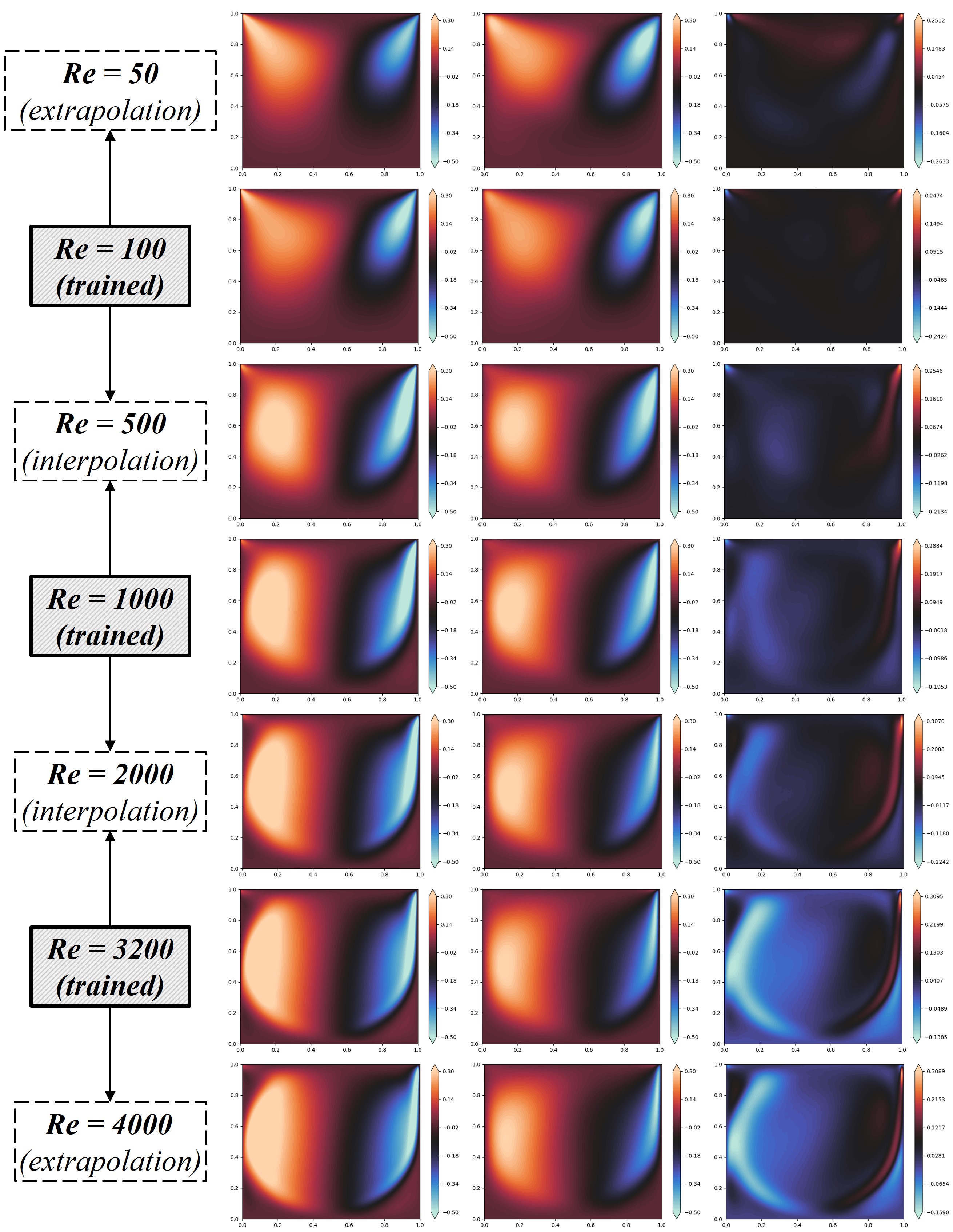}
        
    \caption{Predicted y-velocity flow fields by DD-PINN for parametric NS equations. Note that only $Re=100$, $1000$, $3200$ are trained and other Reynolds numbers are test cases. For three-column figures: (left column) DNS with $160\times160$ grids, (middle column) DD-PINN, (right column) error.}
    \label{fig:ParamResults_DD}
\end{figure*}

Their results are compared with DNS results from $160\times160$ grids with respect to various $Re$ values, and RMSE values are given in Table \ref{tab:Param_comparison}. Herein, to evaluate both interpolation and extrapolation performance, the trained models are tested at $Re=500$, $2000$ (interpolation), and $Re=50$, $4000$, $5000$, $6000$ (extrapolation). It can be confirmed that the conventional data-free PINN, which has the largest errors, fails to predict realistic flow fields as Fig. \ref{fig:failed}. However, conventional data-driven NN and DD-PINN succeed in reconstructing reasonable y-velocity flow fields at various $Re$, not only for the interpolation tasks but also for the extrapolation tasks. In more detail, conventional data-driven NN shows better accuracy at the trained $Re$: $100$, $1000$, and $3200$. On the other hand, DD-PINN shows better performance in interpolation and extrapolation tasks. The reason for this trend is that the conventional data-driven NN focuses only on learning the flow fields at the labeled $Re$ without considering the physics contained in the PDEs, and therefore shows better accuracy than DD-PINN only at the $Re$ used for training. In contrast, DD-PINN is able to learn physical properties as it considers not only the guide data, but also the physics within PDEs in different $Re$ regions, and thus achieves superior generality at $Re$ not used during training. For a more intuitive comparison, the flow fields predicted by purely data-driven NN and the proposed DD-PINN are compared in Fig. \ref{fig:ParamResults_nDD} and Fig. \ref{fig:ParamResults_DD}, respectively. As discussed, the purely data-driven NN predicts accurate flow fields only at trained $Re$: for example, it interpolates the flow field at $Re=500$ to be more similar to that at $Re=100$ than at $Re=1000$, while DNS indicates that the flow at $Re=500$ should be more similar to that at $Re=1000$. Furthermore, especially at the higher Reynolds numbers, it shows much worse quality than DD-PINN: see the predicted flow fields at $Re=2000$ and $Re=4000$.

\renewcommand{\arraystretch}{1.2}
\begin{table}[htb!]
\centering
\caption{RMSE of y-velocity for different NN architectures. Below the Reynolds numbers, \textit{tr} indicates $Re$ used for the training, while \textit{ext} and \textit{int} indicate $Re$ values to be tested by extrapolation and interpolation, respectively.} \label{tab:Param_comparison}
    \begin{NiceTabular*}{0.95\columnwidth}{@{\extracolsep{\fill}}c|ccccccccc}
    \hline
    \Block[c]{3-1}{} & \Block[c]{1-9}{Reynolds number} \\
    & $50$ & $100$ & $500$ & $1000$ & $2000$ & $3200$ & $4000$ & $5000$ & $6000$\\
    & (\textit{ext}) & (\textbf{\textit{tr}}) & (\textit{int}) & (\textbf{\textit{tr}}) & (\textit{int}) & (\textbf{\textit{tr}}) & (\textit{ext}) & (\textit{ext}) & (\textit{ext}) \\ \cline{1-10}
    Purely data-driven NN & 0.0237 	&	\textbf{0.0062} 	&	0.0642 	&	\textbf{0.0232} 	&	0.0652 	&	0.0490 	&	0.0742 & 0.1047 & 0.1327 \\
    Data-free PINN & 0.0940 	&	0.1060 	&	0.1698 	&	0.1890 	&	0.1984 	&	0.1994 	&	0.1998 & 0.1979 & 0.1945 \\
    DD-PINN & 0.0229 	&	0.0093 	&	0.0367 	&	0.0269 	&	0.0454 	&	0.0530 	&	0.0565 	&	0.0589 	&	0.0623 \\
    \hline
    MF-DD-PINN (case 1) & 0.0264 	&	0.0093 	&	0.0363 	&	0.0261 	&	0.0281 	&	0.0280 	&	0.0295 	&	0.0339 	&	0.0429 \\
    MF-DD-PINN (case 2) & 0.0202 	&	0.0096 	&	\textbf{0.0274} 	&	0.0265 	&	0.0271 	&	0.0248 	&	0.0260 	&	0.0332 	&	0.0470  \\
    MF-DD-PINN (case 3) & \textbf{0.0194} 	&	0.0090 	&	0.0341 	&	0.0258 	&	\textbf{0.0241} 	&	0.0219 	&	0.0233 	&	0.0318 	&	0.0497  \\
    MF-DD-PINN (case 4) & 0.0196 	&	0.0087 	&	0.0306 	&	0.0256 	&	\textbf{0.0241} 	&	\textbf{0.0217} 	&	\textbf{0.0214} 	&	\textbf{0.0248} 	&	\textbf{0.0364} \\
    \cline{1-10}

    \end{NiceTabular*}
\end{table}

%\subsection{Comparison with conventional data-driven NNs}

\clearpage
\subsection{Further extension: multi-fidelity DD-PINNs}\label{sec:Param_MF}

In Section \ref{sec:Param_results}, OpenFOAM results obtained from $40\times40$ grids are used as guide data for all $Re$ to train DD-PINN. This section focuses on the fact that it is not always the case that the guide datasets are obtained from identical grids (or sensors) in DT scenarios. Similarly, one does not have to use the datasets from identical sources. That is, since prediction is more difficult at higher Reynolds numbers, one can use higher fidelity guide data (denser CFD grid or higher quality experimental sensors) for the higher $Re$ values. This is referred to as multi-fidelity DD-PINNs in our study, and this section aims to validate their effectiveness. According to Appendix \ref{sec:openfoam}, it is shown that $40\times40$ grids are sufficient for $Re=100$, $1000$, while $Re=3200$ requires at least $80\times80$ grid: therefore, multi-fidelity DD-PINNs take corresponding guide data from each grid for each Reynolds number. This framework intends to exploit more accurate guide data at the higher $Re$, expecting superior accuracy, while reducing the computational cost at lower $Re$ where the guide data from coarse grid brings sufficient accuracy.

Several multi-fidelity DD-PINNs are trained with different guide data combinations: model architectures of them remain the same as DD-PINN in Section \ref{sec:Param_results} and descriptions of their guide data are summarized in Table \ref{tab:MFDGPINN_cases}. Case 1 denotes the baseline multi-fidelity DD-PINN (MF-DD-PINN) where for lower Reynolds (100 and 1,000), $40\times40$ grid is used as guide data and $80\times80$ grid is used for $Re=3200$. All of the weight terms for each data-driven loss, $w_{data}$, are set as 10. In case 2, $w_{data}$ for $Re=3200$ increases to 15. In case 3, the $160\times160$ grid is additionally utilized with $w_{data}=15$, and therefore it can be regarded that three different fidelities are utilized in this case. However, only $5*5$ points are leveraged assuming those high-fidelity but expensive data are obtained from the sparse sensors (as in Section \ref{sec:numeffect}, equal spacing is applied for selecting 25 points). Lastly, case 4 increases $w_{data}$ of $5*5$ points to 20 in order to verify the effect of the $160\times160$ grid's guide data as its impact intensifies.

% \usepackage{multirow}
\begin{table}[htb!]
\centering
\caption{Details of the guide data used to train multi-fidelity DD-PINN models. Note that the guide data remains the same for $Re=100$ and $1000$, while only the guide data at $Re=3200$ varies. Additionally, for cases 3 and 4, which use a $160\times160$ grid, only $5*5$ points of it are used, assuming they are obtained from sparse but expensive sensors.}\label{tab:MFDGPINN_cases}
\begin{tabular}{c|c|c|c}
\hline
\multirow{2}{*}{MF-DD-PINN models} & \multicolumn{3}{c}{Reynolds number}\\
\cline{2-4} 
 & 100 & 1000 & 3200 \\
\hline
Case 1 & \multicolumn{2}{c|}{\multirow{6}{*}{\thead{$40\times40$ grid \\ ($w_{data}=10$)}}} & $80\times80$ grid ($w_{data}=10$) \\
\cline{1-1} \cline{4-4} 
Case 2 & \multicolumn{2}{c|}{} & $80\times80$ grid ($w_{data}=15$) \\
\cline{1-1} \cline{4-4} 
\multirow{2}{*}{Case 3} & \multicolumn{2}{c|}{}                  & $80\times80$ grid ($w_{data}=15$) \\ 
&\multicolumn{2}{c|}{}& $160\times160$ grid ($w_{data}=15$, only $5*5$ points are used) \\
\cline{1-1} \cline{4-4} 
\multirow{2}{*}{Case 4} & \multicolumn{2}{c|}{}                  & $80\times80$ grid ($w_{data}=15$) \\ 
&\multicolumn{2}{c|}{}& $160\times160$ grid ($w_{data}=20$, only $5*5$ points are used) \\
\hline
% % \cline{1-1} \cline{4-4} 
%                   & \multicolumn{2}{c|}{}                  &                   \\
% \multirow{2}{*}{} & \multicolumn{2}{c|}{}                  & \multirow{2}{*}{} \\
%                   & \multicolumn{2}{c|}{}                  &                  
\end{tabular}
\end{table}

Training the models from case 1 to case 4 takes about 4,000 seconds for the first two models and about 4,800 seconds for the next two models. Considering that DNS takes 4,231 seconds for the $160\times160$ grid just for the single case $Re=3200$, the computational efficiency of DD-PINNs in parametric NS is remarkable, since parametric DD-PINNs can be used to predict any untrained $Re$. The final results can be found in Table \ref{tab:Param_comparison} and they demonstrate that the MF-DD-PINN case 4 achieves the highest accuracy at high Reynolds numbers from 2,000 to 6,000 among all trained models due to the highest-fidelity guide data at $Re=3200$: it has half the RMSE compared to the single fidelity DD-PINN. Also, since four cases of MF-DD-PINNs are designed to have higher case numbers as the influence of the high-fidelity guide data at $Re=3200$ increases, there is a clear decreasing trend of the RMSE as the case number increases at $Re=3200,4000,5000$. The notable point here is that although case 3 and case 4 use only 25 additional guide data from the $160\times160$ grid, they show an impressive improvement over case 1 and case 2, underscoring the effectiveness of the multi-fidelity approach in DD-PINNs. Even in the low Reynolds region, MF-DD-PINN case 4 shows remarkable performance, indicating its general predictive performance over the different Reynolds numbers, both in interpolation and extrapolation tasks. Note that all MF-DD-PINN models do not show significantly better accuracies than DD-PINN in low $Re$ regions, since their multi-fidelity approaches only aim to increase fidelity in high $Re$ regions. To examine the effects of different multi-fidelity data on prediction accuracy, one-way ANOVA is performed within four cases of MF-DD-PINNs. Table \ref{tab:ANOVA_MF} shows its results: the different types of data used for each case obviously affect the accuracy of MF-DD-PINNs in the high Reynolds region ($Re=3200$, $4000$, $5000$). However, at $Re=6000$, the p-value is greater than 0.5, indicating that the extrapolation of $Re$ is too far from the training region in this case, and therefore the meaningful difference between the four cases disappears.

\renewcommand{\arraystretch}{1.2}
\begin{table}[htb!]
\centering
\caption{One-way ANOVA results between four MF-DD-PINN cases: only p-values at high Reynolds indicate statistically significant differences among them.} \label{tab:ANOVA_MF}
    \begin{NiceTabular*}{0.95\columnwidth}{@{\extracolsep{\fill}}cc|ccccccccc}
    \hline
    \Block[c]{3-1}{} && \Block[c]{1-9}{Reynolds number} \\
    && $50$ & $100$ & $500$ & $1000$ & $2000$ & $3200$ & $4000$ & $5000$ & $6000$\\
    && (\textit{ext}) & (\textbf{\textit{tr}}) & (\textit{int}) & (\textbf{\textit{tr}}) & (\textit{int}) & (\textbf{\textit{tr}}) & (\textit{ext}) & (\textit{ext}) & (\textit{ext}) \\ \cline{1-11}
    \Block[c]{2-1}{ANOVA} & F statistic & 2.1441 &	1.1710 	&	0.8682 	&	0.1433 	&	2.4070 	&	5.7041 	&	6.4805 	&	3.6144 	&	0.7275 \\
    & p-value & 0.0942 	&	0.3205 	&	0.4576 	&	0.9339 	&	0.0668 	&	\textbf{0.0008} 	&	\textbf{0.0003} 	&	\textbf{0.0134} 	&	0.5360  \\
    \cline{1-11}

    \end{NiceTabular*}
\end{table}

\begin{figure*}[htb!]
    \centering
    
        \includegraphics[width=.9\textwidth]{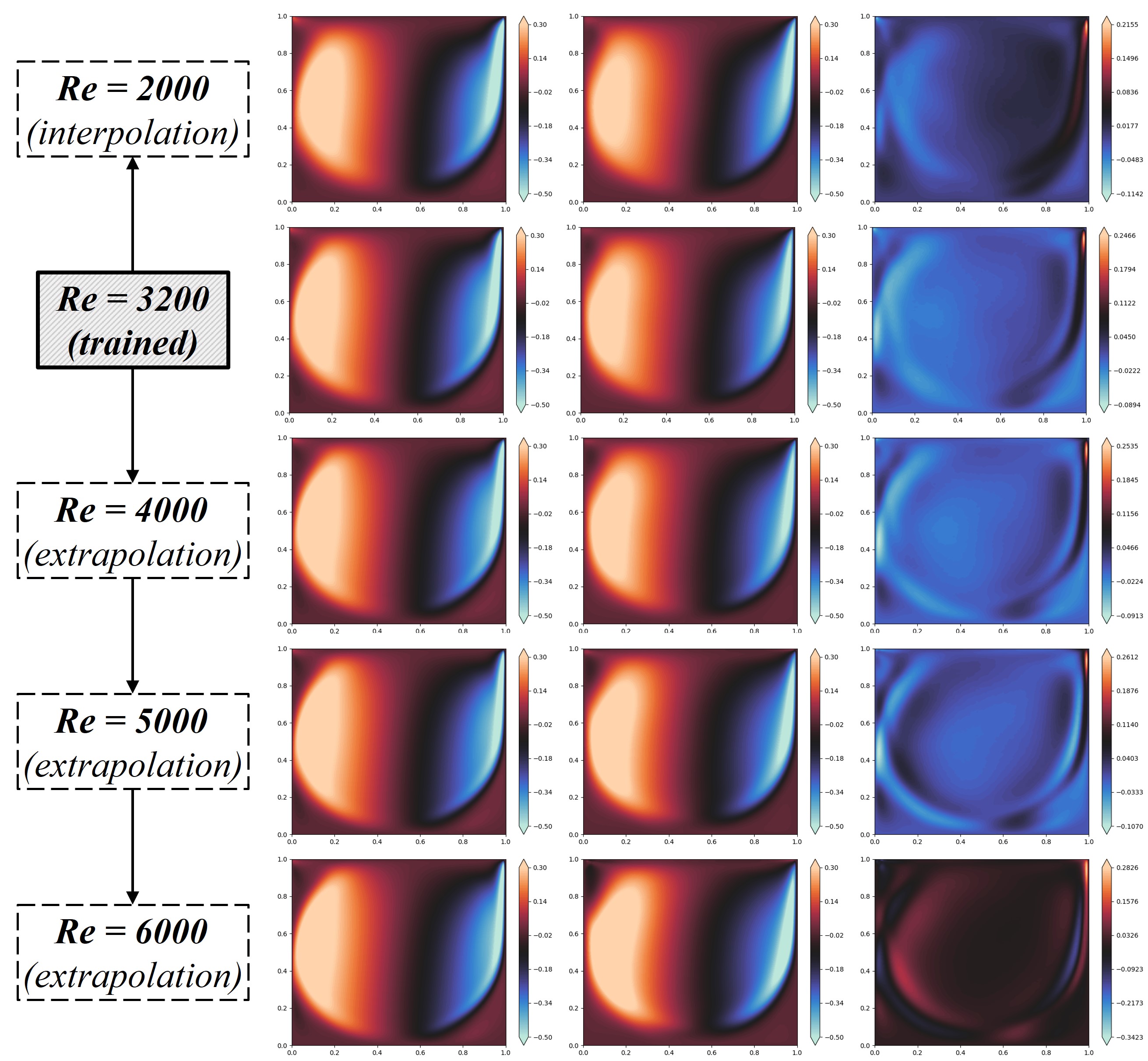}
        
    \caption{Predicted y-velocity flow fields by MF-DD-PINN case 4 for parametric NS equations. Note that only $Re=100$, $1000$, and $3200$ are trained and other Reynolds numbers are test cases. For three-column figures: (left column) DNS with $160\times160$ grids, (middle column) MF-DD-PINN case 4, (right column) error.}
    \label{fig:ParamResults_MFDD}
\end{figure*}

\begin{figure*}[htb!]
    \begin{minipage}[t]{\textwidth}
        \centering
        \begin{subfigure}[t]{0.65\textwidth}
            \centering
            \includegraphics[trim=0 0 0 30,clip, width=\linewidth]{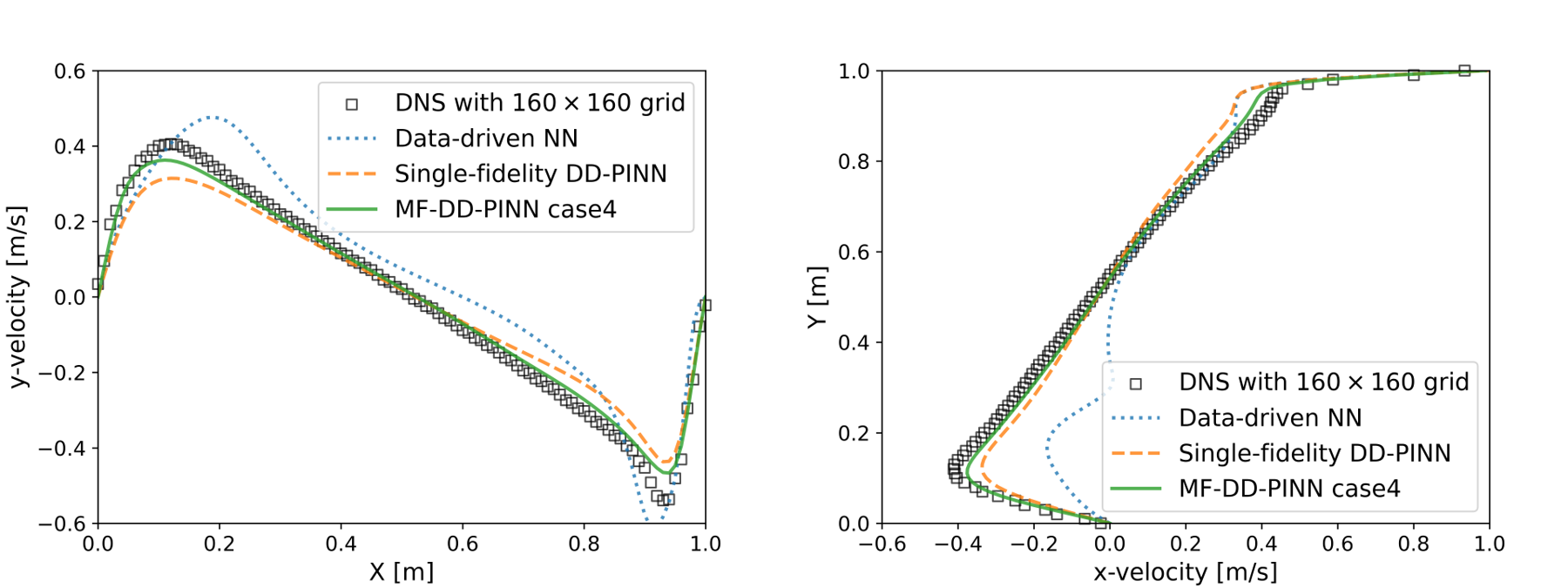}
            \caption{$Re=2000$ \textit{(int)}}
            \label{fig:slice_a}
        \end{subfigure}
        
        \vfill
        
        \begin{subfigure}[t]{0.65\textwidth}
            \centering
            \includegraphics[trim=0 0 0 30,clip, width=\linewidth]{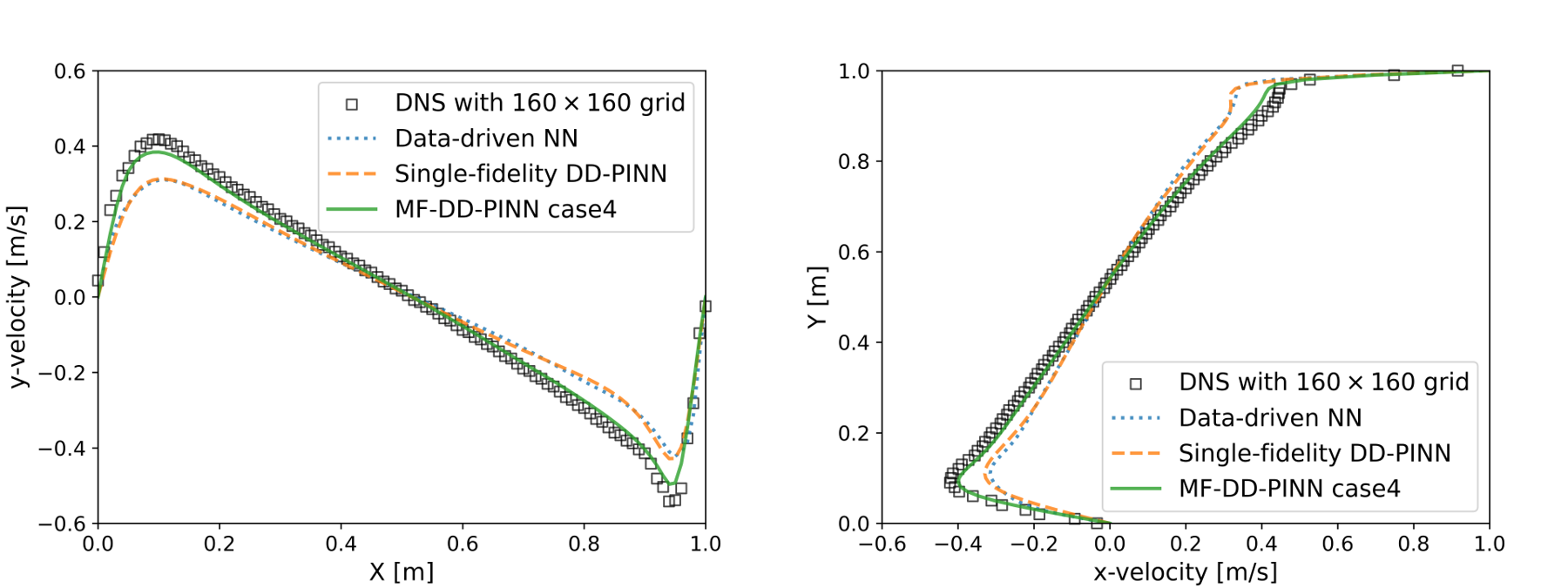}
            \caption{$Re=3200$ \textit{(tr)}}
            \label{fig:slice_b}
        \end{subfigure}
       
        \vfill
        
        \centering
        \begin{subfigure}[t]{0.65\textwidth}
            \centering
            \includegraphics[trim=0 0 0 30,clip, width=\linewidth]{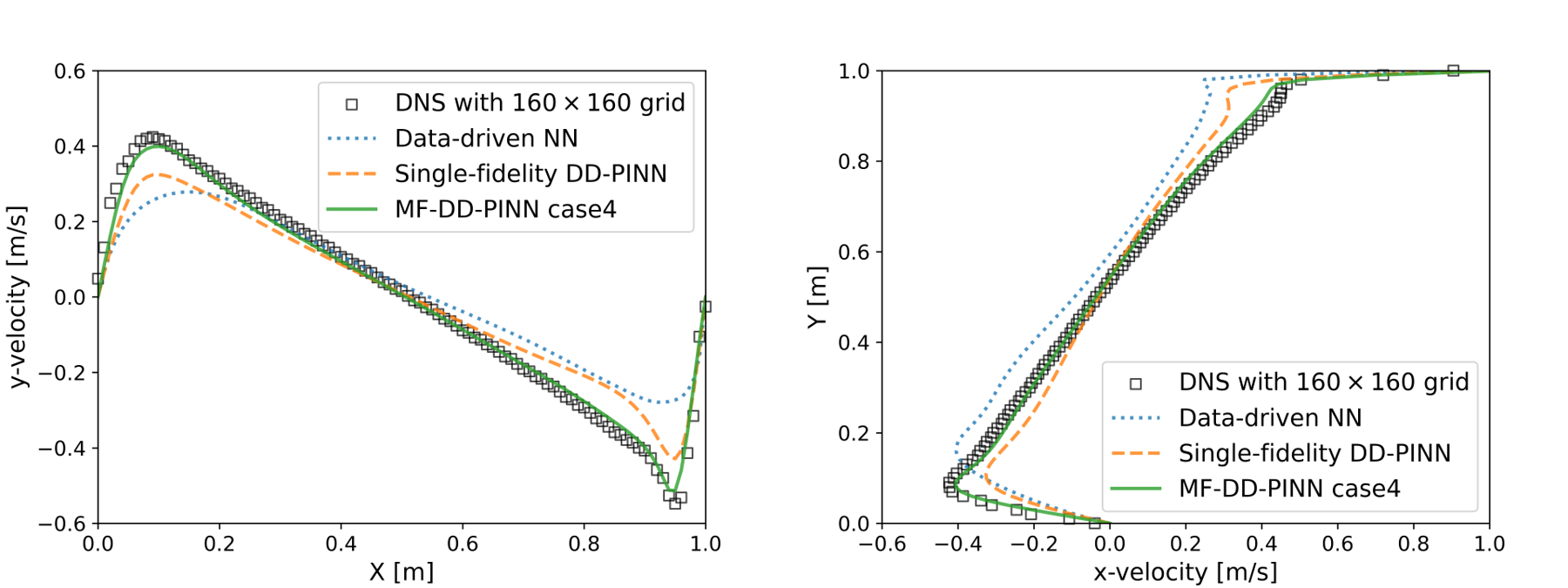}
            \caption{$Re=4000$ \textit{(ext)}}
            \label{fig:slice_c}
        \end{subfigure}

        \vfill
        
        \centering
        \begin{subfigure}[t]{0.65\textwidth}
            \centering
            \includegraphics[trim=0 0 0 30,clip, width=\linewidth]{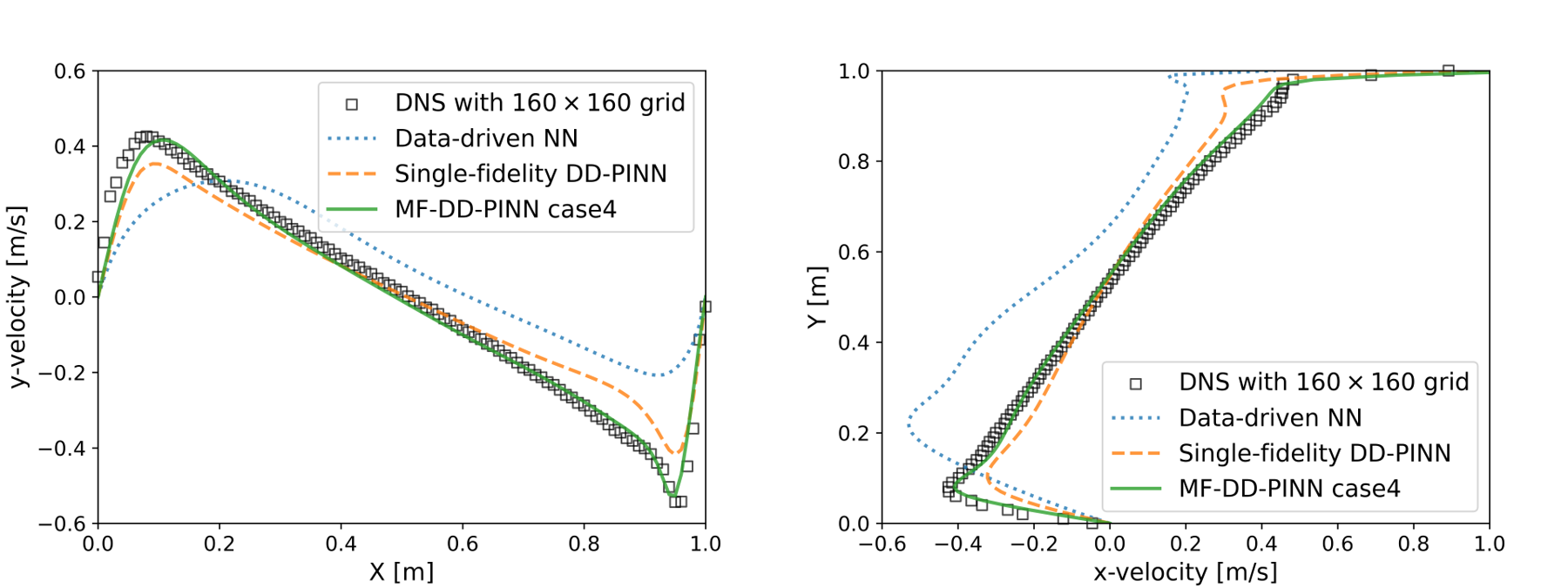}
            \caption{$Re=5000$ \textit{(ext)}}
            \label{fig:slice_d}
        \end{subfigure}

        \vfill
        % \vspace{0mm}
        
        \centering
        \begin{subfigure}[t]{0.65\textwidth}
            \centering
            \includegraphics[trim=0 0 0 30,clip, width=\linewidth]{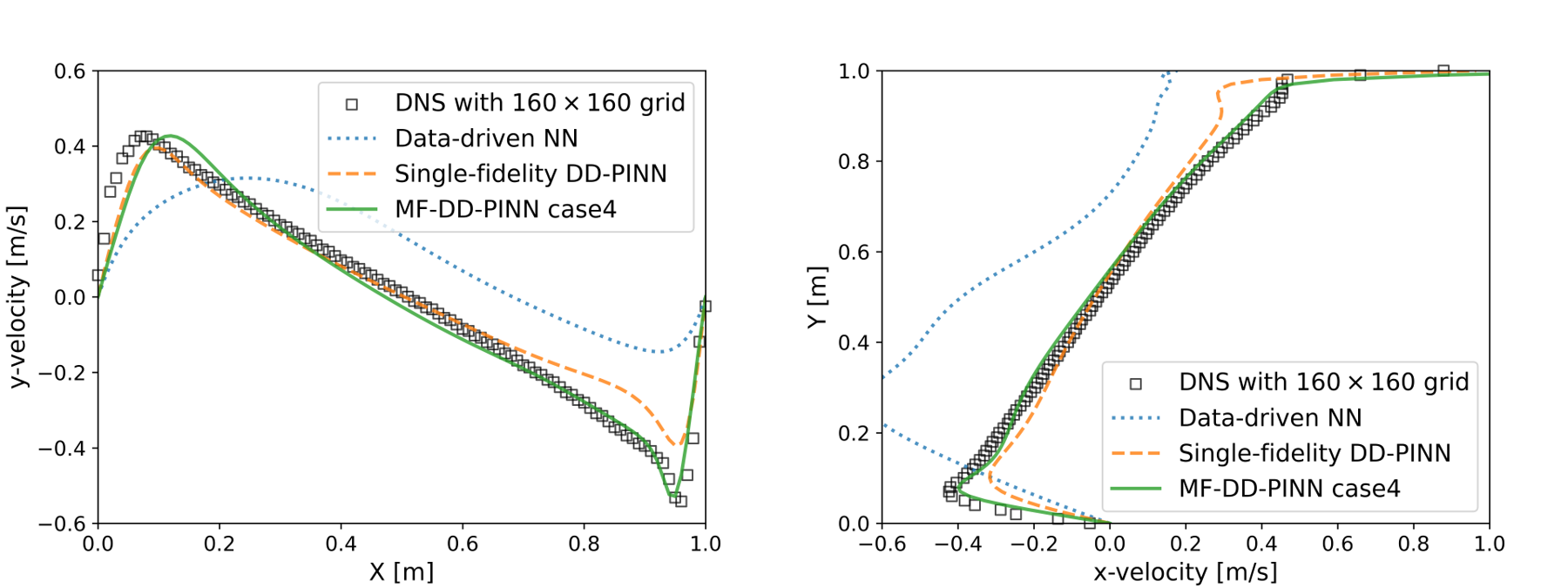}
            \caption{$Re=6000$ \textit{(ext)}}
            \label{fig:slice_e}
        \end{subfigure}
        
    \end{minipage}
    
    \caption{Comparison of MF-DD-PINN case 4 with OpenFOAM DNS ($160\times160$ grid) results: (left) y-velocity profile at $y=0.5$, (right) x-velocity profile at $x=0.5$. The results of data-driven NN and single fidelity DD-PINN in Section \ref{sec:Param_results} are also shown.}
    \label{fig:slice}
\end{figure*}

The flow fields of MF-DD-PINN case 4 are visualized in Fig. \ref{fig:ParamResults_MFDD}: when compared to DD-PINN in Fig. \ref{fig:ParamResults_DD}, they clearly show much more similar predictions with DNS at high $Re$ values (since they maintain similar accuracy at low $Re$ with DD-PINN, only high Reynolds regions are shown). The most fascinating point is that although only the range of $Re\in[50, 4000]$ is explored using collocation points for training MF-DD-PINN (as noted in Section \ref{sec:Param_results}), it successfully extrapolates the flow fields beyond $Re=4000$, as can be seen in the flow fields at $Re=5000$, $6000$. For more specific investigation, velocity profiles at $x=0.5$ and $y=0.5$ from MF-DD-PINN case 4 are compared with DNS results, data-driven NN, and single-fidelity DD-PINN in Fig. \ref{fig:slice}. Multi-fidelity DD-PINN gives the most accurate results at the training point, $Re=3200$, even when the velocity changes abruptly. On the other hand, data-driven NN and single-fidelity DD-PINN failed at local fitting even in the case of $Re=3200$. As the Reynolds number increases towards the extrapolation region, data-driven NN fails to capture the global trend of the velocity profiles, while single-fidelity DD-PINN still follows the global trend of DNS --- again highlighting that incorporating physics using PDE losses improves the generalized prediction performance. Meanwhile, MF-DD-PINN shows the best results: its y-velocity profile (left column) shows only slight deviation from DNS even at $Re=6000$, while the x-velocity profile (right column) shows remarkable accuracy\footnote{This is the reason why the RMSE of y-velocity is primarily used throughout this study to measure the accuracy of PINNs: y-velocity profiles are more difficult to predict than x-velocity profiles.}. In summary, the above results demonstrate the scalability of DD-PINN, the framework newly proposed in our study: it can be extended to solve parametric PDEs, and furthermore, the multi-fidelity approach can even allow flexible fusion of guide data from different data sources, with impressive extrapolation performance than other approaches.

\clearpage
\subsection{Uncertainty quantification for multi-fidelity DD-PINNs}\label{sec:Param_UQ}

\begin{figure*}[htb!]
    \centering
    
        \includegraphics[width=.6\textwidth]{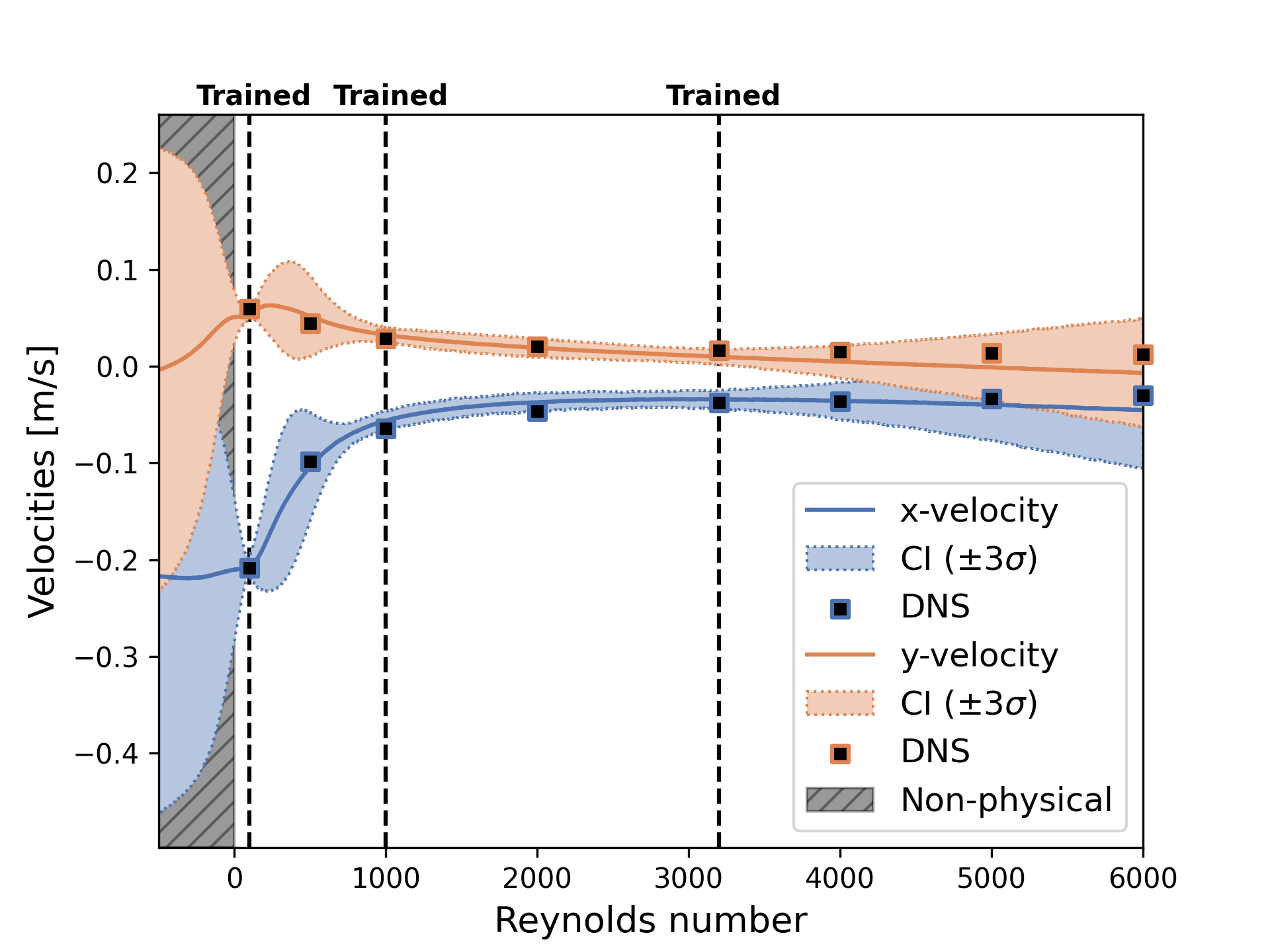}
        
    \caption{Confidence intervals (CIs) of velocities at location $(x=0.5, y=0.5)$ based on MF-DD-PINN case 2 model. At $Re=100$, $1000$, $3200$, it shows a narrow CI because these points are used as training data, while the CI diverges in the unexplored $Re$ region due to the absence of guide data. Also, the non-physical region ($Re<0$) is hatched in gray, showing CIs that diverge much more than the other regions.}
    \label{fig:UQ}
\end{figure*}

In this section, we explore the potential of multi-fidelity DD-PINNs for quantifying predictive uncertainty, which is one of the most fundamental requirements for DT. For this purpose, we use the deep ensemble approach \cite{yang2023towards}, which exploits the ensemble of NNs to quantify the predictive uncertainty of NNs: a total of four MF-DD-PINN case 2 models are trained with different random initializations. Then, the predictive value is determined by taking the average of the multiple models, while the epistemic uncertainty is calculated as the standard deviation within the models \cite{morimoto2022assessments}. It is worth noting that for the efficient UQ of PINN models, the Monte Carlo dropout (MC-dropout) approach can be also easily incorporated \cite{gal2016dropout}. However, due to the ongoing debate surrounding its interpretation as a true Bayesian inference method \cite{yang2023towards, osband2016risk,hron2017variational,hron2018variational,folgoc2021mc}, we have chosen not to adopt this approach in the present study. Finally, the confidence intervals (CIs) estimated by the ensemble approach are plotted in Fig. \ref{fig:UQ} and there are two noteworthy points: one is that the uncertainty in terms of both velocity components shows an apparent diverging trend as it moves away from the trained Reynolds numbers (100, 1000, and 3200). However, the region between $Re=100$ and $Re=1000$ shows much more uncertainty than the region between $Re=1000$ and $Re=3200$: the reason seems to be that the change of the flow fields of the former is much more dramatic than the latter (see DNS results in Fig. \ref{fig:ParamResults_DD}). The second point is that as $Re$ enters the region with negative values, the CI diverges dramatically compared to the other untrained regions (such as from $Re=3200$ to $Re=6000$). Considering that the multi-fidelity DD-PINNs explore only the physically realistic phenomena with the varying magnitude of $Re$ for positive values, but not for negative values, the radically divergent CI trend in the non-physical $Re<0$ region seems reasonable. In summary, from the qualitative analysis of the UQ performance of multi-fidelity DD-PINNs, we conclude that they have the strong potential to provide realistic predictive uncertainty, which enhances their scalability in DT. 

\clearpage
\section{Conclusion}\label{sec:conclu}

This study explored the potential of PINNs for the realization of DT from various perspectives. First, vorticity-aware adaptive sampling strategy tailored to fluid dynamics is proposed and comprehensively compared with existing sampling approaches to investigate the practical effectiveness of PINNs' mesh-free property in automating the construction of virtual space. Then, a data-driven physics-informed neural networks (DD-PINNs) framework, which can update the virtual model in a data-driven manner, is scrutinized as a remedy for the failure of data-free PINNs at high Reynolds in lid-driven cavity flow problem. Then, we extended DD-PINN to parametric PDEs --- NS equations with varying $Re$ for this study --- and successfully demonstrated its superior performance than other conventional architectures including data-free PINNs and purely data-driven NNs. Moreover, its scalability to multi-fidelity guide data was also analyzed, considering that the datasets in physical space are often collected from different sources. Finally, multi-fidelity DD-PINN is applied to an ensemble-based UQ task in order to validate its applicability to DT, where an accurate measure of predictive uncertainty is necessary. The key findings of our study can be summarized as follows:

% This study proposed a data-driven physics-informed neural networks (DD-PINNs) framework, which leverages the simulation data as a guide to the exact PDE solutions. Their potential for the realization of aerodynamic DT from various perspectives It is introduced as a remedy for the failure of data-free PINNs in the forward problem of lid-driven cavity flow at $Re = 1000, 3200$. The proposed framework was comprehensively investigated through various adaptive sampling techniques including vorticity-aware sampling, fidelity of the guide data, and size of the guide data. Then, it extended its applicability to parametric PDEs (NS equations with varying $Re$ for this study) and successfully showed its superior performance than other conventional architectures including data-free PINNs and purely data-driven NNs. Moreover, its scalability to various multi-fidelity guide data was also analyzed demonstrating its further improvement over single-fidelity guide data. The key findings of our study can be summarized as follows:

\begin{enumerate}

    \item In data-free PINNs at $Re=100$, vorticity-aware adaptive sampling showed the equivalent or even better performance than other conventional residual or gradient-aware samplings. It indicated that there is room for the development of flow-specific PINNs in that adaptive sampling based on specific flow quantities can be as accurate as existing sampling techniques.

    \item However, as $Re$ increased to 1,000 and 3,200, data-free PINNs failed to predict realistic flow fields, while the proposed DD-PINN framework succeeded. Since both data-free PINN and DD-PINN are designed to have the same model capacity, it can be concluded that the failure of data-free PINN at high Reynolds number is not due to insufficient capacity. Instead, with respect to the loss landscapes, DD-PINNs showed much sharper valley than data-free PINNs, explaining the reasons for their better performance. The superior performance of DD-PINNs highlights their applicability to DT, where data is consistently fed into the virtual space from the physical space. 

    \item In the DD-PINN framework, random sampling techniques that promote uniform collocation points outperformed other sampling approaches due to the need for a global calibrator to compensate for the local regularization effects of the guide data.

    \item When extended to solve parametric PDEs, DD-PINNs outperformed purely data-driven NNs and data-free PINNs. It indicates that conventional data-free PINNs have obvious limitations in predicting more general physics while highlighting the scalability of the DD-PINNs to parametric problems. In this context, their potential to replace a demanding physical space with a real-time but generalized virtual representation in DT scenarios is verified.

    \item The performance of DD-PINNs can be further improved with multi-fidelity guide data, which aims to efficiently train DD-PINNs with guide data from different fidelities. By increasing the weights of the data-driven loss term and increasing the fidelity of the guide data, the multi-fidelity approach achieves greater accuracy than any other compared model, even when high-fidelity guide data is sparse. It highlights its flexibility in utilizing the heterogeneous dataset acquired from physical space in DT.

    \item When the multi-fidelity DD-PINN was used for uncertainty quantification, it provided realistic confidence intervals that were narrow in the trained region while diverging in the untrained regions. It also shows a dramatic divergence in the non-physical $Re < 0$ region, highlighting its potential to provide reliable predictive uncertainty over its prediction.

\end{enumerate}

% The presented DD-PINN framework showed the capability to overcome the failure of conventional data-free PINNs in the forward problem, especially at high Reynolds regions in the flow problems. Moreover, since it was verified to accurately predict flow quantities along with reasonable uncertainty quantification, its worthiness in application to the digital twin seems promising

Although this work successfully demonstrated the flexibility of the DD-PINN framework in DT scenarios from multiple perspectives, the scope of the case study was limited to the lid-driven cavity flow problem. Therefore, the proposed DD-PINN would be extended to the unsteady forward problem, such as the flow over the cylinder. Nevertheless, as the concept of DD-PINN does not vary regardless of the flow unsteadiness, we believe that the DD-PINNs and their variants proposed in this study would still perform better than the data-free PINNs and leave it as a future study. Furthermore, the proposed PINNs would be tightly coupled with CFD simulations to collect guide data as in the study by \citet{jeon2024residual}, and finally extended to inverse design optimization applications, where the acceleration of flow analysis is a key issue in terms of computational cost \cite{yang2023inverse}. Lastly, although we have focused on investigating the flexibility and scalability of DD-PINNs from a macroscopic point of view, the next step may be more microscopic investigations on them, such as applying state-of-the-art PINN techniques such as weight adaptation, learning rate scheduling, and adaptive activation functions, etc \cite{wang2021understanding, jagtap2020locally, jagtap2020adaptive, mcclenny2020self, wight2020solving}.  

\clearpage
\section*{CRediT authorship contribution statement}
\textbf{S. Yang}: Conceptualization, Methodology, Software, Validation, Formal analysis, Investigation, Data Curation, Writing – Original Draft, Writing – Review \& Editing, Visualization.
\textbf{H. Kim}: Software, Data Curation, Writing – Review \& Editing.
\textbf{Y. Hong}: Conceptualization, Software, Writing – Review \& Editing.
\textbf{K. Yee}: Supervision, Funding acquisition.
\textbf{R. Maulik}: Supervision, Writing – Review \& Editing.
\textbf{N. Kang}: Supervision, Funding acquisition.

\section*{Declaration of competing interest}
The authors declare that they have no known competing financial interests or personal relationships that could have appeared to influence the work reported in this paper.

\section*{Data availability}
Data will be made available on request.

\section*{Acknowledgments}
This work was supported by the National Research Foundation of Korea
(2018R1A5A7025409), and the Ministry of Science and ICT of Korea
(No. 2022-0-00969). 
% \end{linenumbers}
\begin{appendices}

% \clearpage
\section{Importance of capturing vorticity in fluid problems}\label{sec:vorticity}

Vorticity is generated in regions where the flow properties change abruptly. To be more specific, it is concentrated in the region where vortex shedding occurs near the high angle of attack airfoil or in the tip vortices generated at the wingtips. These vortex flows generate induced velocity on the object itself so that when they interact with other objects, they not only affect aerodynamic performance but also have significant effects on subsequent structural and acoustic performance. Accordingly, numerous researchers have investigated numerical methods to accurately capture the strength of vortices with conventional CFD approaches by adding meshes in regions of high vorticity. To name a few, there were studies utilizing vorticity-related parameters such as Q-criterion or $\lambda_2$ \cite{kamkar2011feature, murayama2001simulation, jung2014assessment}. In their work, flow analysis is performed first with a grid constructed without prior knowledge on the flow field. Vorticity-related parameters are then used to identify regions of concentrated vorticity, and new grid points are added to these regions accordingly. Finally, the flow analysis is performed again with the updated grid and this iterative process is repeated until satisfactory results are obtained.

\clearpage
\section{Hyperparameter tuning results at \textit{Re=100\&3200}}
\label{sec:app_hyp}

\begin{table}[htb!]
\centering
    % \caption{Global caption}
    \begin{minipage}{.45\linewidth}
      % \caption{}
      \centering
        \renewcommand{\arraystretch}{1.2}
        \caption{RMSE of hyperparameter tuning in \textit{Re=100} case.}\label{tab:100hyp_results}
        \begin{tabular*}{\textwidth}{@{\extracolsep{\fill}}ccccc}
            \cline{1-5}
            \begin{tabular}[c]{@{}c@{}}$N_{layer}$\end{tabular} & \begin{tabular}[c]{@{}c@{}}$N_{node}$\end{tabular} & \begin{tabular}{@{}c@{}}$lr$\end{tabular} & \begin{tabular}[c]{@{}c@{}}$u$\end{tabular} & \begin{tabular}{@{}c@{}}$v$\end{tabular} \\ \cline{1-5}
            \multirow{9}{*}{4} & \multirow{3}{*}{32} & 1e-3 & 1.21e-2 & 4.60e-3 \\ \cline{3-5}
             &  & 1e-4 & 1.58e-2 & 1.20e-2 \\ \cline{3-5}
             &  & 1e-5 & 2.02e-2 & 1.97e-2 \\ \cline{2-5}
             & \multirow{3}{*}{64} & 1e-3 & 1.06e-2 & 1.60e-3 \\ \cline{3-5}
             &  & 1e-4 & 1.37e-2 & 8.00e-3 \\ \cline{3-5}
             &  & 1e-5 & 1.57e-2 & 1.16e-2 \\ \cline{2-5}
             & \multirow{3}{*}{128} & 1e-3 & 1.08e-2 & 1.92e-3 \\ \cline{3-5}
             &  & 1e-4 & 1.28e-2 & 6.22e-3 \\ \cline{3-5}
             &  & 1e-5 & 1.52e-2 & 1.10e-2 \\ \cline{1-5}
            \multirow{9}{*}{\textbf{6}} & \multirow{3}{*}{\textbf{32}} & \textbf{1e-3} & \textbf{1.01e-2} & \textbf{5.20e-4} \\ \cline{3-5}
             &  & 1e-4 & 1.12e-2 & 2.47e-3 \\ \cline{3-5}
             &  & 1e-5 & 1.78e-2 & 1.76e-2 \\ \cline{2-5}
             & \multirow{3}{*}{64} & 1e-3 & 1.09e-2 & 2.10e-3 \\ \cline{3-5}
             &  & 1e-4 & 1.16e-2 & 3.61e-3 \\ \cline{3-5}
             &  & 1e-5 & 1.56e-2 & 1.16e-2 \\ \cline{2-5}
             & \multirow{3}{*}{128} & 1e-3 & 1.05e-2 & 1.04e-3 \\ \cline{3-5}
             &  & 1e-4 & 1.21e-2 & 4.79e-3 \\ \cline{3-5}
             &  & 1e-5 & 1.49e-2 & 1.06e-2 \\ \cline{1-5}
            \multirow{9}{*}{8} & \multirow{3}{*}{32} & 1e-3 & \textbf{1.01e-2} & \textbf{5.20e-4} \\ \cline{3-5}
             &  & 1e-4 & 1.12e-2 & 2.47e-3 \\ \cline{3-5}
             &  & 1e-5 & 1.78e-2 & 1.76e-2 \\ \cline{2-5}
             & \multirow{3}{*}{64} & 1e-3 & 1.09e-2 & 2.10e-3 \\ \cline{3-5}
             &  & 1e-4 & 1.16e-2 & 3.61e-3 \\ \cline{3-5}
             &  & 1e-5 & 1.56e-2 & 1.16e-2 \\ \cline{2-5}
             & \multirow{3}{*}{128} & 1e-3 & 1.05e-2 & 1.04e-3 \\ \cline{3-5}
             &  & 1e-4 & 1.21e-2 & 4.79e-3 \\ \cline{3-5}
             &  & 1e-5 & 1.49e-2 & 1.06e-2 \\ \cline{1-5}
        \end{tabular*}

    \end{minipage}%
    \hspace{.05\linewidth}
    \begin{minipage}{.45\linewidth}
      \centering
        \renewcommand{\arraystretch}{1.2}
        \caption{RMSE of hyperparameter tuning in \textit{Re=3200} case.}\label{tab:3200hyp_results}

        \begin{tabular*}{\textwidth}{@{\extracolsep{\fill}}ccccc}
        \cline{1-5}
        \begin{tabular}[c]{@{}c@{}}$N_{layer}$\end{tabular} & \begin{tabular}[c]{@{}c@{}}$N_{node}$\end{tabular} & \begin{tabular}{@{}c@{}}$w_{data}$\end{tabular} & \begin{tabular}[c]{@{}c@{}}$u$\end{tabular} & \begin{tabular}{@{}c@{}}$v$\end{tabular} \\ \cline{1-5}
        \multirow{9}{*}{5} & \multirow{3}{*}{32} & 1 & 1.42e-1 & 1.16e-1 \\ \cline{3-5}
         &  & 5 & \textbf{1.35e-1} & 1.08e-1 \\ \cline{3-5}
         &  & 10 & 1.38e-1 & 1.06e-1 \\ \cline{2-5}
         & \multirow{3}{*}{64} & 1 & 1.41e-1 & 1.17e-1 \\ \cline{3-5}
         &  & 5 & 1.37e-1 & 1.06e-1 \\ \cline{3-5}
         &  & 10 & 1.37e-1 & 1.05e-1 \\ \cline{2-5}
         & \multirow{3}{*}{128} & 1 & 1.38e-1 & 1.15e-1 \\ \cline{3-5}
         &  & 5 & 1.37e-1 & 1.04e-1 \\ \cline{3-5}
         &  & 10 & 1.37e-1 & 1.04e-1 \\ \cline{1-5}
        \multirow{9}{*}{\textbf{7}} & \multirow{3}{*}{\textbf{32}} & 1 & 1.42e-1 & 1.19e-1 \\ \cline{3-5}
         &  & 5 & 1.38e-1 & 1.08e-1 \\ \cline{3-5}
         &  & \textbf{10} & 1.36e-1 & \textbf{1.03e-1} \\ \cline{2-5}
         & \multirow{3}{*}{64} & 1 & 1.44e-1 & 1.25e-1 \\ \cline{3-5}
         &  & 5 & 1.40e-1 & 1.05e-1 \\ \cline{3-5}
         &  & 10 & 1.37e-1 & 1.07e-1 \\ \cline{2-5}
         & \multirow{3}{*}{128} & 1 & 1.39e-1 & 1.12e-1 \\ \cline{3-5}
         &  & 5 & 1.39e-1 & 1.11e-1 \\ \cline{3-5}
         &  & 10 & 1.38e-1 & 1.05e-1 \\ \cline{1-5}
        \multirow{9}{*}{9} & \multirow{3}{*}{32} & 1 & 1.41e-1 & 1.18e-1 \\ \cline{3-5}
         &  & 5 & 1.36e-1 & 1.04e-1 \\ \cline{3-5}
         &  & 10 & 1.37e-1 & \textbf{1.03e-1} \\ \cline{2-5}
         & \multirow{3}{*}{64} & 1 & 1.39e-1 & 1.13e-1 \\ \cline{3-5}
         &  & 5 & 1.36e-1 & 1.08e-1 \\ \cline{3-5}
         &  & 10 & 1.35e-1 & 1.07e-1 \\ \cline{2-5}
         & \multirow{3}{*}{128} & 1 & 1.51e-1 & 1.34e-1 \\ \cline{3-5}
         &  & 5 & 1.36e-1 & 1.07e-1 \\ \cline{3-5}
         &  & 10 & 1.37e-1 & 1.06e-1 \\ \cline{1-5}
        \end{tabular*}
    \end{minipage} 
\end{table}

\section{Visualization of grids used in OpenFOAM and their validation}
\label{sec:openfoam}

% \begin{figure*}[htb!]
%     \centering
    
%         \includegraphics[width=.8\textwidth]{Grid_visualization.png}
        
%     \caption{Visualization of grids used for DD-PINNs}
%     \label{fig:Grid_visualization}
% \end{figure*}

\begin{figure*}[htb!]
    \begin{minipage}[t]{\textwidth}
        \centering
        \begin{subfigure}[t]{0.24\textwidth}
            \centering
            \includegraphics[trim=10 10 10 10,clip, width=\linewidth]{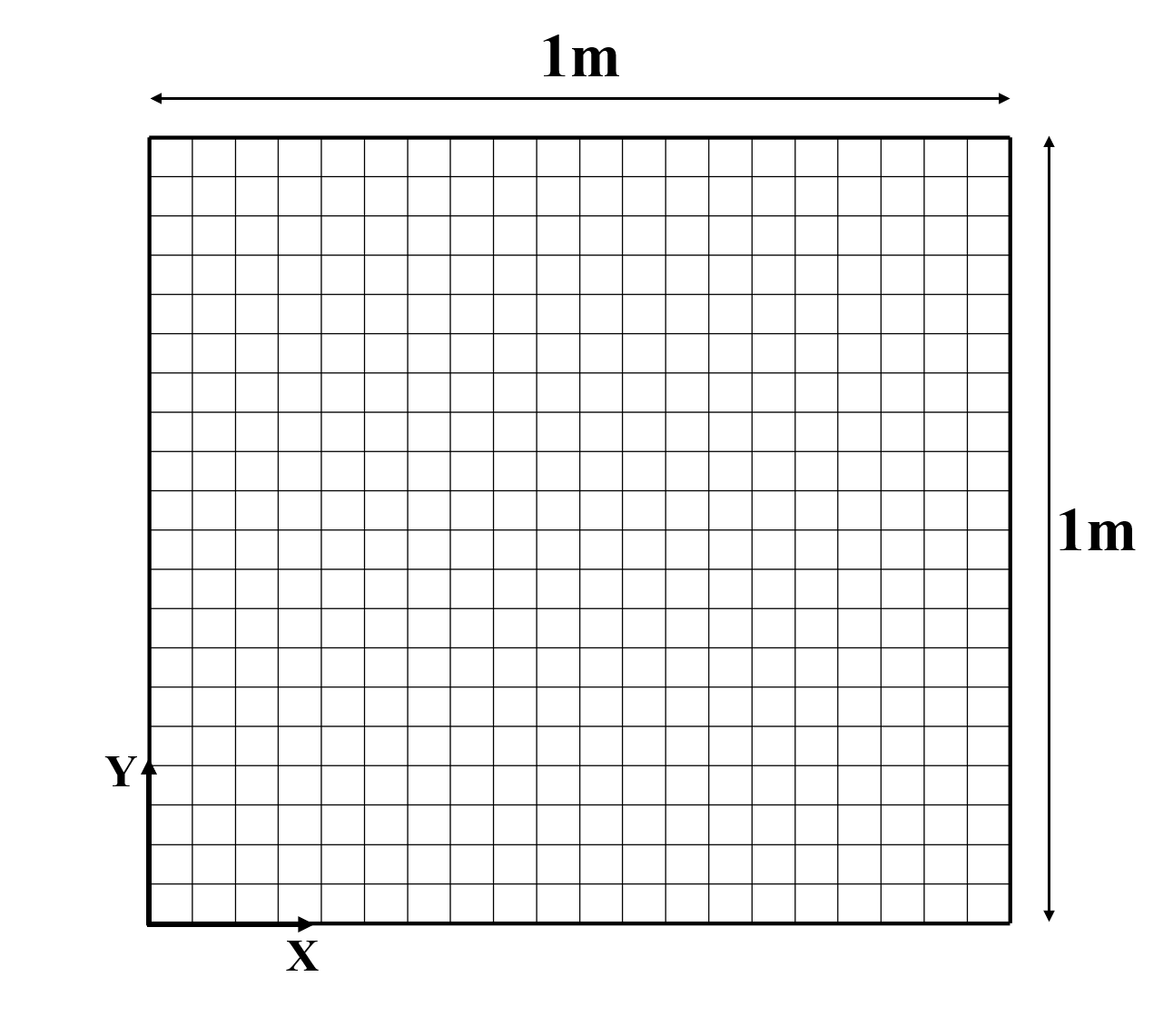}
            \caption{$20\times20$ grid}
            \label{fig:grid_a}
        \end{subfigure}
        \hfill
        \begin{subfigure}[t]{0.24\textwidth}
            \centering
            \includegraphics[trim=20 20 20 20,clip, width=\linewidth]{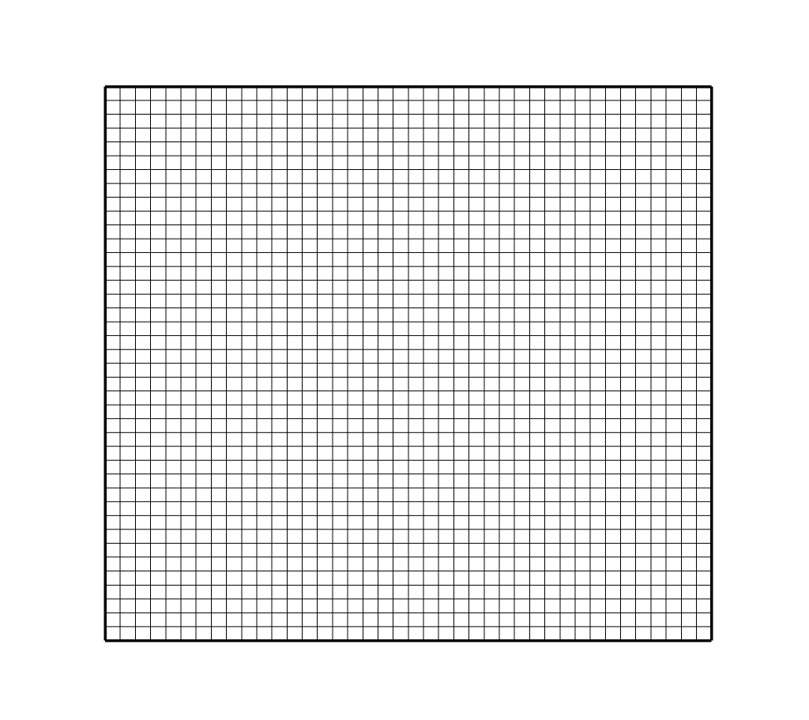}
            \caption{$40\times40$ grid}
            \label{fig:grid_b}
        \end{subfigure}
        \hfill
        \begin{subfigure}[t]{0.24\textwidth}
            \centering
            \includegraphics[trim=20 20 20 20,clip, width=\linewidth]{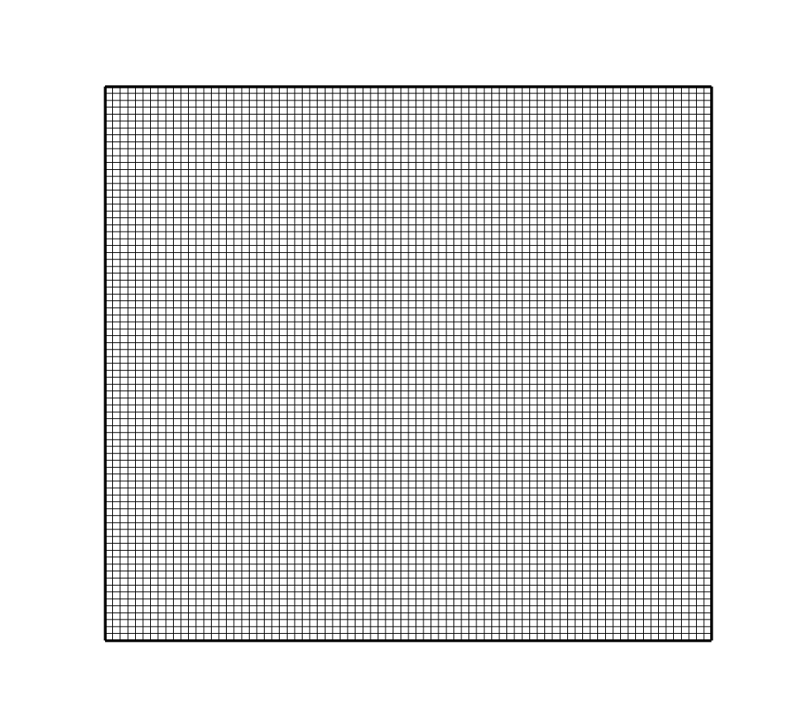}
            \caption{$80\times80$ grid}
            \label{fig:grid_c}
        \end{subfigure}
        \hfill
        \begin{subfigure}[t]{0.24\textwidth}
            \centering
            \includegraphics[trim=20 20 20 20,clip, width=\linewidth]{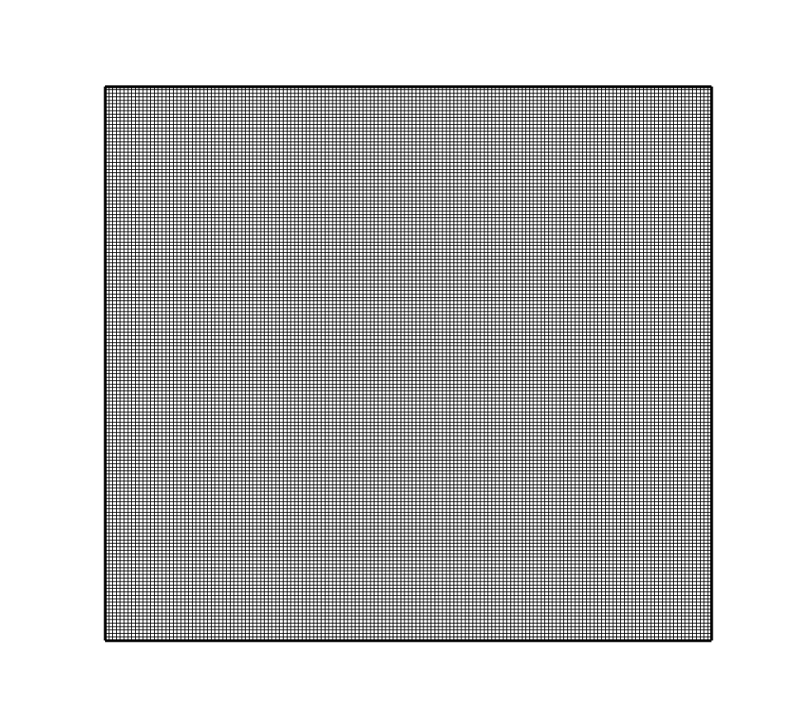}
            \caption{$160\times160$ grid}
            \label{fig:grid_d}
        \end{subfigure}
    \end{minipage}
    
    \caption{Visualization of grids used for OpenFOAM analysis. The results from these grids are also utilized for DD-PINNs as guide data.}
    \label{fig:grid_visualization}
\end{figure*}

\begin{figure*}[htb!]
    \begin{minipage}[t]{\textwidth}
        \centering
        \begin{subfigure}[t]{0.65\textwidth}
            \centering
            \includegraphics[width=\linewidth]{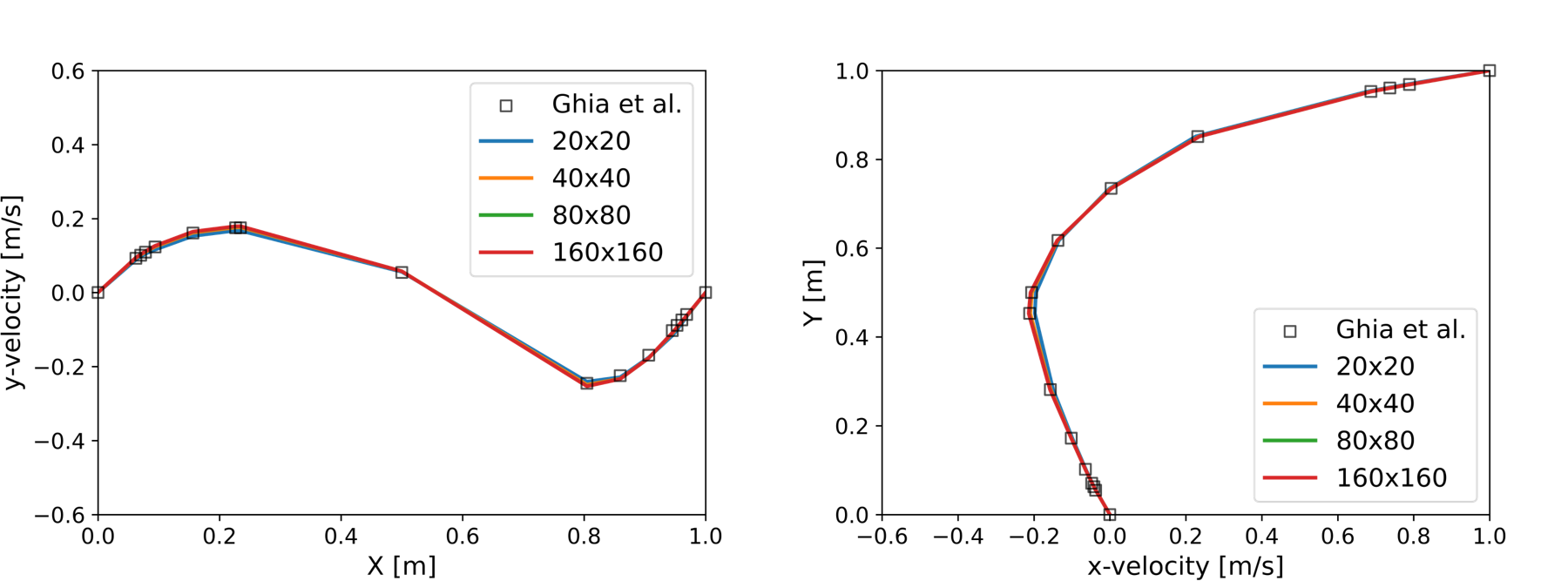}
            \caption{$Re=100$}
            \label{fig:DNS_a}
        \end{subfigure}
        
        \vfill
        
        \begin{subfigure}[t]{0.65\textwidth}
            \centering
            \includegraphics[width=\linewidth]{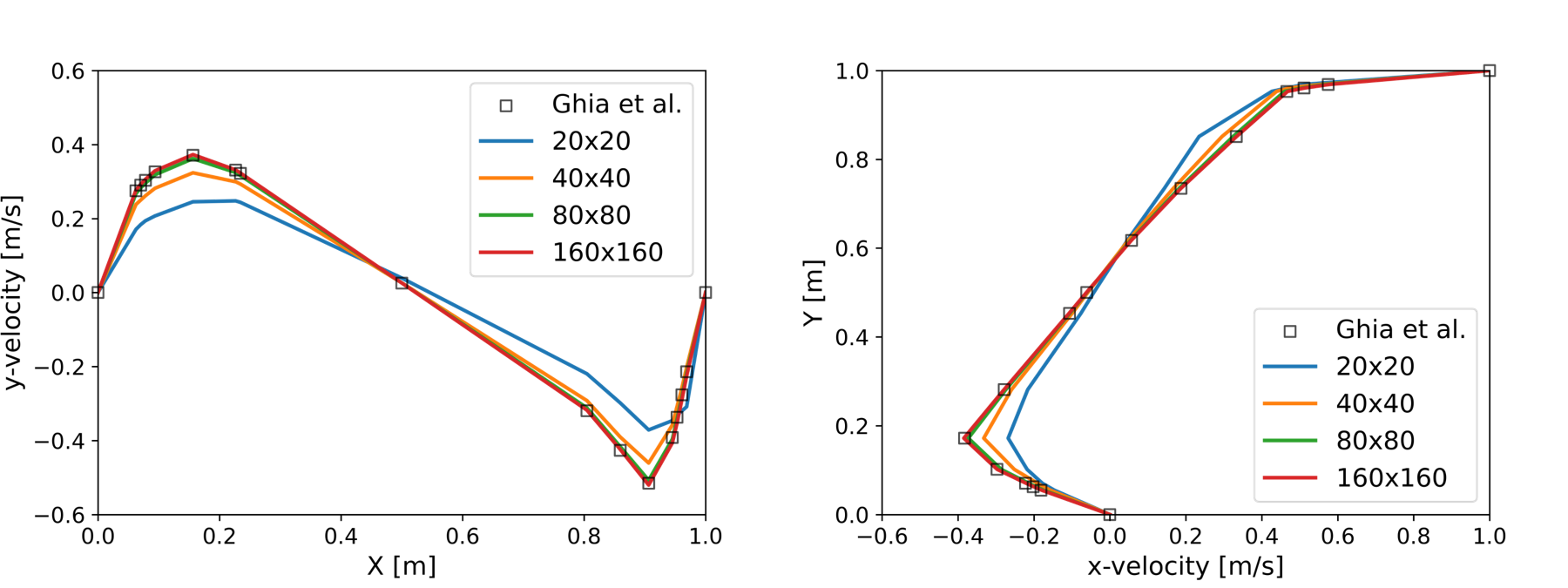}
            \caption{$Re=1000$}
            \label{fig:DNS_b}
        \end{subfigure}
       
        \vfill
        
        \centering
        \begin{subfigure}[t]{0.65\textwidth}
            \centering
            \includegraphics[ width=\linewidth]{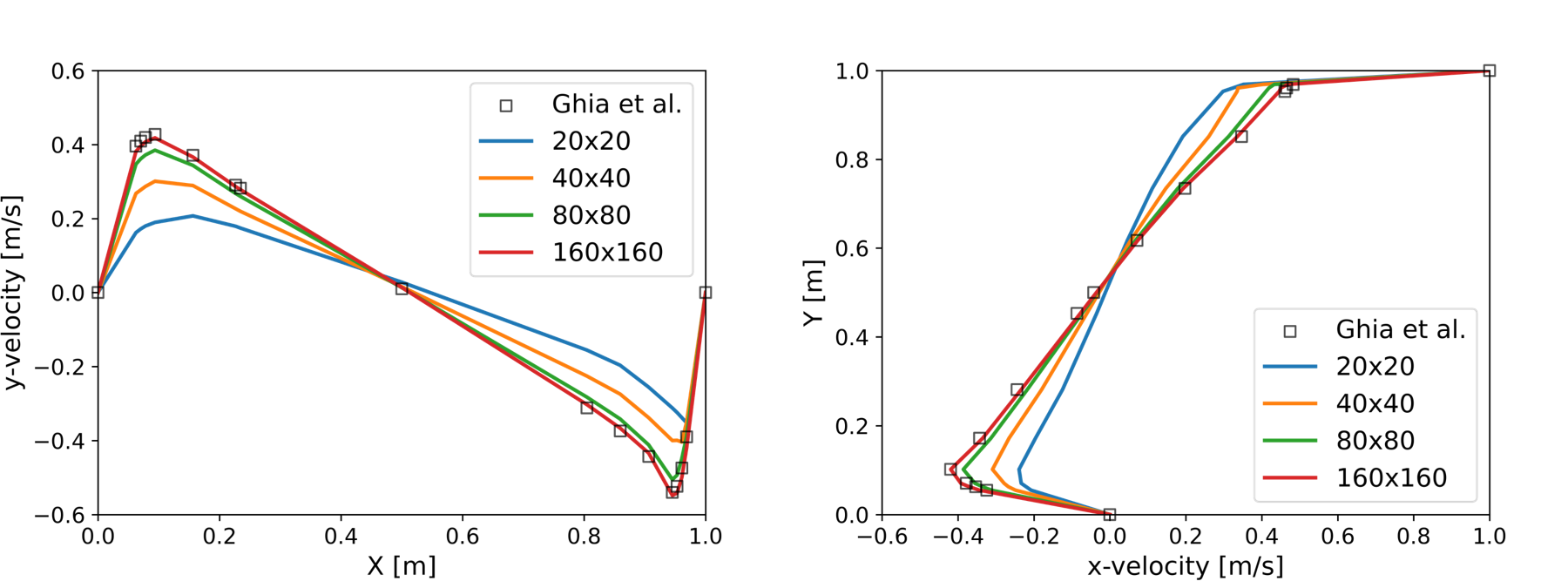}
            \caption{$Re=3200$}
            \label{fig:DNS_c}
        \end{subfigure}
        
    \end{minipage}
    
    \caption{Comparison of OpenFOAM results with \citet{ghia1982high} based on different grid resolutions: (left) y-velocity, (right) x-velocity. Note that each velocity profile is calculated along $y=0.5$ and $x=0.5$, respectively.}
    \label{fig:DNSvali}
\end{figure*}

For validation of CFD simulations, results from OpenFOAM are compared with the data from \citet{ghia1982high}. Uniform grids with four different densities are used as shown in Fig. \ref{fig:grid_visualization}. Then, comparisons of x-velocity profiles along the vertical center line ($x=0.5$) and y-velocity profiles along the horizontal center line ($y=0.5$) can be found in Fig. \ref{fig:DNSvali}. For $Re=100$, even the coarsest grid, $20\times20$, shows a satisfactory result. However, as the Reynolds number increases, results from coarser grids show bad correspondence with the validation data. This indicates that denser grids are required for better prediction of the flow field at higher Reynolds number regions where the thinning of boundary layers near walls becomes more dominant. In addition, at higher $Re$, a larger number of grids are required to accurately capture the near-linear velocity profile around the cavity center, where a uniform vorticity exists. For $Re=1,000$, $80\times80$ and $160\times160$ are indistinguishable from the validation data, while $40\times40$ shows almost the same trend. And for $Re=3,200$, both $80\times80$ and $160\times160$ show good agreement with the validation data.

\clearpage
\section{Additional experiments for explaining the best performance of DD-PINNs with random sampling}
\label{sec:app_adapt}

\begin{table}[htb!]
\centering
\caption{RMSE of y-velocity for different $N/M$ values in adaptive sampling: DD-PINNs at $Re = 3200$ (Section \ref{sec:samplingEffect}).}\label{tab:NM3200}
    \begin{NiceTabular*}{0.5\columnwidth}{@{\extracolsep{\fill}}c|cccc}
    \cline{1-5}
    \Block[c]{2-1}{} & \Block[c]{1-4}{$N/M$} \\ 
    & 0.2 & 0.4 & 0.6 & 0.8 \\ \cline{1-5}
    Van & \textbf{0.0509} 	&	\textbf{0.0509} 	&	\textbf{0.0509} 	&	\textbf{0.0509} \\
    RGV-100 & 0.0530 	&	0.0534 	&	0.0534 	&	0.0516  \\
    RGV-010 & 0.0512 	&	0.0539 	&	0.0529 	&	0.0526  \\
    RGV-001 & 0.0531 	&	0.0528 	&	0.0536 	&	0.0541  \\
    RGV-110 & 0.0525 	&	0.0563 	&	0.0518 	&	0.0520 \\
    RGV-101 & 0.0562 	&	0.0538 	&	0.0533 	&	0.0538 \\
    RGV-011 & 0.0531 	&	0.0544 	&	0.0560 	&	0.0529 \\
    RGV-111 & 0.0540 	&	0.0545 	&	0.0554 	&	0.0527 \\ \cline{1-5}
    % mean &	0.0530 	&	0.0538 	&	0.0534 	&	0.0526 \\
    % \cline{1-5}

    \end{NiceTabular*}
\end{table}

\renewcommand{\arraystretch}{1.2}
\begin{table}[htb!]
\centering
\caption{RMSE of y-velocity for different adaptive sampling techniques in DD-PINNs. Below the Reynolds numbers, \textit{tr} indicates Reynolds used for the training, while \textit{ext} and \textit{int} indicate Reynolds to be tested by extrapolation and interpolation, respectively. Again, as in Table \ref{tab:3200RMSE} and \ref{tab:NM3200}, the random sampling approach (Van) shows the best performance.} \label{tab:DGPINN_adapt}
    \begin{NiceTabular*}{0.85\columnwidth}{@{\extracolsep{\fill}}c|ccccccccc}
    \cline{1-10}
    \Block[c]{3-1}{} & \Block[c]{1-9}{Reynolds number} \\
    & $50$ & $100$ & $500$ & $1000$ & $2000$ & $3200$ & $4000$ & $5000$& $6000$\\
    & (\textit{ext}) & (\textbf{\textit{tr}}) & (\textit{int}) & (\textbf{\textit{tr}}) & (\textit{int}) & (\textbf{\textit{tr}}) & (\textit{ext}) & (\textit{ext}) & (\textit{ext}) \\ \cline{1-10}
    Van & 0.0229 	&	\textbf{0.0093} 	&	0.0367 	&	\textbf{0.0269} 	&	0.0454 	&	\textbf{0.0530} 	&	\textbf{0.0565} & \textbf{0.0589} & \textbf{0.0623} \\
    RGV-100 & 0.0229 	&	0.0096 	&	0.0333 	&	0.0282 	&	0.0533 	&	0.0566 	&	0.0577 &	0.0611 	&	0.0714  \\
    RGV-010 & 0.0216 	&	\textbf{0.0092} 	&	0.0279 	&	0.0289 	&	\textbf{0.0446} 	&	0.0550 	&	0.0611 &	0.0670 	&	0.0724 \\
    RGV-001 & 0.0198 	&	0.0097 	&	0.0400 	&	0.0310 	&	0.0506 	&	0.0579 	&	0.0614 &	0.0654 	&	0.0708 \\
    RGV-110 & 0.0196 	&	0.0112 	&	0.0297 	&	0.0298 	&	0.0470 	&	0.0578 	&	0.0634 &	0.0680 	&	0.0722 \\
    RGV-101 & \textbf{0.0185} 	&	0.0098 	&	\textbf{0.0270} 	&	0.0317 	&	0.0488 	&	0.0575 	&	0.0612 &	0.0633 	&	0.0645 \\
    RGV-011 & 0.0189 	&	0.0102 	&	0.0293 	&	0.0307 	&	0.0480 	&	0.0582 	&	0.0630 &	0.0666 	&	0.0693 \\
    RGV-111 & 0.0216 	&	0.0127 	&	0.0331 	&	0.0344 	&	0.0495 	&	0.0584 	&	0.061 &	0.0719 	&	0.0759 \\
    \cline{1-10}

    \end{NiceTabular*}
\end{table}

\section{Incorporating prior knowledge in DD-PINNs}
\label{sec:app_prior}

The incorporation of prior knowledge in the construction of the DD-PINN framework is crucial for enhancing the model's efficiency and accuracy and there can be three exemplary situations that embedding prior information into the training of DD-PINN can be realized. 1) If one already knows that there will be a vortex at the corners of the square domain (such as lid-driven cavity flow case in this study), one can strategically place sensors at the corners to obtain new guide datasets that can effectively update the DD-PINNs in terms of vortex phenomena. 2) Knowing the characteristics of the flow to be predicted in advance can allow one to apply the appropriate adaptive sampling technique that is effective in that situation. For example, if we know that the adverse velocity gradient occurs along the y-direction, we can focus on using gradient-aware sampling, which only considers the gradient of the velocity with respect to the y-direction. 3) If BCs and ICs are known before the training of DD-PINNs, their information can be incorporated directly rather than relying solely on a data-driven approach to learn the boundary and initial conditions. In fact, in Section \ref{sec:Param_results}, we employed hard BCs and have found that the application of hard BCs actually improves the accuracy of DD-PINNs. Leveraging prior information in these three key aspects enables targeted data acquisition, informed sampling strategies, and the integration of physical constraints, ultimately unlocking the full potential of DD-PINNs in the context of DT.

% \renewcommand{\arraystretch}{1.2}
% \begin{table}[htb!]
% \centering
% \caption{RMSE of y-velocity (v) for different adaptive sampling techniques in multi-fidelity DD-PINN. Below the Reynolds numbers, \textit{tr} indicates $Re$ used for the training, while \textit{ext} and \textit{int} indicate $Re$ values to be tested by extrapolation and interpolation, respectively.} \label{tab:MFDGPINN_adapt}
%     \begin{NiceTabular*}{0.65\columnwidth}{@{\extracolsep{\fill}}c|ccccccc}
%     \cline{1-8}
%     \Block[c]{3-1}{} & \Block[c]{1-7}{Reynolds number} \\
%     & $50$ & $100$ & $500$ & $1000$ & $2000$ & $3200$ & $4000$ \\
%     & (\textit{ext}) & (\textit{tr}) & (\textit{int}) & (\textit{tr}) & (\textit{int}) & (\textit{tr}) & (\textit{ext}) \\ \cline{1-8}
%     Van & 0.0214 	&	\textbf{0.0077} 	&	0.0272 	&	\textbf{0.0253} 	&	\textbf{0.0255} 	&	\textbf{0.0221} 	&	\textbf{0.0227}\\
%     RGV-100 & 0.0211 	&	0.0097 	&	0.0280 	&	0.0285 	&	0.0289 	&	0.0272 	&	0.0278 \\
%     RGV-010 & 0.0203 	&	0.0094 	&	\textbf{0.0240} 	&	0.0270 	&	0.0288 	&	0.0253 	&	0.0250 \\
%     RGV-001 & 0.0196 	&	0.0090 	&	0.0285 	&	0.0307 	&	0.0316 	&	0.0299 	&	0.0307 \\
%     RGV-110 & 0.0192 	&	0.0108 	&	0.0245 	&	0.0291 	&	0.0326 	&	0.0307 	&	0.0307 \\
%     RGV-101 & 0.0207 	&	0.0113 	&	0.0283 	&	0.0308 	&	0.0328 	&	0.0304 	&	0.0310 \\
%     RGV-011 & \textbf{0.0183} 	&	0.0108 	&	0.0254 	&	0.0296 	&	0.0320 	&	0.0307 	&	0.0313 \\
%     RGV-111 & 0.0206 	&	0.0116 	&	0.0287 	&	0.0312 	&	0.0330 	&	0.0314 	&	0.0325 \\
%     \cline{1-8}

%     \end{NiceTabular*}
% \end{table}

\end{appendices}
% \end{linenumbers}
\clearpage
%Bibliography
\bibliographystyle{unsrtnat}  
\bibliography{references}  

\end{document}